\documentclass[a4paper,11pt]{article}
\pdfoutput=1 

\usepackage{jheppub} 

\usepackage[T1]{fontenc} 

\usepackage{amsmath,amsfonts,amsbsy,amssymb,array,accents,dsfont}
\usepackage{enumerate,latexsym,graphicx}
\usepackage{xcolor,tikz}
    \usetikzlibrary{decorations.markings, positioning, shadows}
\usepackage{tcolorbox}
\usepackage{mathtools}
\usepackage{braket}
\usepackage{subfigure}
\usepackage{stmaryrd}
\usepackage{cleveref}

\newcommand{\llangle}{\left\langle\mkern-5mu\left\langle}
\newcommand{\rrangle}{\right\rangle\mkern-5mu\right\rangle}

\def\cA{\mathcal {A}}  \def\cC{\mathcal {C}}
  
  \def\cI{\mathcal {I}}
  
 \def\cN{\mathcal {N}} \def\cO{\mathcal {O}}

\newcommand{\beq}{\begin{equation}}
\newcommand{\eeq}{\end{equation}}
\newcommand{\bea}{\begin{eqnarray}}
\newcommand{\eea}{\end{eqnarray}}

\newcommand{\la}{\lambda}

\newcommand{\der}{\partial}

\newcommand{\nn}{\nonumber}

\newcommand{\n}{{\mathsf n}}
\newcommand{\m}{{\mathsf m}}

\usepackage{braket}
\usepackage{tikz}
\usetikzlibrary{matrix,calc,positioning,decorations.markings,decorations.pathmorphing,decorations.pathreplacing}
\usetikzlibrary{arrows,cd}
\usepackage{bbding}
\usetikzlibrary{positioning}
\tikzset{>=stealth}

\newcommand{\bz}{\bar{z}}

\newcommand{\qqquad}{\;, \quad\qquad}  

\newcommand{\RR}{\mathbb{R}}

\newcommand{\Ii}{\mathcal{I}}

\newcommand{\Oo}{\mathcal{O}}

\DeclareMathOperator{\Tr}{Tr}

\newcommand{\R}{\ensuremath{\mathbb{R}}}

\newcommand{\zb}{\bar{z}}

\newcommand{\xb}{\bar{x}}

\newcommand{\w}{\omega}

\usepackage{color}

\usepackage[normalem]{ulem}  

\definecolor{darkred}{rgb}{0.6,0,0}
\definecolor{darkblue}{rgb}{0,0,0.6}

\newcommand\p{\partial}

\newcommand{\be}{\begin{equation}}
\newcommand{\ee}{\end{equation}}

\usepackage[bbgreekl]{mathbbol}
\usepackage{amsfonts}
\DeclareSymbolFontAlphabet{\mathbb}{AMSb} 
\DeclareSymbolFontAlphabet{\mathbbl}{bbold} 

\newcommand{\G}{\Gamma} 

\title{\boldmath Knizhnik--Zamolodchikov equations and the analytic structure of strings in AdS$_3$
}

\author[a,b]{Sergio Iguri}
\author[c,d]{, Nicolas Kovensky}
\author[a,e]{and Juli\'an H.~Toro}

\affiliation[a]{Instituto de Astronomía y Física del Espacio (IAFE), CONICET - Universidad de Buenos Aires, Facultad de Ciencias Exactas y Naturales,  Ciudad Universitaria, 1428 Buenos Aires, Argentina.}
\affiliation[b]{Universidad Abierta Interamericana, Facultad de Arquitectura, Buenos Aires, Argentina.}
\affiliation[c]{Instituto de F\'isica de La Plata - CONICET, 
Diagonal 113 e/ 63 y 64, 1900 - La Plata, Argentina.}
\affiliation[d]{Departamento de Física, Facultad de Ciencias Exactas y Naturales, Universidad de Buenos Aires, Ciudad Universitaria, 1428 Buenos Aires, Argentina.}
\affiliation[e]{Instituto de Investigaciones Matemáticas Luis A. Santaló (IMAS), CONICET - Universidad de Buenos Aires, Facultad de Ciencias Exactas y Naturales, Ciudad Universitaria, 1428 Buenos Aires, Argentina.}
\emailAdd{siguri@iafe.uba.ar}
\emailAdd{nicolas.kovensky@iflp.unlp.edu.ar}
\emailAdd{jtoro@dm.uba.ar}

\abstract{ The pole structure of string correlators in AdS$_3$ encodes essential information about the AdS$_3$/CFT$_2$ duality. The corresponding residues for spectrally flowed three-point functions have been extracted from their exact expressions and shown to match the predictions of the holographic CFT. Away from the tensionless point, the latter is defined perturbatively in terms of a deformed symmetric orbifold. For $n$-point functions, such a matching has so far only been achieved for a specific subset of poles, and only by means of free-field techniques. Here we develop an alternative approach. We show that, starting from the localization properties of worldsheet  correlators, one can solve the Knizhnik--Zamolodchikov (KZ) equations on the corresponding locus and perform the integration over worldsheet moduli explicitly for any $n$ and for arbitrary values of the spectral flow charges. In this way we reproduce, directly from the exact worldsheet theory and including the normalization, the residues previously obtained with free-field methods. We expect this to provide a systematic framework for computing the remaining residues and, ultimately, for determining the full analytic structure of string amplitudes in AdS$_3$.
}

\newcommand{\of}[1]{\left(#1\right)}
\newcommand{\off}[1]{\left[#1\right]}
\newcommand{\offf}[1]{\left\{#1\right\}}
\newcommand{\vev}[1]{\left\langle #1\right\rangle}
\newcommand{\vvev}[1]{\llangle #1\rrangle}

\newif\ifshowsugg
\showsuggtrue
\newcounter{sugg}
\definecolor{sugold}{rgb}{0.75,0,0.6}
\definecolor{sugnew}{rgb}{0.85,0,0}
\definecolor{sugcom}{rgb}{0,0.25,0.75}

\begin{document}
\maketitle
\flushbottom


\section{Introduction}

Holography underlies many of the recent (and not so recent) advances that have allowed us to deepen our understanding of strongly coupled quantum field theories and of quantum gravity. At the quantitative level, our intuition comes mostly from the  realization of holographic dualities in string theory, which provides many concrete AdS$_{d+1}$/CFT$_{d}$ dual pairs. Through the detailed study of these examples, an overwhelming body of evidence indicates that strings propagating in anti-de Sitter (AdS) spacetimes and certain conformal field theories (CFTs) are but two sides of the same coin. Nevertheless, apart from a handful of topological examples, the AdS/CFT correspondence remains conjectural.

A full non-perturbative proof of AdS/CFT is currently out of reach, mainly due to the strong/weak nature of the duality and the limitations of our knowledge of string dynamics beyond perturbation theory. A slightly less ambitious, yet extremely impactful, goal would be to show that all correlation functions on both sides agree order by order in the genus expansion. A series of recent developments indicates that this may be achievable in the near future in the context of the AdS$_3$/CFT$_2$ duality, at least when the gravitational background is supported by Neveu--Schwarz (NSNS) fluxes.

In fact, in the so-called tensionless limit, such a holographic matching has essentially been demonstrated. Although in this regime the AdS$_3$ radius becomes string-sized and, at least for the AdS$_3\times S^3\times T^4$ case, the RNS formalism breaks down, the worldsheet theory can be formulated in terms of free fields using the hybrid formalism of \cite{Berkovits:1999im,Eberhardt:2018ouy}. The holographic CFT living on the boundary is a free theory of the symmetric orbifold type. The all-loop perturbative spectrum \cite{Gaberdiel:2018rqv, Eberhardt:2018ouy, Eberhardt:2019ywk} and the string partition function \cite{Eberhardt:2020bgq, Eberhardt:2021jvj} were computed on both sides and shown to match exactly. (To be precise, on the boundary one actually needs to consider a grand canonical ensemble of CFTs with different central charges \cite{Kim:2015gak}.) As for the worldsheet correlators, it was established in \cite{Eberhardt:2019ywk, Eberhardt:2020akk, Dei:2020zui, Knighton:2020kuh} that they localize on configurations where certain ramified holomorphic covering maps exist. These are, of course, the same covering maps used in the computation of correlation functions in symmetric orbifold CFTs \cite{Lunin:2000yv,Dei:2019iym}, and thus constitute one of the crucial ingredients in this remarkably explicit holographic duality. The genus-zero holographic matching for generic correlation functions was derived more recently in \cite{Dei:2023ivl} directly from the path integral.

The story is considerably more complicated away from the tensionless limit, and in particular in the supergravity regime, where the AdS$_3$ radius becomes large. From the worldsheet point of view, the spectrum is qualitatively different. Long strings are now allowed to move into the bulk along the radial direction, and short strings must also be included. It was proposed in \cite{Eberhardt:2021vsx} that short string states should be understood as bound states in the language of the holographic CFT, which is deformed by a non-trivial potential. The deforming operator also has twist two, which further breaks the symmetric orbifold structure. Correspondingly, on the worldsheet side, correlators do \textit{not} localize anymore, and spectrally flowed operators become quite difficult to handle \cite{Maldacena:2001km}. The so-called local Ward identities of \cite{Eberhardt:2019ywk}, combined with the $y$-basis formalism \cite{Dei:2021xgh,Iguri:2022eat}, have led to the exact derivation of two- and three-point functions \cite{Dei:2021xgh,Bufalini:2022toj}. This has allowed, for instance, the computation of the exact AdS$_3$/CFT$_2$ chiral ring in the supersymmetric model \cite{Iguri:2023khc}. The exact analysis is, however, very difficult to extend to higher-point functions. Still, in the four-point case it has led to the conjecture of \cite{Dei:2021yom}, later supported by the results of \cite{Iguri:2024yhb,Barone:2025vww} regarding the corresponding factorization structure.

A non-perturbative completion of the boundary theory proposed in \cite{Eberhardt:2021vsx} is still lacking. Given that general correlation functions are hard to compute exactly on both sides, the duality remains unproven. Nevertheless, this proposal has passed a series of highly non-trivial checks. In its simplest large-$\text{N}$ form, the proposal states that perturbative bosonic strings in AdS$_3$ with $k$ units of NSNS flux, times an internal compact manifold $X$, should be equivalent to a holographic CFT of the form\footnote{Interestingly, it was recently argued that the model in \eqref{HCFT def} can be derived from the long string partition function \cite{Knighton:2024pqh}, see also \cite{Seiberg:1999xz}.}
\begin{equation}
\label{HCFT def}
    {\rm Sym}^{\text{N}}(\R_Q \times X) \, + \, {\cal O}^2_{\alpha} \,\,  {\rm deformation} \,.
\end{equation}
Here $\R_Q$ represents a linear-dilaton-type CFT for a scalar field $\varphi$ with background charge $Q$, while ${\cal O}^2_{\alpha}$ is a twist-two operator exponential in $\varphi$ with momentum $\alpha$. The latter is chosen so that ${\cal O}^2_{\alpha}$ is marginal. Boundary correlators have simple poles in momentum space and, as happens in Liouville theory \cite{Zamolodchikov:1995aa}, the Coulomb gas description allows one to compute (some of) the corresponding residues. In \cite{Eberhardt:2021vsx}, it was shown that the exact analytic structure of the three-point functions of \cite{Dei:2021xgh,Bufalini:2022toj} was consistent with the boundary residues. A similar conclusion was reached for certain four-point functions in \cite{Dei:2022pkr}.

Given that exact formulas for worldsheet $n$-point functions with $n\geq 4$ are extremely hard to derive, it was proposed in \cite{Hikida:2023jyc,Knighton:2023mhq} that one could study the residues directly using free-field techniques. This is formulated in terms of the Wakimoto fields $\phi$, $\gamma$ and $\beta$, together with their barred counterparts. The scalar field $\phi$ describes the radial direction, and is closely related to the boundary field $\varphi$ \cite{Martinec:2021vpk}. In the near-boundary region, that is, at large $\phi$, the Wakimoto interaction can be ignored, hence $\gamma$ and $\beta$ are holomorphic. As shown in \cite{Knighton:2024qxd}, free-field correlators can then be studied using two different screening operators. One is the integrated version of the Wakimoto interaction, while the other is an operator with spectral flow charge $-1$, denoted by $D$. The authors of \cite{Knighton:2024qxd} make use of the fact that,  for $n$-point functions where the SL(2,$\R$) spins $j_i$ with $i=1,\dots,n$ satisfy
\begin{equation}
\label{j condition intro with m}
    \sum_{i=1}^n j_i = 1+\frac{k-2}{2}(n-\m-2) \, ,
\end{equation}
for some non-negative integer $\m$, only the second type of screening is involved, and the resulting path integral localizes on configurations where $\gamma(z)$ takes the form of one of the covering maps mentioned above. This map must have poles at the insertions of the screening $D$, and its ramification points correspond not only to the original vertex operators, but also to $\m$ additional twist-two boundary insertions. Computing the free-field correlator and further integrating over the worldsheet moduli (which can be quite non-trivial \cite{Maldacena:2001km,Barone:2025vww}) leads to results that precisely match the perturbative predictions from the holographic CFT \cite{Eberhardt:2021vsx}.

Boundary correlators, however, have a much richer analytic structure than that captured by Eq.~\eqref{j condition intro with m} \cite{Eberhardt:2021vsx}. The full set of poles is characterized by two non-negative integers, one of which is set to zero in the cases studied in \cite{Knighton:2024qxd}. This is because more general configurations force one to include the Wakimoto interaction in the analysis, which greatly complicates the path integral derivation. It is therefore necessary to devise an alternative method for computing the residues of string correlators in AdS$_3$. In this paper we propose that such a method can be developed based on two main ingredients:
\begin{enumerate}
    \item the localization property of exact correlators derived in \cite{Eberhardt:2019ywk}, and
    \item the Knizhnik--Zamolodchikov (KZ) equations of  the SL(2,$\R$) WZW model.
\end{enumerate}
It should be stressed that the KZ equations take a highly unconventional form for correlators involving spectrally flowed operators, since the latter are Virasoro primaries, but not affine primaries \cite{Ribault:2005ms,Dei:2021yom}.

As a proof of concept, we recover the results of \cite{Hikida:2023jyc,Knighton:2023mhq,Knighton:2024qxd} from the exact formalism, thus avoiding any reference to free-field representations, the path-integral formalism, or the near-boundary regime. This is done in three steps. First, we focus on $n$-point functions with arbitrary spectral flow charges which satisfy the condition in Eq.~\eqref{j condition intro with m} but with $\m=0$, i.e.
\begin{equation}
\label{j condition intro}
    \sum_{i=1}^n j_i = 1+\frac{k-2}{2}(n-2) \, .
\end{equation}
As discussed in \cite{Eberhardt:2019ywk}, correlators localize according to
\begin{align}
\Bigl\langle \prod_{i=1}^n V_{j_ih_i}^{\w_i}(x_i,z_i) \Bigr\rangle =\sum_{\Gamma} \prod_{i=1}^n |a_i|^{-2h_i} \prod_{i=4}^n \delta^{(2)}(x_i-\Gamma(z_i)) \, |W_\Gamma(z_1,\cdots,z_n)|^2 \,,
\label{eq:localisation solution intro}
\end{align}
where the sum runs over the discrete set of covering maps $\Gamma(z)$ defined in Sec.~\ref{sec: covering maps} below.
The authors of \cite{Eberhardt:2019ywk} showed that the ansatz \eqref{eq:localisation solution intro} solves the recursion relations derived from the local Ward identities for any function $W_\Gamma$ depending on the worldsheet insertion points (as well as on all $j_i$'s and $k$).
The precise form of $W_\Gamma$ was left unfixed.
Here we will show that the KZ equations can be used to fix this function completely. Up to a normalization constant, to be discussed in Sec.~\ref{sec: normalization}, the solution takes the form
\begin{equation}
\label{Final WGamma m=0}
\boxed{
W_\Gamma(z_1,\dots, z_n) = \frac{J_{\off{z_i \to \Gamma_i}}}{z_{12} z_{23} z_{13}} \, C_\Gamma^{\frac{k}{2}} \prod_{a=1}^N c_a^{-\frac{k}{2}} \prod_{i=1}^n a_i^{\frac{k}{4}(\w_i-1)}  \prod_{i<j} z_{ij}^{-\frac{2j_ij_j}{k-2}+j_i+j_j-\frac{k}{2}+1} \, .
}
\end{equation}
Here $C_\Gamma$, $c_a$ and $a_i$ are covering map data, while $J_{\off{z_i \to \Gamma_i}}$ is the Jacobian associated with the change of variables from $z_i$ to $\Gamma_i \equiv \Gamma(z_i)$ in the integration over worldsheet moduli. This concludes the first step. The second one consists of analyzing the factorization properties of this solution, which are shown to be fully consistent with the spacetime OPE \cite{Dei:2019iym}. This further allows us to fix the normalization constant recursively.
Finally, the third step is the extension to $\m >0$, which, as will be discussed below, can be achieved by inserting the worldsheet avatar of the deformation operator appearing in \eqref{HCFT def}.

This paper is organized as follows. Section \ref{sec: basic definitions} reviews the relevant properties and basic definitions for the SL(2,$\R$) model, bosonic strings in AdS$_3 \times X$ and the putative holographically dual CFT. We then introduce, in Section \ref{sec: maps and localization}, the main properties of the holomorphic covering maps employed in the computation of correlation functions of the worldsheet and boundary theories. We also re-derive the localization property obtained originally in \cite{Eberhardt:2019ywk} for correlators of the worldsheet CFT satisfying the constraint \eqref{j condition intro}, which are the main focus of this paper.

In Section \ref{sec: KZ w=0 y w=1} we begin our study of spectrally flowed $n$-point functions of the form \eqref{eq:localisation solution intro} based on the corresponding Knizhnik--Zamolodchikov equations. We first discuss how unflowed four-point functions described in \cite{Maldacena:2001km} can be encompassed in the present formalism by means of the series identifications \eqref{seriesidentifXbasis}. We then show how to turn KZ constraints for singly-flowed vertex operators into differential equations for $n$-point functions containing only these types of insertions, and obtain the solution for the function $W_\Gamma$ for this specific family of correlators, which makes the holographic matching of momentum-space residues manifest.

Sections \ref{sec: general derivation} and \ref{sec: general results} describe the main technical aspects of our method, and contain our main results. We first combine the localization property for correlators with arbitrary spectral flow charges with their KZ constraints in order to transform the latter into a series of differential equations in Sec.~\ref{sec: general derivation}. These equations are close relatives of those obtained in \cite{Dei:2019iym} in the context of symmetric orbifold CFTs, the main difference being given by a series of terms that contain an unknown function associated with the insertion of one of the symmetry currents into the primary correlator. We then derive this function from the symmetry constraints, and show that it plays a crucial role by introducing additional factors needed to deal with the presence of the reparametrization ghosts and the integration over the worldsheet moduli that define the final string amplitude. This allows us to derive the general solution for the functions $W_\Gamma$ appearing in \eqref{eq:localisation solution intro}, up to a normalization factor.

Finally, we fix the normalization by studying the spacetime factorization limit of string $n$-point functions in Sec.~\ref{sec: normalization}, and extend the analysis to all correlators satisfying the more general condition \eqref{j condition intro with m} in Sec.~\ref{sec: extension to m and matching}. Including these additional ingredients then leads to an exact holographic matching, which also agrees with the recent free-field analysis discussed in \cite{Hikida:2023jyc,Knighton:2023mhq,Knighton:2024qxd}. Our conclusions are presented in Sec.~\ref{sec: conclusions}, along with a broader discussion about future research directions and applications. Explicit examples of the factorization limit are
worked out in Appendix A.

\section{The worldsheet theory}
\label{sec: basic definitions}
We start by setting up the notation and introducing the basic definitions of the bosonic SL(2,$\R$) WZW model at level\footnote{See \cite{Eberhardt:2025sbi} for recent results on the $k=3$ case.} $k>3$. We then review the discussion of \cite{Eberhardt:2019ywk,Dei:2021xgh,Bufalini:2022toj} regarding the so-called local Ward identities, and the complicated constraints they imply for correlation functions involving spectrally flowed insertions. Finally, we introduce the string correlators of interest and the proposed holographic CFT.

\subsection{Basics and conventions for the SL(2,$\R$) WZW model}

The main building block of the worldsheet theory is the AdS$_3$ sector (see \cite{Kovensky:2026usc} for a recent review). We will be agnostic about the internal CFT describing the remaining compact directions, and concentrate on the holomorphic sector. The current algebra is captured by the operator product expansions (OPEs)
\begin{equation}
    J^-(z)J^+(w) \sim \frac{k}{(z-w)^2} + \frac{ 2 J^3(w)}{z-w}  \, , \quad
    J^3(z)J^\pm(w) \sim  \frac{ \pm J^\pm(w)}{z-w}  \, , \quad
    J^3(z)J^3(w) \sim  \frac{-k/2}{(z-w)^2}   \, ,
    \label{OPEjSL2}
\end{equation}
where $k$, the level of the Kac--Moody algebra, sets the squared AdS$_3$ radius in string units. The spectrum includes the usual affine primaries, and also states with non-trivial spectral flow charges $\w > 0$. In the basis that is most useful for holographic applications, they depend on the boundary coordinate $x$, the (unflowed) SL(2,$\R$) spin $j$, and the spacetime weight $h$, related to the spin projection before spectral flow $m$ by
\begin{equation}
\label{hdef}
     h = m + \frac{k}{2}\w \, .
\end{equation}
These states have weight
\begin{equation}
    \Delta = -\frac{j(j-1)}{k-2} - h\w + \frac{k}{4}\w^2\,,
\end{equation}
with respect to the Virasoro generators obtained from the affine current modes $J^a_n$ by means of the Sugawara construction. The values of $\w$, $j$ and $h$ (or equivalently $m$ if $\w$ is nonzero) depend on the SL(2,$\R$) representation. Long strings belong to the continuous representations ${\cal{C}}_j^{\alpha,\w}$. They are physical for $\w>0$, $j \in \frac{1}{2} + i \R$ and $m \in \alpha + \mathbb{Z}$ for some $\alpha \in [0,1)$. Short strings belong to the discrete highest/lowest-weight representations ${\cal{D}}_j^{\pm,\w}$ where $j$ takes real values. These have $\w \geq 0$ and  $m = \pm(j + n)$ with $n \in \mathbb{N}_0$.

The corresponding vertex operators, which will be denoted by $V^{\w}_{jh}(x,z)$, where $z$ is the worldsheet coordinate, are defined by OPEs of the form
\begin{subequations}
\begin{eqnarray}
   J^+(w)  V^{\w}_{jh}(x,z) &\sim &
   \sum_{n=1}^{\w+1} \frac{\left(J_{n-1}^+
    V_{jh}^\w\right) (x,z) }{(w-z)^n}
     \, , \\ [1ex]
    J^3(w)  V^{\w}_{jh}(x,z) &\sim &
    x \sum_{n=2}^{\w+1} \frac{\left(J_{n-1}^+
    V_{jh}^\w\right) (x,z) }{(w-z)^n}
    +\frac{\left[J_{0}^3,
    V_{jh}^\w(x,z)\right]
      }{(w-z)}
    \, , \\ [1ex]
    J^-(w)  V^{\w}_{jh}(x,z) &\sim &
    x^2 \sum_{n=2}^{\w+1} \frac{\left(J_{n-1}^+
    V_{jh}^\w\right) (x,z) }{(w-z)^n}
    +\frac{\left[J_{0}^-,
    V_{jh}^\w(x,z)\right]
      }{(w-z)}
      \, ,
\end{eqnarray}
\label{JVxOPE}
\end{subequations}
up to regular terms. Here the zero modes act as differential operators in $x$,
\begin{subequations}
\begin{eqnarray}
    \left(J_{0}^+V_{jh}^\w\right) (x,z) &=& \der_x V_{jh}^\w (x,z) \, ,
    \label{J0+dx}\\[1ex]
    \left[J_{0}^3,
    V_{jh}^\w(x,z)\right] &=& (x\der_x+h) V_{jh}^\w (x,z) \, , \\[1ex]
    \left[J_{0}^-,
    V_{jh}^\w(x,z)\right] &=& (x^2\der_x + 2hx) V_{jh}^\w (x,z) \, ,
\end{eqnarray}
\label{diffopsx}
\end{subequations}
\hspace{-0.2cm} while\footnote{Here we follow the conventions of \cite{Eberhardt:2019ywk,Knighton:2023mhq,Knighton:2024qxd}, where, compared to  \cite{Dei:2021xgh,Dei:2022pkr,Bufalini:2022toj}, a reflection $j \to 1-j$ is implemented in the flowed sector.}
\begin{equation}
    \left(J_{\pm \w}^\pm V_{jh}^\w\right) (x,z) = \left(h-\frac{k}{2} \w \pm j \right) V_{j,h\pm 1}^\w (x,z)  \, .
    \label{JwVx}
\end{equation}

It will also be useful to translate the currents in $x$-space by defining
\begin{equation}
J^+(x,z)=J^+(z)
\,, \quad
    J^3(x,z) = J^3(z) - x J^+(z) \, , \quad J^-(x,z) = J^-(z) - 2 x J^3(z)+x^2 J^+(z) \, .
\label{defJx}
\end{equation}
From the so-called $m$-basis perspective one finds that
\begin{subequations}
\begin{eqnarray}
    J^3(x,w)  V^{\w}_{jh}(x,z) & = &
    \frac{h
      }{(w-z)} V_{jh}^\w(x,z) + \cdots
    \, , \label{J3xOPE}\\ [1ex]
    J^-(x,w)  V^{\w}_{jh}(x,z) & = &
    (w-z)^{\w-1} \left(J_{-\w}^-V_{jh}^\w\right) (x,z) + \cdots
    \, ,
    \label{JmxOPE}
\end{eqnarray}
\end{subequations}
where the ellipsis indicates higher order terms in powers of $(z-w)$.

It turns out that not all spectrally flowed sectors are independent. More explicitly, flowed highest/lowest-weight states with adjacent spectral flow charges are identified according to \cite{Maldacena:2001km}
\begin{equation}
    V_{j,h= -j + \frac{k}{2}\w}^{\w} (x,z) \sim V_{\frac{k}{2}-j,h  = \frac{k}{2}-j + \frac{k}{2}(\w-1)}^{\w-1}(x,z) \, , \qquad \w\geq 1.
    \label{seriesidentifXbasis}
\end{equation}
These so-called series identifications were crucial in deriving the exact formula for three-point functions with arbitrary spectral flow charges \cite{Bufalini:2022toj}.
Additionally, there is a reflection symmetry in the SL(2,$\R$) WZW model under $j \to 1-j$. For $x$-basis states in the unflowed sector, this reads \begin{equation}
    V_{j}(x,z) = B(j) \int d^{2}x' \, |x-x'|^{-4j}V_{1-j}(x',z) \,,
    \label{Reflection unflowed}
\end{equation}
while for spectrally flowed operators one simply has
\begin{equation}
V_{jh}^\w(x,z) = R(1-j,h,\w) V_{1-j,h}^\w(x,z) \, .
    \label{Reflection flowed}
\end{equation}
The proportionality constants introduced in Eqs.~\eqref{Reflection unflowed}--\eqref{Reflection flowed} are defined as \begin{equation}
    B(j)=\frac{2j-1}{\pi}
    \frac{\Gamma[1-b^2(2j-1)]}{\Gamma[1+b^2(2j-1)]} \, \nu^{1-2j} \, ,  \quad  b^2 = (k-2)^{-1}
     \label{defBj} \, ,
\end{equation}
where $\nu$ can be taken as a free parameter,  and
\begin{equation}
\label{def N and R}
       R(j,h,\w) = \frac{ \pi \gamma \left(h-\frac{k}{2}\w+j\right) B(j)}{\gamma(2j) \gamma\left(h-\frac{k}{2}\w+1-j\right)}  \, ,
\end{equation}
with $\gamma(x) \equiv \Gamma(x)/\Gamma(1-\bar{x})$. The parameter $\nu$ can be absorbed into a rescaling of the vertex operators and of the string coupling, and is thus arbitrary, see \cite{Teschner:1999ug,Dabholkar:2007ey,Eberhardt:2021vsx}.

\subsection{Local Ward identities}

\label{sec:recursions}

Correlation functions involving spectrally flowed insertions must satisfy a highly non-trivial set of constraints \cite{Maldacena:2001km,Eberhardt:2019ywk}. To begin with, one has selection rules for $n$-point functions of the form
\begin{equation}
\label{defF}
    F =\vev{\prod_{j=1}^n V_{j_j,h_j}^{\w_j}(x_j,z_j)} \, ,
\end{equation}
given by
\begin{equation}
\label{selection rules}
    \w_i - \sum_{j\neq i}\w_j \leq n-2  \qquad \forall \, i = 1, \dots n \, .
\end{equation}
In other words, correlators that do not satisfy this condition must vanish.
We also have the usual global Ward identities, which encode conformal covariance in both $x$ and $z$.  Finally, there are also the so-called local Ward identities. In order to formulate them, we define
\begin{equation}
    F_\ell^{a,i} =\left\langle
    \left(J_\ell^aV_{j_i,h_i}^{\w_i}\right)(x_i,z_i)
    \prod_{j\neq i} V_{j_j,h_j}^{\w_j}(x_j,z_j)\right\rangle,
    \label{defFin}
\end{equation}
so that, in particular,
\begin{equation}
    F_0^{+,i} = \der_{x_i} F\, , \qquad
    F_0^{3,i} = h_i F\, ,
    \label{F0+ and F03}
\end{equation}
and
\begin{equation}
    F_{\pm \w_i}^{\pm,i} = \left(h_i-\frac{k}{2} \w_i \pm j_i\right) \left\langle V_{j_i, h_i\pm 1}^{\w_i}(x_i,z_i) \prod_{j\neq i} V_{j_j,h_j}^{\w_j}(x_j,z_j)\right\rangle\,,
    \label{Fwi}
\end{equation}
see Eqs.~\eqref{J0+dx}, \eqref{JwVx} and \eqref{J3xOPE}. Now, the $F^{+,i}_\ell$ with $\ell=1,\dots, \w_i-1$ involve terms in the OPEs \eqref{JVxOPE} which have no simple expression in terms of primary operators and their derivatives.

The $JV$ OPEs imply that, upon inserting the currents, one has
\begin{subequations}
\label{currents OPEs inside correlators}
\begin{eqnarray}
    \left\langle J^+(z) \prod_{j=1}^n V_{j_j,h_j}^{\w_j}(x_j,z_j)\right\rangle &=& \sum_{i=1}^n \left[
    \frac{\der_{x_i} F}{z-z_i} + \sum_{\ell=1}^{\w_i}
    \frac{F_\ell^{+,i}}{(z-z_i)^{\ell+1}}
    \right] \, , \\[1ex]
    \left\langle J^3(z) \prod_{j=1}^n V_{j_j,h_j}^{\w_j}(x_j,z_j)\right\rangle &=& \sum_{i=1}^n \left[
    \frac{(h_i+x_i\der_{x_i}) F}{z-z_i} + \sum_{\ell=1}^{\w_i}
    \frac{x_i F_\ell^{+,i}}{(z-z_i)^{\ell+1}}
    \right]\, , \\[1ex]
    \left\langle J^-(z) \prod_{j=1}^n V_{j_j,h_j}^{\w_j}(x_j,z_j)\right\rangle &=& \sum_{i=1}^n \left[
    \frac{(2h_i x_i + x_i^2\der_{x_i} )F}{z-z_i} + \sum_{\ell=1}^{\w_i}
    \frac{x_i^2 F_\ell^{+,i}}{(z-z_i)^{\ell+1}}
    \right].
\end{eqnarray}
\end{subequations}
Combining these expressions gives
\begin{equation}
\label{Gjdef}
    G_j(z) \equiv \left\langle J^-(x_j,z) \prod_{i=1}^n V_{j_i,h_i}^{\w_i}(x_i,z_i)\right\rangle =
    \sum_{i\neq j} \left[
    \frac{(2h_i x_{ij} + x_{ij}^2\der_{x_i} )F}{z-z_i} + \sum_{\ell=1}^{\w_i}
    \frac{x_{ij}^2 F_{\ell}^{+,i}}{(z-z_i)^{\ell+1}}
    \right]\,,
\end{equation}
with  $x_{ij} = x_i - x_j$. However, the function $G_j(z)$ is severely restricted by Eq.~\eqref{JmxOPE}. More explicitly, as $z \to z_j$ it must behave as
\begin{equation}
     G_j(z) = (z-z_j)^{\w_j-1} F_{-\w_j}^{-,j} + {\cal{O}}\off{(z-z_j)^{\w_j}} \, .
    \label{Gjconditions}
\end{equation}
These regularity conditions for the functions $(z-z_j)^{1-\w_j}G_j(z)$ at $z=z_j$ for all $j=1,\dots,n$ provide enough information to solve for all the unknown $F_\ell^{+,i} $ in terms of $F$, its $x_i$-derivatives, and finally $F_{\w_i}^{+,i}$. Further inserting the resulting expressions into Eq.~\eqref{Gjconditions} leads, in principle, to an additional set of $n$ conditions which take the form of recursion relations between primary correlators with shifted values of $h_i$ and their $x_i$-derivatives.

These recursion relations are hard to derive in full generality for arbitrary $n$. Nevertheless, this has been achieved for three-point functions, leading to a fully explicit (integral) formula for exact correlators with arbitrary spectral flow charges \cite{Dei:2021xgh,Iguri:2022eat,Bufalini:2022toj}. This is best done by working in what is known as the $y$-basis, which combines operators with different values of $h$ into a formal power series by introducing an auxiliary complex variable $y$. This turns the recursion relations into first-order linear differential equations, which can then be solved explicitly. Analogous techniques have also led to the conjecture of \cite{Dei:2021yom} regarding the relation between the flowed conformal blocks and the unflowed ones, for which additional evidence was collected in \cite{Iguri:2024yhb,Barone:2025vww}. In both cases, the resulting expressions involve a crucial ingredient: ramified holomorphic covering maps from the worldsheet to the AdS$_3$ boundary. These are discussed in detail in Sec.~\ref{sec: covering maps} below.

\subsection{String correlators in AdS$_3$ and holographic CFT}
\label{sec: string correlators def}

The SL(2,$\R$) WZW model, which has central charge
\begin{equation}
    c = \frac{3k}{k-2} \, ,
\end{equation}
is the main ingredient for describing bosonic string propagation in AdS$_3\times X$. Here we  remain agnostic about the internal manifold $X$, and simply assume that the corresponding compact CFT has central charge $c_X = 26 - c$.  This guarantees that, once the $bc$ ghost system is included, the total central charge vanishes. Worldsheet vertex operators are then built by multiplying the SL(2,$\R$) ones and those of the CFT on $X$, and the Virasoro condition reads
\begin{equation}
\label{Virasoro cond general}
    -\frac{j(j-1)}{k-2} - h \w + \frac{k}{4}\w^2 + \Delta_X = 1 \, .
\end{equation}
For $\w\neq0$, solving for $h$ gives the spacetime weight
\begin{equation}
    h = -\frac{j(j-1)}{\w (k-2)} + \frac{k}{4}\w + \frac{\Delta_X- 1}{\w} \, .
\end{equation}
Perturbative string amplitudes are computed by including the ghost contributions and  integrating over the worldsheet moduli. Here we focus on the genus-zero case, where these moduli are simply the worldsheet insertion points, or, more precisely, $n-3$ of them. For simplicity, we consider only operators that are trivial in the internal sector. The leading-order amplitude then reads
\begin{equation}
    {\cal A}_n^{\rm string} = C_{S^2} \int d^2z_4 \dots d^2z_n \vev{\prod_{i=1}^3 c(z_i)\bar{c}(\zb_i)V_{j_i,h_i}^{\w_i}(x_i,\xb_i,z_i,\zb_i) \prod_{j\geq 4} V_{j_j,h_j}^{\w_j}(x_j,\xb_j,z_j,\zb_j)}\,.
\end{equation}
where $C_{S^2}$ is the normalization of the string path integral.

By defining\footnote{Here we follow the conventions of \cite{Knighton:2024qxd}, which are slightly different from the ones used originally  in \cite{Eberhardt:2021vsx}.}
\begin{equation}
\label{qi def}
     Q = \sqrt{2}\of{b-b^{-1}} = (3-k)\sqrt{\frac{2}{k-2}} \, , \qquad q  = \sqrt{\frac{2}{k-2}}\of{j+1-\frac{k}{2}}\, ,
\end{equation}
one can rewrite this as
\begin{equation}
    h = \frac{k(\w^2-1)}{4\w} + \frac{\Delta_X}{\w} + \frac{q(Q-q)}{2\w} \, .
\end{equation}
Moreover, for states in the continuous representations (which have $\w \geq 1$)  $j \in \frac{1}{2}+i\R$, and the spectrum is symmetric under $j \leftrightarrow 1-j$. In terms of $q$, this translates into
\begin{equation}
    q \in \frac{Q}{2}+i\R \,, \qquad q \leftrightarrow Q-q \, .
\end{equation}
This shows that the long string spectrum coincides with that of a symmetric orbifold CFT of the form
\begin{equation}
\label{sym orbifold}
        {\rm Sym}^\text{N}(\R_Q \times X)\,,
\end{equation}
where $\text{N}$ is taken to infinity, and $\w$ is identified with the twist. This has central charge
\begin{equation}
    c_{\rm st} = \text{N} (1+3Q^2 + c_X) = 6 \text{N} k \,.
\end{equation}
However, as can be seen directly from the worldsheet, spacetime correlators are \textit{not} of the symmetric orbifold type. This has led to the proposal of \cite{Eberhardt:2021vsx}, according to which, at least at the perturbative level, the holographic CFT can be understood as a perturbation of \eqref{sym orbifold}, generated by a marginal twist-two operator ${\cal O}^2_{\alpha}$ with  momentum
\begin{equation}
\label{def alpha}
\alpha = q\of{j=k-2} = \sqrt{(k-2)/2} \, .
\end{equation}
The associated coupling constant will be denoted by $\mu$. This leads to the definition introduced in Eq.~\eqref{HCFT def}. On top of breaking the orbifold structure, the presence of ${\cal O}^2_{\alpha}$ creates an exponential wall for the non-compact scalar field $\varphi$. Short string states are then interpreted as bound states in this language.

Let us briefly discuss correlators in this context. We are interested in $n$-point functions of the form
\begin{equation}
{\cal A}_n^{\rm HCFT} =
\vev{{\cal O}_{q_1}^{\w_1}(x_1,\xb_1) \cdots {\cal O}_{q_n}^{\w_n}(x_n,\xb_n)}  \, ,
\end{equation}
where each operator ${\cal O}_{q}^{\w}$ is constructed by combining an exponential in $\varphi$ with charge $q$, a twist operator $\sigma_\w$, and an identity operator in $X$. The fact that, besides the twist $\w$,  there is only one quantum number $q$ as opposed to two, namely $j$ and $h$, is consistent with the relation imposed by the Virasoro condition \eqref{Virasoro cond general}. Now, as happens in Liouville theory \cite{Zamolodchikov:1995aa}, one can argue that, if a suitable non-perturbative definition of this holographic CFT exists, the resulting correlation functions must have poles when a charge conservation condition of the form
\begin{equation}
\label{charge conservation}
    \sum_{i=1}^n q_i = Q - \m \, \alpha
\end{equation}
is satisfied for some non-negative integer $\m$. Using \eqref{qi def} and the definition of $\alpha$, this is easily shown to be equivalent to Eq.~\eqref{j condition intro with m}.
The corresponding residues can be computed using the Coulomb gas formalism, giving (omitting the antiholomorphic dependence)
\begin{equation}
\label{HCFT residues def}
    \mathop{\mathrm{Res}}_{\sum_{i=1}^nq_i =Q- \m \, \alpha}
    \vev{\prod_{i=1}^n{\cal O}_{q_i}^{\w_i}(x_i)}_\mu = \frac{(-\mu)^\m}{\m!} \int d^2 \xi_1 \dots d^2\xi_\m \vev{\prod_{i=1}^n{\cal O}_{q_i}^{\w_i}(x_i) \prod_{p=1}^\m {\cal O}^2_{\alpha}(\xi_p)}_{\mu=0}'\,,
\end{equation}
where the expectation value on the RHS is computed in the \textit{undeformed} theory, and the prime indicates that the momentum-conservation delta function is stripped off.

At this level, the statement of the holographic duality is that, once the relative normalizations are taken into account, the pole structures of ${\cal A}_n^{\rm string}$ and ${\cal A}_n^{\rm HCFT}$, together with the corresponding residues, should match.  In this paper we show how to derive this matching, i.e., how to reproduce \eqref{HCFT residues def} directly from the symmetry constraints of the worldsheet theory (without using free-field methods) for all values of $n$ and for arbitrary spectral flow charges. In particular, when charge conservation holds with $\m=0$, the deformation can be ignored for computing the residue.
In the worldsheet language, these cases correspond to correlators satisfying Eq.~\eqref{j condition intro}. At large $\text{N}$, the covering space method \cite{Lunin:2000yv,Dei:2019iym} then leads to
\begin{equation}
\label{HCFT correlator m=0 unfixed}
        \vev{\prod_{i=1}^n{\cal O}_{q_i}^{\w_i}(x_i)}_{\mu=0}'=\text{N}^{1-\frac{n}{2}}\prod_{i=1}^n\w_i^{\frac{1}{2}-\frac{k}{2}(\w_i+1)} \sum_\Gamma \Big|C_\Gamma^{\frac{k}{2}} \prod_{a=1}^N c_a^{-\frac{k}{2}} \prod_{i=1}^n  a_i^{\frac{k}{4}(\w_i-1)-h_i}  \prod_{i<j} z_{ij}^{-q_iq_j}\Big|^2 \, .
\end{equation}
The sum runs over the possible connected covering maps $\Gamma(z)$ which behave appropriately near each of the insertion points, $z$ being the coordinate on the covering space. In each of these contributions, the values of the auxiliary variables $z_i$ are defined by the condition $\Gamma(z_i) = x_i$. The coefficients $c_a$ and $a_i$, which will be defined shortly, are also obtained from the function $\Gamma(z)$. The value of $N$ is given below in Eq.~\eqref{Ndef}.
Of course, one may also fix the first three insertion points $(z_1,z_2,z_3)$ and $(x_1,x_2,x_3)$ to $(0,1,\infty)$, giving
\begin{equation}
\label{HCFT correlator m=0 fixed}
\begin{aligned}
        &\vev{
     {\cal O}_{q_1}^{\w_1}(0){\cal O}_{q_2}^{\w_2}(1){\cal O}_{q_3}^{\w_3}(\infty)\prod_{i=4}^n{\cal O}_{q_i}^{\w_i}(x_i)}_{\mu=0}' = \\&\quad\text{N}^{1-\frac{n}{2}}\w_3^{\frac{1}{2}+\frac{k}{2}(\w_3+1)} \prod_{i=1,i\neq3}^n\w_i^{\frac{1}{2}-\frac{k}{2}(\w_i+1)} \sum_\Gamma \Big| \prod_{a=1}^{N-\w_3} c_a^{-\frac{k}{2}} \prod_{i=1}^n  a_i^{\frac{k}{4}(\w_i-1)-h_i}  \prod_{i<j} z_{ij}^{-q_iq_j}\Big|^2 \, .
\end{aligned}
\end{equation}
A formula analogous to \eqref{HCFT correlator m=0 unfixed} can of course be written for the correlator with $n+\m$ insertions appearing in the integrand of Eq.~\eqref{HCFT residues def} when $\m >0$.

Having introduced all relevant definitions, we can now write down the precise holographic dictionary for residues of correlation functions that will be discussed in this paper. It reads
\begin{equation}
\label{matching holográfico}
    \mathop{{\rm Res}}_{\sum_{i=1}^n j_i = 1+\frac{k-2}{2}(n-\m-2)}\Bigg[\cA_n^{\rm string}\Bigg] =\prod_{i=1}^n\cC(j_i,\w_i)\mathop{{\rm Res}}_{\sum_{i=1}^n q_i = Q-\m \alpha}\Bigg[\cA_n^{\rm HCFT}\Bigg]\,,
\end{equation}
where $\cC(j_i,\w_i)$ are the relative normalization factors between the vertex operators.

It should be pointed out that this is not enough to describe the full analytic structure of the holographic CFT. On general grounds, one expects  the full set of poles to be parametrized by two non-negative integers. This two-dimensional lattice can be described in terms of the condition
\begin{equation}
    \sum_{i=1}^n q_i = Q - \m \, \alpha  - \frac{\n}{\alpha} \, ,
    \label{charge conservatio two integers}
\end{equation}
which generalizes Eq.~\eqref{charge conservation}. This is analogous to the Liouville case, where both terms are related by the symmetry under $b \to 1/b$. Although the situation is
less straightforward in the present context, the additional poles were argued to come from subleading corrections in \cite{Eberhardt:2021vsx}. Their presence also makes sense from the worldsheet point of view. In the free-field description of \cite{Knighton:2023mhq,Knighton:2024qxd}, correlators satisfying the more restrictive condition \eqref{charge conservation} are confined to the so-called near-boundary region, where the Wakimoto interaction term can be ignored \cite{Giveon:1998ns}. Upon including $\n$ additional insertions of the corresponding screening operator (the integrated version of this interaction term) one obtains a modification of the charge conservation condition for the radial field which is consistent with Eq.~\eqref{charge conservatio two integers}, although in these cases the path integral becomes much more difficult to evaluate. One of the goals of this paper is to develop an alternative method that could allow us to compute the residues associated with poles with $\n > 0$ in the future.

\section{Localization of SL(2,$\R$) correlators}
\label{sec: maps and localization}

It was established in \cite{Eberhardt:2019ywk} that SL(2,$\R$) correlation functions satisfy a localization property when the unflowed spins are related in a certain way. The corresponding locus is defined by the existence of appropriate ramified holomorphic covering maps from the worldsheet to the AdS$_3$ boundary. Here we present the main properties of such covering maps, and provide an alternative  derivation of this localization along the lines of \cite{Bufalini:2022toj,Kovensky:2026usc}.

\subsection{Holomorphic covering maps}
\label{sec: covering maps}

The key idea for deriving and solving the recursion relations that constrain spectrally flowed correlators is to write correlators with current insertions in terms of contour integrals. Given that the currents $J^a(z)$ act very differently near each of the vertex operators involved, it would be very helpful to find a function $\Gamma(z)$ such that the combination
\begin{equation}
    J^-(\Gamma(z),z) = J^-(z)-2\Gamma(z) J^3(z)
    +\Gamma^2(z) J^+(z) \,
    \label{JGamma}
\end{equation}
acts as the mode $J^-_{-\w_i}(x_i)$ on each of the spectrally flowed insertions, up to corrections of order $(z-z_i)^{\w_i}$. More explicitly, one looks for a  function $\Gamma(z)$ satisfying
\begin{equation}
\Gamma(z\sim z_i) \sim x_i + a_i (z-z_i)^{\w_i} + b_i (z-z_i)^{\w_i+1} + \cdots \,, \qquad  i=1,\dots,n \,,
\label{Gamma conditions at zi}
\end{equation}
up to higher-order terms in $(z-z_i)$. When there is no operator inserted at infinity, we impose that $\Gamma(z\to \infty) = \Gamma_\infty$ for some constant $\Gamma_\infty$. On the other hand, if, say, $V_{j_3 h_3}^{\w_3}(x_3,z_3)$ is inserted at $x_3 = z_3 = \infty$, we require that
\begin{equation}
    \Gamma(z\to \infty) = (-1)^{\w_3+1} a_3^{-1} z^{\w_3} + \cdots\,.
\end{equation}

As it turns out, in many cases ---but not all--- such functions do exist. They are known as holomorphic covering maps, and are the subject of Hurwitz theory; see, for instance, \cite{cavalieri_miles_2016}. At genus zero, we have the following necessary conditions \cite{Eberhardt:2019ywk}. First, all $\w_i$ should be positive. Second, the total spectral flow $\w = \sum_{i=1}^n \w_i$ should be even (odd) if $n$ is even (odd). Third, the strict version of the inequality \eqref{selection rules}
should be satisfied.
In the  $n=3$ case these are also sufficient conditions. Moreover, in this case the existence of the map imposes no constraint on the $x_i$, since we can always use Möbius transformations to set the values of $\Gamma(z)$ for three points, say $\Gamma(z_1)=x_1$, $\Gamma(z_2)=x_2$ and $\Gamma(z_3)=x_3$.
However, for $n\geq 4$ this does not guarantee that the map exists: since there is no freedom left, the candidate solution $\Gamma(z)$ is fixed by the values at the three chosen points, the rest of the worldsheet insertion points $z_i$,  with $i=4,\dots,n$, and the corresponding values of $\w_i$, hence it will be able to satisfy the constraint in Eq.~\eqref{Gamma conditions at zi} if and only if the values of the boundary insertion points $x_{4},\dots,x_n$ satisfy
\begin{equation}
   x_i =  \Gamma_i \equiv \Gamma(z_i)   \,, \qquad i=4,\dots,n \, .
\end{equation}

When the covering map exists, it satisfies a number of interesting properties. We now record those that will be useful later on, assuming for simplicity that no operator is inserted at infinity. The corresponding generalization is straightforward.
The function $\Gamma(z)$ is rational and generically has only simple poles, hence it can be written as
\begin{equation}
\label{Gamma as sum with poles}
    \G (z) = \Gamma_\infty + \sum_{a=1}^N \frac{c_a}{z-\lambda_a} \, ,
\end{equation}
where
\begin{equation}
\label{Ndef}
    N=1+ \frac{1}{2}\sum_{i=1}^n(\w_i-1) \, ,
\end{equation}
according to the Riemann--Hurwitz formula\footnote{Setting $x_3=z_3=\infty$ sends some of these poles to infinity, leading to the replacement $N \to N-\w_3$.}.
Note that, once the $\w_i$ are chosen, the poles $\lambda_a$ and residues $c_a$ should be understood as functions of the worldsheet insertion points $z_i$ (and of the values of $x_{j}$ with $j=1,2,3$). The same goes for the coefficients $a_i$ and $b_i$ appearing in \eqref{Gamma conditions at zi}, and also for the $\Gamma_i$ with $i\geq 4$.

A simple counting argument shows that a more explicit expression can be given for the derivative of the covering map, namely
\begin{equation}
\label{dGamma}
    \der \Gamma(z) = C_\Gamma \frac{\prod_{i=1}^n (z-z_i)^{\w_i-1}}{\prod_{a=1}^N (z-\lambda_a)^2} \, ,
\end{equation}
for some constant $C_\Gamma$ (which also depends on the $z_i$).
This immediately implies that the residues $c_a$ and parameters $a_i$ can be written as
\begin{equation}
\label{ca expression}
c_a = -C_\Gamma\frac{\prod_{i=1}^n (\lambda_a-z_i)^{\w_i-1}}{\prod_{b \neq a}^N \lambda_{ab}^2} \qquad  \forall \ a = 1,\dots,N \, ,
\end{equation}
and
\begin{equation}
\label{ai expression}
    \w_i a_i = C_\Gamma \frac{\prod_{j\neq i} z_{ij}^{\w_j-1}}{\prod_{a=1}^N (z_i-\lambda_a)^2} \qquad  \forall \ i = 1,\dots,n \, ,
\end{equation}
respectively, with $z_{ij} = z_i-z_j$ and $\lambda_{ab}=\lambda_a-\lambda_b$.
Also, the quantities $\lambda_a$ must be solutions of the so-called scattering equations
\begin{equation}
\label{scattering equations}
    \sum_{i=1}^n\frac{\w_i-1}{z_i-\lambda_a} +
    \sum_{b \neq a} \frac{2}{\lambda_{ab}} =0 \qquad  \forall \ a = 1,\dots,N \, ,
\end{equation}
which follow from the fact that $\der \Gamma(z)$ cannot have first-order poles. Finally, from the expansion of $\der \Gamma(z)$ near each of the insertion points we can derive an expression for the second coefficient appearing in \eqref{Gamma conditions at zi}, giving
\begin{equation}
\label{Identity bi 1}
\frac{b_i (\w_i+1)}{a_i\w_i} = \sum_{j\neq i} \frac{\w_j-1}{z_{ij}} - \sum_{a=1}^{N} \frac{2}{z_i-\lambda_a} \qquad  \forall \ i = 1,\dots,n \, .
\end{equation}

As highlighted above, the coefficients that define the function $\Gamma(z)$ depend on the value of the worldsheet insertion points. It will be useful to gather some identities that characterize this dependence. For the first-order coefficients $a_i$ we have
\begin{align}
    \frac{\p_{z_i}a_j}{a_j} =  \frac{\w_i-1}{z_{ij}} +\sum_{a=1}^N \frac{2 \der_{z_i} \lambda_a}{z_j-\lambda_a} + \frac{\partial_{z_i}C_\Gamma}{C_\Gamma} & \qquad  \forall \, i = 1,\dots,n \, , \,  j \neq i \, ,
    \label{daj/dzi}
\end{align}
and
\begin{align}
    \frac{\p_{z_i}a_i}{a_i} =  \sum_{j\neq i}\frac{\w_j-1}{z_{ij}} +\sum_{a=1}^N \frac{2\of{\der_{z_i} \lambda_a-1}}{z_i-\lambda_a}+ \frac{\partial_{z_i}C_\Gamma}{C_\Gamma} & \qquad  \forall \, i = 1,\dots,n \, . \label{dai/dzi}
\end{align}
An additional relation of the form
\begin{equation}
\label{Identity bi 2}
    \frac{2 b_i (\w_i^2-1)}{a_i\w_i} = \sum_{j=1}^n (\w_j-1)\frac{\p_{z_i}a_j}{a_j} - \sum_{a=1}^{N} 2\frac{\p_{z_i}c_a}{c_a} + 2 \frac{\p_{z_i}C_\G}{C_\G}\qquad  \forall \, i = 1,\dots,n \, .
\end{equation}
was derived in appendix A of \cite{Dei:2019iym}\footnote{The extra factor $C_\G^{-1}\p_{z_i}C_\G$ in \cref{Identity bi 2} appears when  no operator is inserted at infinity.}. The derivative of the covering map further satisfies
\begin{equation}
\label{dlogdGamma/dzi}
    \der_{z_i} \log \der \Gamma(z) = - \frac{\w_i-1}{z-z_i} + \sum_{a=1}^N \frac{2\der_{z_i} \lambda_a}{z-\lambda_a} +  \frac{\partial_{z_i}C_\Gamma}{C_\Gamma}  \qquad  \forall \ i = 1,\dots,n \, .
\end{equation}
As for the function $\Gamma(z)$ itself, we have
\begin{equation}
\label{dGammadzi near zj}
    \der_{z_i} \Gamma(z\sim z_j) \sim \der_{z_i} \Gamma_j + \der_{z_i} a_j (z-z_j)^{\w_j} + \der_{z_i} b_j (z-z_j)^{\w_j+1} + \cdots   \qquad  \forall \, i \neq j \, .
\end{equation}
Near the insertion $z_i$, on the other hand, one gets
\begin{equation}
\label{dGammadzi near zi}
    \der_{z_i} \Gamma(z\sim z_i) \sim \der_{z_i} \Gamma_i  - a_i \w_i (z-z_i)^{\w_i-1} + \off{\der_{z_i} a_i - b_i (\w_i+1)} (z-z_i)^{\w_i} + \cdots    \, ,
\end{equation}
while,  near the singularities,
\begin{equation}
\label{dGammadzi near lambda}
    \der_{z_i} \Gamma(z\sim \lambda_a) \sim  \frac{c_a \der_{z_i}\lambda_a}{(z-\lambda_a)^2}  + \frac{\der_{z_i}c_a}{z-\lambda_a} + \cdots  \,.
\end{equation}

As discussed above, one can either keep $(z_1,z_2,z_3)$ and $(x_1,x_2,x_3)$ generic or fix both sets to $(0,1,\infty)$. The resulting covering maps, which we denote by $\G$ and $\hat{\G}$, respectively, are related to each other. Using the conformal maps given by
\begin{align}
    f(z) &= \frac{(z-z_1)(z_3-z_2)}{(z_3-z)(z_2-z_1)} \,, &
    g(x) &= \frac{(x-x_1)(x_3-x_2)}{(x_3-x)(x_2-x_1)}\,,
\end{align}
this relation can be written as
\begin{equation}
    \G(z)\equiv g^{-1}(\hat{\G}(f(z)))\,.
\end{equation}
It follows that the map $\G(z)$ and, by extension, its values at the insertion points $\G_j$ with $j \geq 4$ are actually functions of the cross-ratios $f(z_p)$ with $p \geq 4$. Applying the chain rule, one finds that, for $j \geq 4$,
\begin{equation}
    \der_{z_1} \G_j = 
    \sum_{p=4}^n \frac{z_{p2}z_{3p}}{z_{21}z_{31}}\partial_{z_p} \G_j\label{eq:chainrule1}\,,
\qquad
\der_{z_2}\G_j= \sum_{p=4}^n \frac{z_{1p}z_{3p}}{z_{12}z_{23}}\partial_{z_p} \G_j\,, \qquad
    \der_{z_3}\G_j = \sum_{p=4}^n \frac{z_{1p}z_{2p}}{z_{13}z_{32}}\partial_{z_p} \G_j\,.
\end{equation}
However, the coefficients $a_i$ are \textit{not} only functions of these cross-ratios. Indeed, they transform as
\begin{equation}
    a_i = \left[f'(z_i)\right]^{\omega_i} \left[g'\left(\G_i\right)\right]^{-1}\hat{a}_i\,,
    \label{ai aihat}
\end{equation}
where $\hat{a}_i$ is the corresponding coefficient of the map $\hat{\G}$ evaluated at $f(z_i)$.
The last two factors on the RHS of \eqref{ai aihat} depend only on the cross-ratios, but this is not true for $f'(z_i)$. This leads to equations of the form
\begin{equation}
    \partial_{z_1}a_j = \omega_j a_j \der_{z_1} \ln f'(z_j)
    + \left[f'(z_j)\right]^{\omega_j}
    \sum_{p=4}^n \frac{z_{p2}z_{3p}}{z_{21}z_{31}}
    \partial_{z_p}\left\{ \left[g'\left(\G_j\right)\right]^{-1}\hat{a}_j \right\}\,,
\label{ai derivative z123 eq1}
\end{equation}
and similarly for the derivatives with respect to $z_2$ and $z_3$ (with a proper regularization of  $f'(z_3)$).
Now, when looking at the derivatives of the first three coefficients, namely $a_j$ with $j = 1,2,3$, we can further use that $f'(z_j)$ depends only on the first three insertion points, hence
\begin{equation}
    \der_{z_p}f'(z_j) =0 \qquad \forall \, j < 4 \leq p \ .
\end{equation}
It follows that in such cases Eq.~\eqref{ai derivative z123 eq1} simplifies considerably, giving
\begin{subequations}
\begin{align}
    \partial_{z_1}a_1  &= \sum_{p=4}^n \frac{z_{p2}z_{3p}}{z_{21}z_{31}}\partial_{z_p}a_1 + \omega_1 a_1\left(\frac{1}{z_{21}}+\frac{1}{z_{31}}\right)\label{eq:chainrule_a1}\,, \\
    \partial_{z_2}a_2
    &= \sum_{p=4}^n \frac{z_{1p}z_{3p}}{z_{12}z_{23}}\partial_{z_p}a_2 + \omega_2 a_2\left(\frac{1}{z_{12}}+\frac{1}{z_{32}}\right)\,,\label{eq:chainrule_a2}\\
    \partial_{z_3}a_3
    &= \sum_{p=4}^n \frac{z_{1p}z_{2p}}{z_{13}z_{32}}\partial_{z_p}a_3 + \omega_3 a_3\left(\frac{1}{z_{31}}+\frac{1}{z_{32}}\right)\label{eq:chainrule_a3}\,.
\end{align}
\end{subequations}

\subsection{Correlators satisfying the $j$-constraint}
\label{sec: localization m=0}

We now show that, when the spectral flow charges are such that one or more covering maps $\Gamma(z)$ \textit{can} exist, and the unflowed spins $j_i$ are related by the condition
\be
\sum_{i=1}^n j_i=1+\frac{k-2}{2}(n-2)\,,  \label{j-constraint}
\ee
which we refer to as the $j$-constraint, the recursion relations discussed in Sec.~\ref{sec:recursions} are solved by
\begin{align}
\Bigl\langle \prod_{i=1}^n V_{j_ih_i}^{\w_i}(x_i,z_i) \Bigr\rangle =\sum_{\Gamma} \prod_{i=1}^n |a_i|^{-2h_i} \prod_{i=4}^n \delta^{(2)}(x_i-\Gamma(z_i)) \, |W_\Gamma(z_1,\cdots,z_n)|^2 \,,
\label{eq:localisation solution}
\end{align}
where the function $W_\Gamma$ must be independent of the boundary insertion points $x_i$ and the spacetime weights $h_i$, but is otherwise arbitrary. We will loosely refer to the correlation functions in \eqref{eq:localisation solution} as primary correlators.

\medskip

\noindent \textbf{Preliminary considerations}

\medskip

\noindent Let us consider the function
\begin{equation}
    G(z) \equiv \vev{J^-(\G (z),z)\prod_{j=1}^n V_{j_j,h_j}^{\w_j}(x_j,z_j)} \, .
    \label{G(z) def}
\end{equation}
This must vanish as $1/z^{2}$ at large $z$. Moreover, Eq.~\eqref{JGamma}
implies that  $G(z)$ must have double poles located at the same points as the poles of the covering map. More precisely, we have
\begin{equation}
    G(z \sim \lambda_a) = \frac{c_a^2}{(z-\lambda_a)^2}\vev{J^+(\lambda_a)\prod_{j=1}^n V_{j_j,h_j}^{\w_j}(x_j,z_j)}
 + \cdots    \qquad   \forall \ a = 1,\dots,N \, .
\end{equation}
Now, if the primary correlator is to vanish outside of the locus $x_i = \Gamma_i$, from Eq.~\eqref{Gamma conditions at zi} one might naively  conclude that $G(z)$ should have zeros of order $\w_i-1$ at $z= z_i$ for all values of $i$.
Indeed, near any of the $z_i$ we can replace
\begin{equation}
    J^-(\Gamma(z),z) = J^-(\Gamma(z)-\Gamma_i+x_i,z)-2(\Gamma_i-x_i)J^3(\Gamma(z),z)-(\Gamma_i-x_i)^2 J^+(z) \,,
    \label{J- wrong map}
\end{equation}
and it might seem that the last two terms vanish for a solution of the form \eqref{eq:localisation solution} due to the presence of the Dirac delta function $\delta(x_i-\Gamma_i)$.
However, this conclusion is premature. The reason is that the insertion of the current modes will, in some cases, translate into the action of differential operators such as \eqref{diffopsx}, which will then produce terms proportional to the derivative of one of these delta functions. The distributional identity
\begin{equation}
    (\Gamma_i-x_i) \delta'(x_i-\Gamma_i) = \delta(x_i-\Gamma_i)
    \label{deltap identity}
\end{equation}
then implies that terms proportional to
$(\Gamma_i-x_i)$ cannot be discarded so easily. Nevertheless, given that  $G(z)$ is linear in the currents, there can be at most one such factor $\delta'(x_i-\Gamma_i)$, hence terms proportional to $(\Gamma_i-x_i)^2$ can safely be ignored.

We thus need to analyze the middle term on the RHS of Eq.~\eqref{J- wrong map} more carefully.
For this, we first fix a value of $i$ and consider the auxiliary function
\begin{equation}
    G^{(i)} (z) \equiv (\Gamma_i - x_i)G(z) = (\Gamma_i - x_i)\vev{J^-(\G (z),z)\prod_{j=1}^n V_{j_j,h_j}^{\w_j}(x_j,z_j)} \, .
\end{equation}
On the one hand, we know that $G^{(i)}(z)$ must have, at most, double poles at all $\lambda_a$, and that it must  vanish at infinity as $1/z^2$. On the other hand, near any $z_j$ we can replace $J^-(\Gamma,z) \to J^-(\Gamma-\Gamma_j + x_j,z)$ thanks to the overall factor $(\Gamma_i-x_i)$. Using Eqs.~\eqref{Fwi} and \eqref{Gjconditions}, it follows that  the first potential contribution, namely the term of order $(z-z_j)^{\w_j-1}$, is proportional to a primary correlator (albeit with a shifted value of $h_j$), and hence vanishes upon multiplication by $(\Gamma_i-x_i)$ by assumption. We conclude that $G^{(i)}(z)$ must have zeros of order $\w_j$ at $z=z_j$ for $j = 1,\dots,n$. A simple counting argument involving the total number of potential poles, given in Eq.~\eqref{Ndef}, then allows us to conclude that $G^{(i)}(z)$ must vanish identically. Of course, this holds for any value of $i$.

Let us now go back to the original computation, and consider the functions
\begin{equation}
    H^{(i)} (z) \equiv  (\Gamma_i - x_i)\vev{J^3(\G (z),z)\prod_{j=1}^n V_{j_j,h_j}^{\w_j}(x_j,z_j)} \, .
\end{equation}
As in the previous cases, this must vanish at infinity at least as fast as $1/z^2$. $H^{(i)}(z)$ also has potential \textit{single} poles at $z=\lambda_a$ for $a = 1,\dots, N$. However, the corresponding residues must actually vanish given that, at  first non-trivial order we have
\begin{equation}
    H^{(i)}(z \sim \lambda_a) = -\frac{c_a (\Gamma_i - x_i)}{z-\lambda_a}\vev{J^+(\lambda_a)\prod_{j=1}^n V_{j_j,h_j}^{\w_j}(x_j,z_j)} = -\frac{G^{(i)}(z)}{\Gamma(z)}\Bigg|_{z\sim \lambda_a} = 0 \, .
\end{equation}
As for the worldsheet insertion points of the vertex operators $V_{j_jh_j}^{\w_j}(x_j,z_j)$, we may use that terms quadratic in $(\Gamma_i - x_i)$ can be set to zero, hence near each $z_j$ we can replace
\begin{equation}
    (\Gamma_i-x_i)J^3(\Gamma(z),z) \to   (\Gamma_i-x_i)J^3(\Gamma(z)-\Gamma_j+x_j,z) \, .
\end{equation}
Eqs.~\eqref{J3xOPE} and \eqref{Gamma conditions at zi} then imply that $H^{(i)}(z)$ has at most single poles at each $z=z_j$ and, as before, the potential residues must actually vanish since linear combinations of primary correlators multiplied by $(\Gamma_i-x_i)$ do so. We thus finally conclude that the second term on the RHS of Eq.~\eqref{J- wrong map} can also be set to zero for correlators of the form \eqref{eq:localisation solution}.

The discussion above guarantees that, even though the presence of Dirac delta functions in the Ansatz \eqref{eq:localisation solution} makes it slightly more complicated to prove, for $n\geq 4$ the function $G(z)$ defined by inserting $J^-(\Gamma(z),z)$ into the primary correlator behaves as in the three-point case \cite{Eberhardt:2019ywk,Bufalini:2022toj}: it has double poles at each of the $\lambda_a$ for all values of $a$, zeros of order $\w_j-1$ at each of the $z_j$, and must behave as $1/z^2$ as one takes $z\to \infty$. In other words, there must exist a coefficient $\alpha$ such that
\begin{equation}
    G (z) = \alpha \der \Gamma(z) \, .
\end{equation}
This coefficient can be derived as follows. We have
\begin{equation}
    \alpha N = -\sum_{a=1}^N \oint_a dz \frac{G(z)}{\Gamma(z)}\, ,
\end{equation}
where the contours encircle each of the $\lambda_a$. Upon inserting Eqs.~\eqref{currents OPEs inside correlators} and turning the contour around we obtain
\begin{align}
\begin{aligned}
    \alpha N &= - \sum_{a=1}^N \sum_{j=1}^n \sum_{\ell=0}^{\w_j}
    \oint_a dz \frac{\Gamma(z) F_\ell^{+,j}}{(z-z_j)^{\ell+1} } = \sum_{j=1}^n \sum_{\ell=0}^{\w_j}
    \oint_{z_j} dz \frac{\Gamma(z) F_\ell^{+,j}}{(z-z_j)^{\ell+1} } \\
    & = \sum_{j=1}^n \off{ \of{\Gamma_j - \Gamma_\infty} F_0^{+,j} + a_j F_{\w_j}^{+,j}} \\
    & = \sum_{j=1}^n \off{ \of{\Gamma_j - x_j} F_0^{+,j} + \of{x_j - \Gamma_\infty} F_0^{+,j} + a_j F_{\w_j}^{+,j}}\, .
\end{aligned}
\end{align}
Here, the first term in the last line vanishes for $j=1,2,3$, but contributes a factor $1$ for each of the remaining factors in the sum because $F_0^{+,j}$ encodes the action of $J_0^+$, which acts on the $j$-th vertex as $\der_{x_j}$. Additionally, the global Ward identities imply  that
\begin{equation}
    \sum_{j=1}^n F_0^{+,j} = 0 \, , \qquad \sum_{j=1}^n \of{h_j F + x_j F_0^{+,j}} = 0 \, ,
\end{equation}
where $F$ stands for the original primary correlator \eqref{defF}.  Hence, we find that
\begin{equation}
   \alpha N = \sum_{j=1}^n \of{a_j F_{\w_j}^{+,j} - h_j F} + (n-3)F \, ,
\end{equation}
from which the value of $\alpha$ can be read off.

\medskip

\noindent \textbf{Solving the recursion relations}

\medskip

\noindent Using the results above, it is now possible to prove that  \eqref{eq:localisation solution} solves the recursion relations for spectrally flowed $n$-point functions satisfying the $j$-constraint \eqref{j-constraint}. For this, we compute the contour integral
\begin{equation}
    \oint_{z_i}  dz \frac{G(z)}{(z-z_i)^{\w_i}}
\end{equation}
in two different ways.
On the one hand, we have
\begin{align}
\begin{aligned}
\label{Rec rel 1}
    \oint_{z_i}  dz \frac{G(z)}{(z-z_i)^{\w_i}} &= \oint_{z_i}  dz \frac{\alpha \der \G(z)}{(z-z_i)^{\w_i}} = \oint_{z_i}  dz \frac{\alpha\off{a_i \w_i (z-z_i)^{\w_i-1}+ \cdots}}{(z-z_i)^{\w_i}} = \alpha \, a_i \w_i
    \\
    &=  \frac{a_i\w_i}{N} \off{\sum_{j=1}^n \of{a_j F_{\w_j}^{+,j} - h_j F} + (n-3)F} \, .
\end{aligned}
\end{align}
On the other hand, from the (first term of the) expansion \eqref{J- wrong map}, and using the notation introduced in Eq.~\eqref{defFin}, we get
\begin{equation}
\label{Rec rel 2}
    \oint_{z_i}  dz \frac{G(z)}{(z-z_i)^{\w_i}} = F_{-\w_i}^{-,i} -2 a_i F_{0}^{3,i} + a_i^2 F_{\w_i}^{+,i} \, .
\end{equation}
The constraints imposed by the recursion relations are thus equivalent to
\begin{equation}
    F_{-\w_i}^{-,i} -2 a_i F_{0}^{3,i} + a_i^2 F_{\w_i}^{+,i} = \frac{a_i\w_i}{N} \off{\sum_{j=1}^n \of{a_j F_{\w_j}^{+,j} - h_j F} + (n-3)F} \qquad \forall \quad i = 1,\dots, n \, .
    \label{Rec rel final}
\end{equation}
The terms $F_0^{3,i}$ and $F_{\pm\w_j}^{\pm,j}$ were computed in Eqs.~\eqref{F0+ and F03} and \eqref{Fwi}. Inserting the proposal \eqref{eq:localisation solution} ---or,  more precisely, the contribution  from a given covering map--- into \eqref{Rec rel final}, we see that the shifts $h_i \to h_i \pm 1$ appearing in $F_{\pm\w_j}^{\pm,j}$ are compensated by the corresponding factor of $a_i$, giving
\begin{equation}
  F_0^{3,i} = h_i F \, , \qquad
  a_i^{\pm 1} F_{\pm \w_i}^{\pm,i} = \of{h_i - \frac{k}{2}\w_i \pm j_i} F \, .
\end{equation}
Hence, assuming the primary correlator does not vanish, Eq.~\eqref{Rec rel final} holds if and only if
\begin{equation}
   - N k = \sum_{i=1}^n \of{j_i - \frac{k}{2}\w_i} + (n-3) \, ,
\end{equation}
which is equivalent to \eqref{j-constraint}. This establishes the localization property: as long as the $j$-constraint is satisfied, the Ansatz \eqref{eq:localisation solution} solves the recursion relations derived from the local Ward identities  \textit{for any function} $W_\Gamma(z_i)$.

\section{Knizhnik--Zamolodchikov equations and spectral flow}
\label{sec: KZ w=0 y w=1}

Here we discuss general aspects of KZ equations in the SL(2,$\R$) WZW model. We begin by briefly reviewing the unflowed case, focusing on four-point functions for concreteness. We then write down the KZ equations for spectrally flowed $n$-point functions in a compact way that constitutes the starting point for the subsequent derivations. Finally, we derive the solutions for the correlators of singly-flowed vertex operators, and discuss their holographic interpretation.

\subsection{KZ equations in the unflowed case}

Let us first review some basic facts about the unflowed four-point functions
\begin{equation}
    \langle V_{j_1}(x_1,z_1)V_{j_2}(x_2,z_2)V_{j_3}(x_3,z_3)V_{j_4}(x_4,z_4)\rangle. \label{unflowed4pt}
\end{equation}
As usual, the global Ward identities can be used to fix the insertion points at $(z_1,z_2,z_3,z_4)$ $= (0,1,\infty,z)$ and $(x_1,x_2,x_3,x_4) = (0,1,\infty,x)$, where we have introduced the cross-ratios
\begin{equation}
    z = \frac{z_{32}z_{14}}{z_{12}z_{34}}\qqquad x = \frac{x_{32}x_{14}}{x_{12}x_{34}} \, ,
\end{equation}
giving
\begin{align}
\label{4points 01inf}
    &\langle V_{j_1}(x_1,z_1)V_{j_2}(x_2,z_2)V_{j_3}(x_3,z_3)V_{j_4}(x_4,z_4)\rangle= \nn\\
    & \qquad \qquad \Bigg|\frac{x_{12}^{-j_1-j_2+j_3-j_4}x_{13}^{-j_1+j_2-j_3+j_4}x_{23}^{j_1-j_2-j_3+j_4}x_{34}^{-2j_4}}{z_{12}^{\Delta_1+\Delta_2-\Delta_3+\Delta_4}z_{13}^{\Delta_1-\Delta_2+\Delta_3-\Delta_4}z_{23}^{-\Delta_1+\Delta_2+\Delta_3-\Delta_4}z_{34}^{2\Delta_4}} \Bigg|^2 \mathcal{F}_{0}(x,z) \,,
\end{align}
with
\begin{equation}
    \mathcal{F}_{0}(x,z) \equiv \langle V_{j_1}(0,0)V_{j_2}(1,1)V_{j_3}(\infty,\infty)V_{j_4}(x,z)\rangle \, .
\end{equation}
The Sugawara construction of the worldsheet Virasoro generators implies that $\mathcal{F}_{0}(x,z)$  must be a solution of the KZ equation. Indeed, affine primaries are annihilated by the operator
\begin{equation}
    L_{-1} - \frac{1}{k-2} \left(J^+_{-1} J^-_0 + J^-_{-1} J^+_0-
    2 J^3_{-1} J^3_0\right) \, .\label{unflowed KZ}
\end{equation}
This turns into a differential equation upon acting with the zero modes $J_0^a$, and then evaluating the action of the modes $J^a_{-1}$ using contour integrals in combination with the defining OPEs  \cite{DiFrancesco:1997nk}. More explicitly, since
\begin{equation}
    \of{J_{-1}^a V_{j_i}}(x_i,z_i) = \oint_{z_i}  \frac{dz}{z-z_i} J^a(x_i,z)V_{j_i}(x_i,z_i) \, ,
\end{equation}
and
\begin{equation}
\label{J(xi) to J(xj)}
    J^3(x_i,z) = J^{3}(x_{j},z)-x_{ij}J^{+}(z) \, , \quad
    J^-(x_i,z) = J^{-}(x_{j},z)-2x_{ij}J^{3}(x_j,z)+ x_{ij}^2J^{+}(z)\,,
\end{equation}
while $J^+(x_i,z) = J^{+}(z)$, turning the contour around and making use of \eqref{diffopsx} gives
\begin{align}
    \vev{\of{J_{-1}^+V_{j_i}}(x_i,z_i) \prod_{j\neq i} V_{j_j}(x_j,z_j)} &= \of{\sum_{j\neq i} \frac{\der_{x_j}}{z_{ij}}} \vev{\prod_{p=1}^4 V_{j_p}(x_p,z_p)} \, ,
\\
    \vev{\of{J_{-1}^3V_{j_i}}(x_i,z_i) \prod_{j\neq i} V_{j_j}(x_j,z_j)} &= \of{\sum_{j\neq i} \frac{j_j+x_{ji} \der_{x_j}}{z_{ij}}} \vev{\prod_{p=1}^4 V_{j_p}(x_p,z_p)} \, ,
\\
\vev{\of{J_{-1}^-V_{j_i}}(x_i,z_i) \prod_{j\neq i} V_{j_j}(x_j,z_j)} &= \of{\sum_{j\neq i} \frac{2x_{ji}j_j+x_{ji}^2 \der_{x_j}}{z_{ij}}} \vev{\prod_{p=1}^4 V_{j_p}(x_p,z_p)}\, .
\end{align}
This leads to a KZ equation of the form
\begin{equation}
\label{KZ eq w=0}
    \left[\der_z - \frac{1}{k-2} \left(\frac{{\cal{P}}}{z}+\frac{{\cal{Q}}}{z-1}\right)\right] \mathcal{F}_{0}(x,z)=0 \, ,
\end{equation}
where
\begin{subequations}
\label{def P and Q ops}
\begin{align}
 & \hspace{-0.1cm}   {\cal{P}} =
    x^2(x-1)\der_x^2 - \left[(\kappa-1)x + 2j_1  - 2 j_4 (x-1)\right]x \der_x - 2 j_4(j_1 + \kappa x) \, ,
    \\
& \hspace{-0.1cm}
    {\cal{Q}} =
    -x(1-x)^2\der_x^2 + \left[(\kappa-1)(1-x) + 2j_2 + 2 j_4 x \right](1-x)\der_x - 2 j_4[j_2 + \kappa (1-x)] \, ,
\end{align}
\end{subequations}
with $\kappa = j_3-j_1-j_2-j_4$.

As it turns out, beyond the usual singularities at $z=0,1,\infty$ and $x=0,1,\infty$, unflowed four-point functions have an additional singularity at $z=x$ \cite{Maldacena:2001km}. More precisely, we have
\begin{equation}
    \mathcal{F}_{0}(x\sim z) \sim |x-z|^{2 (k-j_1-j_2-j_3-j_4)} \, .
\end{equation}
This fits nicely with the analysis of the previous section. To see this, one must first use the series identifications \eqref{seriesidentifXbasis} to interpret the unflowed four-point function as a spectrally flowed one with quantum numbers $\w_i'=1$ and $j_i' = \frac{k}{2}-j_i$ for $i=1,2,3,4$. This can always be done since the only physical states in the unflowed sector are short strings.  Using the formula
\begin{equation}
    {\rm Res}_{\alpha=-1}|x-z|^{2\alpha} = \pi \delta^{(2)} (x-z) \, ,
\end{equation}
we find that there is a momentum-space singularity at
\begin{equation}
\sum_{i=1}^4 j_i' = k-1 \, ,
\end{equation}
which is nothing but the $j$-constraint \eqref{j-constraint} for the case $n=4$. As expected, the corresponding residue localizes. In this case the relevant covering map is simply $\Gamma(z)=z$, hence the delta function in Eq.~\eqref{eq:localisation solution} is precisely $\delta^{(2)} (x-z)$.

\subsection{Singly-flowed correlators and the untwisted sector of the boundary theory}
\label{sec: untwisted}

Spectrally flowed states are not affine primaries. To study the KZ equations for correlators involving flowed insertions we thus need the full expression of the worldsheet Virasoro mode $L_{-1}$ in terms of the currents, namely
\begin{equation}
    L_{-1} = \frac{1}{k-2} \sum_{m = 0}^{\infty} \left(-2J^3_{-1-m}J^3_{m} + J^+_{-1-m}J^-_{m} + J^-_{-1-m}J^+_{m}\right)\,.
\end{equation}
For $m > 0$, the first two terms inside the sum annihilate states with $\w \geq 1$ but the third one does not. This was first studied in \cite{Ribault:2005ms} (in a different basis), and revisited more recently in \cite{Dei:2021yom}.  The KZ equation for flowed vertex operators can be obtained either by using the OPEs \eqref{JVxOPE} or by performing a spectral flow transformation to the unflowed equation \eqref{unflowed KZ}. This gives
\begin{align}
    \label{KZ flowed states v1}    (k-2)\der_z V_{jh}^\w(x,z) & =
    -2 (h-\w) \of{J^3_{-1}V_{jh}^\w}(x,z) \\[1ex]
    & \hspace{-1cm}
    + \of{h-\frac{k}{2}\w + j}\of{J^-_{-\w-1}V_{j,h+1}^\w}(x,z) + \of{h-\frac{k}{2}\w - j}\of{J^+_{\w-1}V_{j,h-1}^\w}(x,z) \,. \nn
\end{align}
Note that the first two terms on the RHS involve modes of the currents $J^3(x,z)$ and $J^{-}(x,z)$, defined in Eq.~\eqref{defJx}.

For reasons that will become clear below, it will be useful to rewrite \eqref{KZ flowed states v1} as follows. The correlators we are interested in are only non-vanishing on the locus where an appropriate covering map exists. Although there might be more than one such map, here we focus on a particular contribution to the sum in \eqref{eq:localisation solution}, i.e.~a particular function $\Gamma(z)$. Let $a_i$  be the coefficient of the leading non-constant term in the expansion of $\Gamma(z)$ near  $z=z_i$. We define the shorthands 
\begin{subequations}
\label{J modes zero with ai}
    \begin{align}
    J^3_n(a_i)&=J^3_n(x_i)-a_iJ^+_{n+\w_i}
    \label{def J3(a)}\\
    J^-_{-\w_i+n}(a_i)&=J^-_{-\w_i+n}(x_i)-2a_i\off{J^3_{n}(x_i)-\delta_{n,0} \frac{k}{2}\w_i}+a_i^2J^+_{\w_i+n}\,,
    \label{def J-(a)}
\end{align}
\end{subequations}
for any integer $n$, and omitting the $x_i$-dependence  on the LHS to avoid cluttering the notation.  The KZ condition \eqref{KZ flowed states v1} for the $i$-th operator in our correlation function can then be expressed as
\begin{align}
\label{KZ flowed states v2} &(k-2)\off{L_{-1}+a_i\w_iJ^+_{\w_i-1}}V_{j_ih_i}^{\w_i}(x_i,z_i) \\[1ex]
    & \quad = \left\{J^+_{\w_i-1}J^-_{-\w_i}(a_i) -2J^3_{-1}(a_i)\off{J^3_0(a_i)-\w_i}+J^-_{-\w_i-1}(a_i)J^+_{\w_i} \right\}V_{j_ih_i}^{\w_i}(x_i,z_i)\,, \nn
\end{align}
This is a natural way to write the KZ constraint since the combinations \eqref{J modes zero with ai} arise when considering contour integrals involving current insertions that make use of our covering map, namely  $J^3(\Gamma(z),z)$ and $J^-(\Gamma(z),z)$. Some examples of these manipulations already appeared  when studying the recursion relations satisfied by flowed correlators in the previous section, see for instance  Eqs.~\eqref{Rec rel 1} and \eqref{Rec rel 2}.

We deal with the general case where the spectral flow charges are arbitrary in the following section. For now, we focus on the simplest non-trivial situation and set $\w_i=1$ for all $i=1,\dots,n$, i.e.~we consider the correlator
\begin{equation}
    \vev{\prod_{i=1}^n V_{j_i,h_i}^{1}(x_i,z_i)}
    \,.
\end{equation}
In the language of \cite{Eberhardt:2021vsx}, and assuming our string states are trivial in the internal manifold $X$, this corresponds to the Liouville-like part of the seed theory in Eq.~\eqref{HCFT def}, that is, the untwisted sector of the holographic CFT  briefly discussed in the introduction. From the worldsheet point of view, the computation simplifies because all the terms in the OPEs \eqref{JVxOPE} that are relevant for the explicit evaluation of \eqref{KZ flowed states v2} can be written in terms of (Virasoro) primary operators and their derivatives. This is nevertheless an interesting example, which will allow us to showcase some important aspects of the computation and draw a number of relevant conclusions regarding the boundary theory.

For simplicity, we set $z_1=x_1=0$, $z_2=x_2=1$ and $z_3=x_3=\infty$. The only covering map that solves all required conditions is then simply $\Gamma(z) = z$. This means that
\begin{equation}
    \Gamma_i = z_i \, , \qquad a_i = 1 \, , \qquad b_i = 0 \, \qquad \forall \, i = 1,\dots,n \, .
\end{equation}
The KZ equation \eqref{KZ flowed states v2} thus reads
\begin{align}
\label{KZ winding 1} &(k-2)\off{L_{-1}+J^+_{0}}V_{j_ih_i}^{1}(x_i,z_i) \\[1ex]
    & \quad = \left\{J^+_{0}J^-_{-1}(1) -2J^3_{-1}(1)\off{J^3_0(1)-1}+J^-_{-2}(1)J^+_{1} \right\}V_{j_i,h_i}^{1}(x_i,z_i)\, . \nn
\end{align}
Assuming that the $j$-constraint \eqref{j-constraint} holds, the Ansatz \eqref{eq:localisation solution} tells us that this must be satisfied for any operator inserted in the correlator
\begin{equation}
\Bigl\langle V_{j_1h_1}^{1}(0,0)V_{j_2h_2}^{1}(1,1)V_{j_3h_3}^{1} (\infty,\infty)\prod_{i=4}^n V_{j_ih_i}^{1}(x_i,z_i) \Bigr\rangle
=\prod_{i=4}^n \delta^{(2)}(x_i-z_i) \, |W_\Gamma(z_4,\dots, z_n)|^2\ ,
\label{eq:localisation solution wi=1 a}
\end{equation}
which will allow us to derive $n-3$ constraints on the function $W_\Gamma$ for this particular case.

We first analyze the LHS of Eq.~\eqref{KZ winding 1}. The Virasoro mode $L_{-1}$ and the current zero mode $J_0^+$ act on $V_{j_ih_i}^{1}(x_i,z_i)$ as the derivatives $\der_{z_i}$ and $\der_{x_i}$, respectively. The combination $\der_{z_i}+\der_{x_i}$ that we get is particularly interesting in light of the formula \eqref{eq:localisation solution wi=1 a} because the delta function $\delta(x_i-z_i)$ is invariant under simultaneous translations of $z_i$ and $x_i$. Since $W_\Gamma$ depends only on the worldsheet insertion points and, in this example, all the other Dirac delta functions $\delta(x_j-z_j)$ with $j \neq i$ are independent of these particular variables, we have
\begin{equation}
    (\der_{z_i}+\der_{x_i}) \off{\prod_{j=4}^n \delta^{(2)}(x_j-z_j) \, |W_\Gamma(z_4,\dots, z_n)|^2 } =  \prod_{j=4}^n \delta^{(2)}(x_j-z_j) \der_{z_i}|W_\Gamma(z_4,\dots, z_n)|^2 \, .
\end{equation}
As a consequence, the KZ constraints \eqref{KZ winding 1} bypass the delta functions, and will easily be interpreted as a set of differential equations that fix the functional form of $W_\Gamma(z_4,\dots,z_n)$. This will also be the case for $\w_i>1$, although the technical derivation is slightly more involved.

Let us now consider the RHS of Eq.~\eqref{KZ winding 1}. All states have unit spectral flow charge, hence
\begin{equation}
    \of{J^3_0 V_{j_i,h_i}^1}(x_i,z_i) = h_iV_{j_i, h_i}^1(x_i,z_i) \, , \quad
    \of{J^\pm_{\pm 1} V_{j_i,h_i}^1}(x_i,z_i) = \of{h_i - \frac{k}{2} \pm j_i} V_{j_i, h_i \pm 1}^1(x_i,z_i) \, .
\end{equation}
The analysis simplifies by noting that the expression on the RHS of \eqref{eq:localisation solution wi=1 a} is actually independent of the value of the spacetime weights, which is a consequence of having $a_i=1$ for all $i$. We can thus ignore the shifts $h_i\pm 1$ coming from the action of $J^+_1$ and $J^-_{-1}$, although we must keep track of the corresponding coefficients.
This immediately implies that the term involving the action of $J^-_{-1}(a_i=1)$ vanishes. Indeed, from \eqref{def J-(a)} we find that inside the singly-flowed correlator \eqref{eq:localisation solution wi=1 a} we can replace
\begin{align}
\begin{aligned}
    J^-_{-1}(a_i=1) &= J^-_{-1}(x_i) -2 \off{J^3_0(x_i) - \frac{k}{2}} + J^+_1 \\
    & \to
    \of{h_i - \frac{k}{2} - j_i} -2 \of{h_i-\frac{k}{2}} + \of{h_i - \frac{k}{2} + j_i} = 0\, .
\end{aligned}
\end{align}
Importantly, the presence of the additional mode $J^+_0$ in this term does not affect this conclusion since it acts as a simple derivative. This will \textit{not} be the case in more general situations.

We now deal with the remaining contributions, which contain the action of
\begin{equation}
    J^3_{-1}(1) = J^3_{-1}(x_i) - J^+_0 = J^3_{-1}-x_i J^+_{-1} - J^+_0
\end{equation}
and
\begin{equation}
    J^-_{-2}(1) = J^-_{-2}(x_i) - 2 J^3_{-1}(x_i) + J^+_0 =
    J^-_{-2}-2x_i J^3_{-2}+x_i^2 J^+_{-2} - 2 J^3_{-1}+2 x_i J^+_{-1} + J^+_0 \, .
\end{equation}
Since there are no unknown terms in the OPEs, these terms can be analyzed by expressing the negative modes as contour integrals and using Eqs.~\eqref{J(xi) to J(xj)} and \eqref{JVxOPE} as usual. We get
\begin{eqnarray}
    &&\vev{\of{J^-_{-2}(a_i=1)J^+_1 V_{j_i,h_i}^1}(x_i,z_i) \prod_{p \neq i} V_{j_ph_p}^1(x_p,z_p)} \nn \\
     && = \of{h_i-\frac{k}{2}+j_i}\vev{\of{J^-_{-2}(a_i=1) V_{j_i, h_i+1}^1}(x_i,z_i) \prod_{p \neq i} V_{j_ph_p}^1(x_p,z_p)} \\
    && = - \of{h_i-\frac{k}{2}+j_i} \sum_{j \neq i} \oint_{z_j}
    dz  \vev{ \off{
    \frac{J^-(x_i,z)}{(z-z_i)^2} - \frac{2J^3(x_i,z)}{z-z_i} + J^+(z)
    } \prod_{p=1}^n V_{j_p,h_p+\delta_{pi}}^1(x_p,z_p)} \nn \\
    && = - \of{h_i-\frac{k}{2}+j_i}\sum_{j \neq i}\left\{ \off{\frac{(x_{ij}-z_{ij})^2\der_{x_j} -2 h_j (x_{ij}-z_{ij})}{z_{ij}^2}} \vev{ \prod_{p=1}^n V_{j_p,h_p+\delta_{pi}}^1(x_p,z_p)}
    \right. \nn  \\
    && \qquad\qquad \left.
    -2\of{h_j-\frac{k}{2}+j_j}\frac{x_{ij}(x_{ij}-z_{ij})}{z_{ij}^3}\vev{ \prod_{p=1}^n V_{j_p,h_p+\delta_{pi}+\delta_{pj}}^1(x_p,z_p)}
    \right\}  = 0 \nn \, .
\end{eqnarray}
The final equality follows from the fact that all factors $x_{ij} - z_{ij} = (x_i-z_i) - (x_j-z_j)$ vanish upon inserting the solution \eqref{eq:localisation solution} due to the presence of the Dirac delta functions. An analogous computation for the remaining term gives
\begin{align}
\begin{aligned}
   & \vev{\of{J^3_{-1}(a_i=1)\off{J^3_0-J^+_1-1} V_{j_i,h_i}^1}(x_i,z_i)  \prod_{p \neq i} V_{j_ph_p}^1(x_p,z_p)}  \\
     & \hspace{4cm} = \sum_{j \neq i } \frac{\of{\frac{k}{2}-j_i-1}\of{\frac{k}{2}-j_j-1}}{z_{ij}} \vev{ \prod_{p=1}^n V_{j_ph_p}^1(x_p,z_p)} \, .
\end{aligned}
\end{align}
By combining all the results above, we obtain the following differential equation:
\begin{equation}
    \der_{z_i} \log W_\Gamma = -\frac{q_1q_i}{z_i}-\frac{q_2q_i}{z_i-1}-\sum_{j\geq4,\,j \neq i} \frac{q_i q_j}{z_{ij}} \, , \qquad \forall \, i=4,\dots,n \, ,
\end{equation}
where $q_i$ was defined in Eq.~\eqref{qi def}. 
We conclude that the residue of the singly-flowed SL(2,$\R$) $n$-point functions  satisfying the $j$-constraint \eqref{j-constraint} takes the form
\begin{align}
    \Bigl\langle V_{j_1h_1}^{1}(0,0)V_{j_2h_2}^{1}(1,1)V_{j_3h_3}^{1} (\infty,\infty)\prod_{i=4}^n V_{j_ih_i}^{1}(x_i,z_i) \Bigr\rangle \propto \prod_{i=4}^n \delta^{(2)}(x_i-z_i)\prod_{i<j} |z_{ij}|^{-2q_iq_j}
\label{eq:localisation solution wi=1} \ ,
\end{align}
up to a $j_i$-independent normalization constant, which will be determined later.

As discussed in Sec.~\ref{sec: string correlators def}, in order to interpret this result in holographic terms we need to embed the SL(2,$\R$) WZW model into AdS$_3$ string theory. For operators which are trivial in the internal sector, the string correlator is obtained by taking the SL(2,$\R$) spins $j_i$ and the spacetime weights $h_i$ to satisfy the Virasoro condition \eqref{Virasoro cond general} with $\Delta_X=0$, and integrating over the worldsheet moduli. In other words, we need to integrate over the worldsheet insertion points $z_4,\dots,z_n$, which is easily done  by using the delta functions $\delta^{(2)}(x_i-z_i)$ in \eqref{eq:localisation solution wi=1}. We thus finally obtain
\begin{equation}
   \int d^2z_4 \dots d^2z_n \Bigl\langle V_{j_1h_1}^{1}(0,0)V_{j_2h_2}^{1}(1,1)V_{j_3h_3}^{1} (\infty,\infty)\prod_{i=4}^n V_{j_ih_i}^{1}(x_i,z_i) \Bigr\rangle \propto \prod_{i<j} |x_{ij}|^{-2q_iq_j} \, ,
   \label{final string result wi=1}
\end{equation}
with $q_i$ as in \eqref{qi def}. This result is to be interpreted as the residue of the boundary correlator at the momentum-space pole defined by the $j$-constraint \eqref{j-constraint}. Even though the normalization will be fixed below for the general case, we can already see that Eq.~\eqref{final string result wi=1} reproduces the expected structure of correlation functions in the untwisted sector of the symmetric orbifold theory \eqref{sym orbifold}. When the $j$-constraint is satisfied, and the contributions from the internal CFT can be ignored, this is simply a correlation function of $n$ exponential operators involving a free non-compact scalar field with the background charge $Q$ given in Eq.~\eqref{qi def}.

\section{Correlators with arbitrary spectral flow charges: derivation}
\label{sec: general derivation}

The contour manipulations performed in the previous section are considerably more involved when operators with $\w>1$ are included. Here we are interested in the particular family of correlators satisfying \eqref{j-constraint}. We will now show that the localization properties of \eqref{eq:localisation solution}, combined with the identity in Eq.~\eqref{KZ flowed states v1}, provide enough leverage for deriving differential equations that completely fix the remaining functions $W_\Gamma(z_i)$ for correlators involving insertions with arbitrary spectral flow charges. The main intermediate steps are discussed in this section, while the final solutions are derived in the following one.

\subsection{KZ equations in flowed sectors: part I}

We consider the KZ equations for correlators satisfying the $j$-constraint \eqref{j-constraint} and for which the spectral flow charges are such that suitable covering maps exist, without imposing any additional restrictions. Starting from the localized expression in Eq.~\eqref{eq:localisation solution}, we focus on the contribution from a particular covering map $\Gamma(z)$.

We first rewrite once more the KZ constraint \eqref{KZ flowed states v2} as
\begin{align}
\label{KZ flowed states v3} &(k-2)\off{L_{-1}+a_i\w_iJ^+_{\w_i-1}+b_i(\w_i+1)J^+_{\w_i}}V_{j_ih_i}^{\w_i}(x_i,z_i) = \left\{ J^+_{\w_i-1}J^-_{-\w_i}(a_i) \right.\\[1ex]
    & \quad \left.-2J^3_{-1}(a_i,b_i)\off{J^3_0(a_i)-\w_i}+\off{J^-_{-\w_i-1}(a_i,b_i)+kb_i(\w_i+1)}J^+_{\w_i} \right\}V_{j_ih_i}^{\w_i}(x_i,z_i)\,, \nn
\end{align}
where $a_i$ and $b_i$ are the coefficients appearing in \eqref{Gamma conditions at zi}, while
\begin{equation}
J^3_{-1}(a_i,b_i)\equiv J^3_{-1}(x_i)-a_iJ^+_{\w_i-1}-b_iJ^+_{\w_i} = J^3_{-1}(a_i) -b_iJ^+_{\w_i}
    \,, \label{def J3(a,b)}
\end{equation}
and
\begin{align}
    \begin{aligned}
        J^-_{-\w_i-1}(a_i,b_i)&\equiv J^-_{-\w_i-1}(x_i)-2a_iJ^3_{-1}(x_i)+a_i^2J^+_{\w_i-1}-2b_i\off{J^3_{0}(x_i) - a_i J^+_{\w_i}} \\
        & =
    J^-_{-\w_i-1}(a_i) -2 b_i J^3_0(a_i)
    \,.
    \label{def J-(a,b)}
    \end{aligned}
\end{align}
These expressions generalize those introduced in Eq.~\eqref{J modes zero with ai}. This is motivated by the fact that the additional factor of $(z-z_i)^{-1}$ needed to compute the modes $J^-_{-\w_i-1}$ and $J^3_{-1}$ from contour integrals involving $J^-(\Gamma(z),z)$ and $J^3(\Gamma(z),z)$ will generically pick up subleading contributions, which are proportional to  $b_i$. We did not have to deal with such contributions in the singly-flowed case discussed in  Sec.~\ref{sec: untwisted} because all such coefficients vanish when $\Gamma(z)=z$.

In the following subsections, we show in detail how to derive explicit expressions for the different terms one obtains upon inserting \eqref{KZ flowed states v3} in a flowed primary correlator. These results are then collected and combined in order to compute the string correlators in Sec.~\ref{sec: general results}.

\subsection{Insertions of $J^-_{-\w_i-1}$ and $J^3_{-1}$}
\label{sec: negative modes J- and J3}

Let us first consider the last term on the RHS of Eq.~\eqref{KZ flowed states v3}. As discussed in Sec.~\ref{sec: localization m=0}, the correlation functions we are interested in satisfy the identity
\begin{equation}
\label{JGamma is dGamma}
\vev{J^-(\Gamma,z) \cdots} = \alpha \der\Gamma(z) \, .
\end{equation}
Here we have omitted the $z$-dependence of the covering map as well as the vertex operator insertions by introducing the shorthand
\begin{equation}
    \vev{J^-(\Gamma,z) \cdots} \equiv \vev{J^-(\Gamma(z),z) \prod_{i=1}^n V_{j_i,h_i}^{\w_i}(x_i,z_i)} \, .
\end{equation}
Moreover, it follows from Eqs.~\eqref{j-constraint} and \eqref{eq:localisation solution} that the relative coefficient $\alpha$ (normalized by the primary correlator) is given by
\begin{equation}
    \frac{\alpha}{\vev{\cdots}} =\frac{1}{N}\off{ n-3 +\sum_{j=1}^n \left(
    a_j \hat{F}_{\w_j}^{+,j} - h_j \right)} = \frac{1}{N} \off{ n-3 +\sum_{j=1}^n \left(
    j_j-\frac{k}{2} \w_j\right)} = - k
\end{equation}
with $\hat{F}_{\ell}^{+,j} = F_{\ell}^{+,j}/\vev{\cdots}$, and $\vev{\cdots} \equiv \langle\prod_{i=1}^n V_{j_i,h_i}^{\w_i}(x_i,z_i)\rangle$.
As we will argue below, this relation has far-reaching consequences. For instance, we may use it to compute
\begin{equation}
   \oint_{z_i}  dz \frac{\vev{J^-(\Gamma,z)\cdots}}{(z-z_i)^{\w_i+1}} = -\oint_{z_i}  dz \frac{k \der \Gamma(z)}{(z-z_i)^{\w_i+1}}\vev{\cdots} = - k b_i (\w_i+1) \vev{\cdots}\, ,
\end{equation}
as can be derived from \eqref{Gamma conditions at zi}. A second expression for this contour integral can be obtained by proceeding as in Sec.~\ref{sec: localization m=0}, i.e.~inserting
\begin{equation}
    J^-(\Gamma,z)= J^-(\Gamma-\Gamma_i+x_i,z)-2(\Gamma_i-x_i)J^3(\Gamma,z)-(\Gamma_i-x_i)^2 J^+(z) \, .
\end{equation}
We have already established that for this particular combination the last two terms on the RHS do not contribute. Hence, using \eqref{Gamma conditions at zi} once again leads precisely to the combination in Eq.~\eqref{def J-(a,b)}, namely
\begin{equation}
    \oint_{z_i}  dz \frac{\vev{J^-(\Gamma,z)\cdots}}{(z-z_i)^{\w_i+1}} = \vev{
    \of{J^-_{-\w_i-1}(a_i,b_i) V_{j_ih_i}^{\w_i}}(x_i,z_i) \prod_{j \neq i} V_{j_j,h_j}^{\w_j}(x_j,z_j)} \, .
\end{equation}
We conclude that
\begin{equation}
\label{RHS J- final}
    \vev{
    \off{\of{J^-_{-\w_i-1}(a_i,b_i)+kb_i(\w_i+1)} V_{j_ih_i}^{\w_i}}(x_i,z_i) \prod_{j \neq i} V_{j_j,h_j}^{\w_j}(x_j,z_j)} = 0 \, .
\end{equation}
Given that the mode $J^+_{\w_i}$ merely shifts $h_i \to h_i +1$ up to a numerical factor, this implies that the final term on the RHS of \eqref{KZ flowed states v3} vanishes.

Let us now move to the middle term on the RHS of \eqref{KZ flowed states v3}. In order to study the action of the mode $J^3_{-1}$ we  compute
\begin{align}
\begin{aligned}
    &\oint_{z_i}  dz \frac{\vev{J^3(\Gamma,z)\cdots}}{z-z_i} \\
    & \qquad = \vev{\of{J^3_{-1}(a_i,b_i)V_{j_i,h_i}^{\w_i}}(x_i,z_i) \prod_{p \neq i} V_{j_p,h_p}^{\w_p} (x_p,z_p)} - (\Gamma_i-x_i) \oint_{z_i}  dz \frac{\vev{J^+(z)\cdots}}{z-z_i} \, .
\end{aligned}
\end{align}
This follows from \eqref{Gamma conditions at zi} and
\begin{equation}
\label{J3Gamma near zi}
J^3(\Gamma,z) = J^3(\Gamma-\Gamma_i+x_i,z) - (\Gamma_i-x_i) J^+(z) \, .
\end{equation}
Turning the contour around leads to an alternative expression for this quantity. Importantly, we pick up contributions not only from the remaining insertion points, but also from the poles of the covering map.  We thus need an auxiliary result which follows directly from the relation \eqref{JGamma is dGamma}. By comparing the first non-trivial contribution to the expansion for both sides near one of the poles of the covering map $\Gamma(z)$, located at  $z=\lambda_a$, one finds that
\begin{equation}
\label{J+ lambda}
    c_a \vev{J^+(\lambda_a) \cdots } = k \vev{\cdots} \, .
\end{equation}
Recall that $c_a$ denotes the residue of $\Gamma(z)$ at $z=\lambda_a$. As a consequence, we have
\begin{equation}
    \oint_{\lambda_a} dz \frac{\vev{J^3(\Gamma,z)\cdots}}{z-z_i} = \frac{c_a \vev{J^+(\lambda_a)\cdots }}{z_i-\lambda_a} =  \frac{k \vev{\cdots}}{z_i-\lambda_a} \, .
\end{equation}
On the other hand, using an expansion analogous to \eqref{J3Gamma near zi} near any of the other insertion points gives
\begin{align}
    &\oint_{z_{j}\neq z_{i}}dz \frac{\vev{J^3(\Gamma,z)\cdots}}{z-z_i} \\
    & \qquad = -\frac{1}{z_{ij}}\vev{\of{J^3_{0}(a_j)V_{j_j,h_j}^{\w_j}}(x_j,z_j) \prod_{p \neq j} V_{j_p,h_p}^{\w_p} (x_p,z_p)} - (\Gamma_j-x_j) \oint_{z_{j}\neq z_{i}} dz \frac{\vev{J^+(z)\cdots}}{z-z_i} \, . \nn
\end{align}
We conclude that, for any of the correlators we are interested in, the negative mode under consideration gives
\begin{align}
\begin{aligned}
\label{RHS J3 final}
&\vev{\of{J^3_{-1}(a_i,b_i)V_{j_i,h_i}^{\w_i}}(x_i,z_i) \prod_{p \neq i} V_{j_p,h_p}^{\w_p} (x_p,z_p)} =  \sum_{j=1}^n (\Gamma_j-x_j) \oint_{z_j} dz \frac{\vev{J^+(z)\cdots}}{z-z_i} \\
& \qquad \qquad + \sum_{j \neq i} \frac{1}{z_{ij}}\vev{\of{J^3_{0}(a_j)V_{j_j,h_j}^{\w_j}}(x_j,z_j) \prod_{p \neq j} V_{j_p,h_p}^{\w_p} (x_p,z_p)} - \sum_{a=1}^N \frac{k \vev{\cdots}}{z_i-\lambda_a} \, .
\end{aligned}
\end{align}

Somewhat surprisingly, in Eq.~\eqref{RHS J3 final} the terms proportional to $(\Gamma_j-x_j)$ \textit{cannot} be discarded, and actually end up giving non-trivial contributions that are crucial for our holographic matching. They will be discussed in detail in  Sec.~\ref{sec: non-zero terms J+} below.

\subsection{An intermediate step}

Before continuing with the discussion of \eqref{KZ flowed states v3} we derive an explicit expression for the correlator $\vev{\der J^+(\lambda_a) \cdots}$. For this, we again consider Eq.~\eqref{JGamma is dGamma} and now focus on the first subleading order of the corresponding expansions near one of the poles of the covering map. By construction, $\der \Gamma(z)$ has no residue at $z = \lambda_a$. It follows that
\begin{equation}
\label{der J+ lambda id1}
    -2 \vev{J^3(\lambda_a) \cdots} + 2 d_a \vev{J^+(\lambda_a) \cdots} + c_a \vev{\der J^+(\lambda_a) \cdots} = 0 \, ,
\end{equation}
where the coefficients $c_a$ and  $d_a$ are defined by the expansion
\begin{equation}
    \Gamma(z\sim \lambda_a) =  \frac{c_a}{z-\lambda_a} + d_a + \cdots \, ,
\end{equation}
up to terms that vanish in the limit $z\to \lambda_a$. We also consider the contour integral
\begin{align}
\begin{aligned}
        \oint_{\lambda_a} \frac{\vev{J^3(\Gamma)\cdots}}{z-\lambda_a} & = \vev{J^3(\lambda_a) \cdots }-d_a \vev{J^+(\lambda_a) \cdots } - c_a \vev{\der J^+(\lambda_a) \cdots }  \, ,
\end{aligned}
\end{align}
where the contour encircles a given pole $\lambda_a$. Therefore, using the identity \eqref{der J+ lambda id1} we have
\begin{equation}
\label{der J+ lambda int}
    \oint_{\lambda_a} \frac{\vev{J^3(\Gamma)\cdots}}{z-\lambda_a}  = - \frac{1}{2}c_a \vev{\der J^+(\lambda_a) \cdots}  \, .
\end{equation}
Upon turning the contour around, we pick up contributions from the other poles of the covering map
and from the insertion points of the vertex operators. The former  give
\begin{equation}
    \sum_{b\neq a} \oint_{\lambda_b} \frac{\vev{J^3(\Gamma) \cdots}}{z-\lambda_a} = \sum_{b\neq a} \frac{c_b \vev{J^+(\lambda_b) \cdots}}{\lambda_{ab}} = \sum_{b\neq a} \frac{k \vev{\cdots}}{\lambda_{ab}} \, ,
\end{equation}
while, as in the previous section,
\begin{align}
    &\sum_{j=1}^n
    \oint_{z_j} \frac{\vev{J^3(\Gamma) \cdots}}{z-\lambda_a} \\
 & = \sum_{j=1}^n \off{
    \frac{1}{z_j-\lambda_a} \vev{\of{J^3_{0}(a_j)V_{j_j,h_j}^{\w_j}}(x_j,z_j) \prod_{p \neq j} V_{j_p,h_p}^{\w_p} (x_p,z_p)} - (\Gamma_j-x_j) \oint_{z_j} \frac{\vev{J^+(z)\cdots}}{z-\lambda_a}
    } \, . \nn
\end{align}
Moreover, for solutions of the form \eqref{eq:localisation solution} we have
\begin{equation}
    \vev{\of{J^3_{0}(a_j)V_{j_j,h_j}^{\w_j}}(x_j,z_j) \prod_{p \neq j} V_{j_p,h_p}^{\w_p} (x_p,z_p)} = \of{\frac{k}{2}\w_j - j_j} \vev{\cdots} \, .
\end{equation}
By combining the above results we conclude that, for each pole $\lambda_a$,
\begin{equation}
    c_a \vev{\der J^+(\lambda_a) \cdots} = \off{\sum_{b\neq a} \frac{2k}{\lambda_{ab}} +  \sum_{j=1}^n  \frac{k\w_j-2j_j}{z_j-\lambda_a}} \vev{\cdots}  - 2\sum_{j=1}^n(\Gamma_j-x_j) \oint_{z_j} \frac{\vev{J^+(z) \cdots}}{z-\lambda_a}\, .
\end{equation}
Finally, we may use the scattering relations \eqref{scattering equations} in order to rewrite this as
\begin{equation}
\label{derJ+ at lambda final}
    c_a \vev{\der J^+(\lambda_a) \cdots} = \sum_{j=1}^n \off{ \frac{k-2j_j}{z_j-\lambda_a}\vev{\cdots} - 2(\Gamma_j-x_j) \oint_{z_j} \frac{\vev{J^+(z)\cdots}}{z-\lambda_a} } \qquad \forall \, a = 1,\dots, N  \, .
\end{equation}
We will use this identity in the following section.

\subsection{Insertions of $J^+_{\w_i-1}$}

We now deal with the insertions of the mode $J^+_{\w_i-1}$, which appears twice in the KZ condition \eqref{KZ flowed states v3}. We use the expansion of the derivative of the covering map with respect to  $z_i$ near that same insertion point, namely Eq.~\eqref{dGammadzi near zi}, to compute the contour integral
\begin{align}
&
    \oint_{z_i}  dz \, \der_{z_i} \Gamma \vev{J^+(z)\cdots}= \der_{z_i} \Gamma_i \,   \der_{x_i}\vev{\cdots}-
    a_i \w_i \vev{\of{ J^+_{\w_i-1} V_{j_i,h_i}^{\w_i}}(x_i,z_i) \prod_{p \neq i} V_{j_p,h_p}^{\w_p} (x_p,z_p)} \nn \\
& \qquad
+ \off{\der_{z_i} a_i- b_i (\w_i+1)} \vev{\of{J^+_{\w_i} V_{j_i,h_i}^{\w_i}}(x_i,z_i) \prod_{p \neq i} V_{j_p,h_p}^{\w_p} (x_p,z_p)}
 \, .
 \label{J+wi id1}
\end{align}
Turning the contour around, Eq. \eqref{dGammadzi near zj} implies that we get contributions from the remaining insertion points of the form
\begin{equation}
     \oint_{z_j\neq z_i}  dz \, \der_{z_i} \Gamma \vev{J^+(z)\cdots}= \der_{z_i} \Gamma_j \,   \der_{x_j}\vev{\cdots}+
    \der_{z_i}a_j\vev{\of{ J^+_{\w_j} V_{j_j,h_j}^{\w_j}}(x_j,z_j) \prod_{p \neq j} V_{j_p,h_p}^{\w_p} (x_p,z_p)} \, . 
\end{equation}
We also have contributions from the poles of the covering map, where the integrand actually has double poles, see Eq.~\eqref{dGammadzi near lambda}. The corresponding residues read
\begin{align}
\begin{aligned}
    &\oint_{\lambda_a}  dz \, \der_{z_i} \Gamma \vev{J^+(z) \cdots}
    = \der_{z_i} c_a \vev{J^+(\lambda_a) \cdots} + c_a \vev{\der J^+(\lambda_a)\cdots } \der_{z_i} \lambda_a \\
    & \qquad = k\frac{\der_{z_i}c_a}{c_a} \vev{\cdots}  + \sum_{j=1}^n \off{ \frac{k-2j_j}{z_j-\lambda_a}\vev{\cdots} - 2(\Gamma_j-x_j) \oint_{z_j} \frac{\vev{J^+(z)\cdots}}{z-\lambda_a} } \der_{z_i}\lambda_a \, .
\end{aligned}
\end{align}
In the last line we have used Eqs. \eqref{J+ lambda} and \eqref{derJ+ at lambda final}.
Putting everything together and using \eqref{JwVx}, we find that, after inserting the LHS of Eq.~\eqref{KZ flowed states v3} in a given primary correlator, we can replace
\begin{align}
\label{J+ term LHS}
& \vev{\off{ \of{ a_i\w_i J^+_{\w_i-1} + b_i(\w_i+1)J^+_{\w_i}}V_{j_i,h_i}^{\w_i}}(x_i,z_i) \prod_{j \neq i} V_{j_j,h_j}^{\w_j} (x_j,z_j)}
= \\
        &\offf{ \sum_{j=1}^n \off{\der_{z_i} \Gamma_j \, \der_{x_j} + \frac{\der_{z_i} a_j}{a_j}  \of{h_j - \frac{k}{2}\w_j + j_j}} + \sum_{a=1}^N  k\frac{\der_{z_i} c_a}{c_a} + \sum_{a=1}^{N}\sum_{j=1}^{n}  \frac{(k-2j_j)\der_{z_i}\lambda_a}{z_j-\lambda_a} }\vev{\cdots}
        \nn \\
        & \qquad \quad - \sum_{a=1}^{N}\sum_{j=1}^{n} \off{  2(\Gamma_j-x_j) \oint_{z_j} \frac{\vev{J^+(z)\cdots}}{z-\lambda_a} } \frac{\der \lambda_a}{\der z_i} \nn \, .
\end{align}
Combined with the results of the previous sections, this shows that all the terms proportional to $b_i$ that were present in the KZ condition as written in Eq.~\eqref{KZ flowed states v3} cancel out.

There is also another instance where $J^+_{\w_i-1}$ appears: the first term on the RHS of \eqref{KZ flowed states v3}. This must be treated carefully. As in the singly- owed case, due to Eq.~\eqref{def J-(a)}, inside a primary correlator of the form  \eqref{eq:localisation solution}  we can replace
\begin{align}
\begin{aligned}
    J^-_{-\w_i}(a_i) &= J^-_{-\w_i}(x_i) -2 a_i \off{J^3_0(x_i) - \frac{k}{2}\w_i} + a_i^2 J^+_{\w_i} \\
    & \to
    a_i \off{\of{h_i - \frac{k}{2}\w_i - j_i} -2 \of{h_i-\frac{k}{2}\w_i} + \of{h_i - \frac{k}{2}\w_i + j_i}} = 0\, .
\end{aligned}
\end{align}
However, $J^-_{-\w_i}(a_i)$ is not acting alone in \eqref{KZ flowed states v3} due to the presence of the mode $J^+_{\w_i-1}$. In this case one \textit{does} pick up a non-zero contribution, which originates from the term proportional to $J^+_{\w_i}$ in Eq.~\eqref{J+wi id1}. Roughly speaking, this is because one cannot replace a shift $h_i \to h_i \pm 1$ with a power of $a_i^{\pm 1}$ if the correlator is not a primary one. More explicitly, we have
\begin{align}
    & J^+_{\w_i} J^-_{-\w_i}(a_i)V_{j_ih_i}^{\w_i} \nn\\
    & = J^+_{\w_i} \off{
    \of{h_i -\frac{k}{2}\w_i-j_i}
    V_{j_i,h_i-1}^{\w_i} - 2 a_i \of{h_i - \frac{k}{2}\w_i}
    V_{j_i,h_i}^{\w_i} + a_i^2
    \of{h_i -\frac{k}{2}\w_i+j_i}
    V_{j_i,h_i+1}^{\w_i}
    } \nn \\
    & = \of{h_i-1 -\frac{k}{2}\w_i+j_i}\of{h_i -\frac{k}{2}\w_i-j_i}
    V_{j_i,h_i}^{\w_i} - 2 a_i \of{h_i -\frac{k}{2}\w_i+j_i}\of{h_i - \frac{k}{2}\w_i}
    V_{j_i,h_i+1}^{\w_i} \nn \\
    &\qquad + a_i^2
    \of{h_i+1 -\frac{k}{2}\w_i+j_i}\of{h_i -\frac{k}{2}\w_i+j_i}
    V_{j_i,h_i+2}^{\w_i}\nn \\
    & \to \left[\of{h_i-1 -\frac{k}{2}\w_i+j_i}\of{h_i -\frac{k}{2}\w_i-j_i}  - 2 \of{h_i -\frac{k}{2}\w_i+j_i}\of{h_i - \frac{k}{2}\w_i} \right. \nn
    \\
    & \qquad \left.+
    \of{h_i+1 -\frac{k}{2}\w_i+j_i}\of{h_i -\frac{k}{2}\w_i+j_i}\right]
    V_{j_i,h_i}^{\w_i} = 2j_i V_{j_i,h_i}^{\w_i} \, ,
\end{align}
where the final replacement holds inside a primary correlator of the form \eqref{eq:localisation solution}. Hence, we find that in our KZ equation
\begin{equation}
\label{RHS J+ final}
    \vev{\off{ J^+_{\w_i-1} {J^-_{-\w_i}}(a_i)V_{j_i,h_i}^{\w_i}}(x_i,z_i) \prod_{p \neq i} V_{j_p,h_p}^{\w_p} (x_p,z_p)} = \off{\frac{\p_{z_i}a_i -b_i({\w_i}+1)}{a_i \w_i}} 2j_i \vev{\cdots} \, .
\end{equation}

\subsection{KZ equations in flowed sectors: part II}

We now combine the results obtained in the previous subsections to see how a differential equation for the function $W_\Gamma$ appearing in Eq.~\eqref{eq:localisation solution} emerges from the KZ condition \eqref{KZ flowed states v3}. For now, we ignore the terms proportional to $(\Gamma_j-x_j)$, as they will be addressed later on.

We first analyze the LHS of \eqref{KZ flowed states v3}. From the identity in Eq.~\eqref{J+ term LHS}, and upon including the Virasoro mode $L_{-1}$, the combination of derivatives acting on the primary correlator is of the form
\begin{equation}
\label{chain rule deltas 1}
    \text{d}_i = \der_{z_i} + \sum_{j=1}^n \offf{\der_{z_i} \Gamma_j \der_{x_j} +h_j\frac{\p_{z_i}a_j}{a_j}}\, .
\end{equation}
Note that the terms involving $\der_{x_j}$ with $j \leq 3$ inside the sum actually vanish since the corresponding $\Gamma_j$ are fixed.
The operator ${\rm d}_i$ generalizes the combination $\der_{z_i} + \der_{x_i}$ appearing in the singly-flowed case. Indeed, by the chain rule we have
\begin{align}
\begin{aligned}
	&\text{d}_i \off{\prod_{j=1}^n|a_j^{\Gamma}|^{-2h_j}\prod_{j=4}^n \delta^{(2)}\of{x_j-\Gamma_j}|W_\Gamma(z_1,\cdots,z_n)|^2} \\
    &\hspace{3cm} = \prod_{j=1}^n|a_j^{\Gamma}|^{-2h_j}\prod_{j=4}^n \delta^{(2)}\of{x_j-\Gamma_j}\p_{z_i}|W_\Gamma(z_1,\cdots,z_n)|^2\,,\label{chain rule deltas 2}
\end{aligned}
\end{align}
where we have used that $W_\Gamma$ does not depend on $x_4,\dots,x_n$.

The remaining contributions in Eq.~\eqref{J+ term LHS} can be rewritten in a way that makes the similarities with the differential equation for symmetric orbifold correlators derived in \cite{Dei:2019iym} manifest. This is achieved by using the covering map relations given in Eqs.~\eqref{Identity bi 1}-\eqref{Identity bi 2}, giving
\begin{align}
\label{rewriting LHS with cov map identities}
\sum_{j=1}^n  \frac{\der_{z_i} a_j}{a_j}  \of{j_j - \frac{k}{2}\w_j } + \sum_{a=1}^N  k\frac{\der_{z_i} c_a}{c_a} &+ \sum_{a,j=1}^{N,n}  \frac{(k-2j_j)\der_{z_i}\lambda_a}{z_j-\lambda_a} =    \\
& \hspace{-1.8cm}  -
    (\w_i-1) \sum_{j \neq i} \frac{\of{\frac{k}{2}-j_j} }{z_{ij}}- \frac{k}{2} \frac{b_i (\w_i^2-1)}{a_i \w_i} - \of{\frac{k}{2}-j_i} \frac{b_i(\w_i+1)}{a_i\w_i}\nn\\
    &\hspace{-1.8cm}
    -\left(\frac{k}{2}+n-3\right)\frac{\p_{z_i}C_\G}{C_\G} - \sum_{j=1}^n  \frac{k}{4}(\w_j-1)\frac{\der_{z_i} a_j}{a_j} + \sum_{a=1}^N  \frac{k}{2}\frac{\der_{z_i} c_a}{c_a}
        \, .\nn
\end{align}
The final three terms on the RHS of this expression are exactly what we expect. Indeed, once integrated they imply that, as anticipated in Eq.~\eqref{Final WGamma m=0}, we must have
\begin{equation}
    W_\Gamma(z_1,\dots,z_n) \propto C_\G^{\frac{k}{2}}\prod_{a=1}^N c_a^{-\frac{k}{2}} \prod_{i=1}^n a_i^{\frac{k}{4}(\w_i-1)} \, .
\end{equation}

It remains to show that the remaining terms on the RHS of Eq.~\eqref{rewriting LHS with cov map identities} cancel most of the contributions coming from the RHS of the KZ equation \eqref{KZ flowed states v3}. For this, we need to use the results obtained in Eqs.~\eqref{RHS J- final}, \eqref{RHS J3 final} and \eqref{RHS J+ final}. For localized correlators such as \eqref{eq:localisation solution}, we obtain
\begin{align}
    &\vev{\off{\of{J^+_{\w_i-1}J^-_{-\w_i}(a_i)-2J^3_{-1}(a_i,b_i)\of{J^3_0(a_i)-\w_i}}V_{j_i,h_i}^{\w_i}}(x_i,z_i) \prod_{p \neq i} V_{j_p,h_p}^{\w_p} (x_p,z_p)}\sim\\
    & \offf{\frac{2j_i}{a_i \w_i} \off{\der_{z_i}a_i- b_i(\w_i+1)} - \off{(k-2)\w_i-2j_i}\off{\sum_{j \neq i} \frac{\of{\frac{k}{2}\w_j-j_j}}{z_{ij}} - \sum_{a=1}^N \frac{k}{z_i-\lambda_a}}} \vev{\cdots} \,,\nn
\end{align}
where the symbol $\sim$ is there to remind the reader that these identities hold only up to terms proportional to $(\Gamma_i-x_i)$. Using once again Eqs.~\eqref{Identity bi 1}-\eqref{Identity bi 2} then allows us to rewrite this as
\begin{align}
    &\vev{\off{\of{J^+_{\w_i-1}J^-_{-\w_i}(a_i)-2J^3_{-1}(a_i,b_i)\of{J^3_0(a_i)-\w_i}}V_{j_i,h_i}^{\w_i}}(x_i,z_i) \prod_{p \neq i} V_{j_p,h_p}^{\w_p} (x_p,z_p)} \sim\nn \\
    & \qquad  \left\{
    -2\sum_{j\neq i} \frac{\of{\frac{k}{2}-j_i-1}\of{\frac{k}{2}-j_j-1}}{z_{ij}} + \frac{2j_i}{\w_i} \of{\frac{\der_{z_i}a_i}{a_i} + \sum_{j\neq i} \frac{\w_i}{z_{ij}}} - \sum_{j \neq i }\frac{k-2}{z_{ij}}
    \right.
    \label{rewriting RHS with cov map identities}\\
    &  \qquad \left. -(k-2)\off{\frac{k}{2}\frac{b_i(\w_i^2-1)}{a_i\w_i} + \of{\frac{k}{2}-j_i} \frac{b_i (\w_i+1)}{a_i\w_i} + \sum_{j \neq i } \frac{\of{\frac{k}{2}-j_j}(\w_i-1)}{z_{ij}}}\right\}\vev{\cdots} \, . \nn
\end{align}
As anticipated above, the terms in the last line of \eqref{rewriting RHS with cov map identities} precisely cancel the first three terms on the RHS of \eqref{rewriting LHS with cov map identities}.

The computation above implies that, after reinstating the terms proportional to \\ $(\Gamma_j-x_j)$, the full KZ equation can be expressed as\footnote{In order to derive this equation, we assumed that the correlator \(\vev{J^+(z)\cdots}\) shares the same $h_i$ dependence as the primary correlator given by \eqref{eq:localisation solution}. Although this is by no means obvious at this point, we will show it is true in Sec.~\ref{sec: general results}.}
\begin{equation}
	\begin{aligned}
		& \offf{\text{d}_i -
			\sum_{j=1}^n \frac{k}{4}(\w_j-1)\frac{\der_{z_i} a_j}{a_j} + \sum_{a=1}^N  \frac{k}{2}\frac{\der_{z_i} c_a}{c_a} + \sum_{j \neq i} \frac{q_i q_j}{z_{ij}}-\of{\frac{k}{2}+(n-3)}\frac{\p_{z_i}C_\G}{C_\G}
		} \vev{\cdots}  \\
		& +
		\sum_{j \neq i} \frac{1}{z_{ij}}\vev{\cdots} +\sum_{j=1}^n (\Gamma_j-x_j) \off{ \w_i \oint_{z_j} dz \frac{\vev{J^+(z)\cdots}}{z-z_i}
			- 2 \sum_{a=1}^N \der_{z_i}\lambda_a \oint_{z_j} dz \frac{\vev{J^+(z)\cdots}}{z-\lambda_a}}
		\\
		& = \frac{2j_i}{k-2} \off{\of{\frac{\der_{z_i}a_i}{\w_i a_i}+\sum_{j \neq i} \frac{1}{z_{ij}}}\vev{\cdots} + \sum_{j=1}^n (\Gamma_j-x_j) \oint_{z_j} dz \frac{\vev{J^+(z)\cdots}}{z-z_i}} ,
	\end{aligned}	\label{KZ almost final}
\end{equation}
where we have used the shorthands  $q_i$ defined in \eqref{qi def}. This identity admits the following interpretation.
We first note that, once \eqref{chain rule deltas 2} is taken into account, the first line of \eqref{KZ almost final} contains precisely all terms that define the boundary correlator. This can be seen by comparing with the results of \cite{Dei:2019iym}. More explicitly, in a symmetric orbifold theory one expects that correlators involving twist field insertions factorize into a product of coefficients associated with the covering map data, and a correlator computed in the seed CFT. In our notation, the former correspond to the terms proportional to $\der_{z_i}a_j$, $\der_{z_i} c_a$ and $\der_{z_i} C_\G$, while the latter corresponds to the factor proportional to $q_iq_j$, which is the appropriate one for the Liouville-type linear dilaton sector appearing in the proposal of
\cite{Eberhardt:2021vsx}, see Sec.~\ref{sec: string correlators def}.


In order to discuss the remaining parts of Eq.~\eqref{KZ almost final} we need to carefully deal with the terms proportional to $(\Gamma_j-x_j)$, which involve contour integrals of correlators with insertions of $J^+(z)$. Somewhat miraculously, it turns out that the third line in \eqref{KZ almost final} adds up to zero, while the second one accounts for the last missing factor in the final solution \eqref{Final WGamma m=0}: the Jacobian associated with the change of variables from $\{z_{4},\dots,z_{n}\}$ to $\{\Gamma_{4},\dots,\Gamma_{n}\}$. This is needed for the (residue of the) string correlation function involving an integration over the worldsheet moduli to give the result obtained from the holographic CFT. Indeed, once the Jacobian is inserted, the integration can be carried out trivially using the product of Dirac delta functions in Eq.~\eqref{eq:localisation solution}.

\section{Correlators with arbitrary spectral flow charges: results}
\label{sec: general results}

Here we finish the analysis for $n$-point functions with arbitrary spectral flow charges satisfying the $j$-constraint. To gain some intuition and make the presentation clear, we discuss three- and four-point functions first, and then derive the solution to the KZ equations in the general case.

\subsection{Three-point functions}
\label{sec: three-point functions KZ}

Correlation functions involving three flowed insertions are known exactly \cite{Dei:2021xgh,Bufalini:2022toj}. This was done without mentioning KZ equations because the dependence on the worldsheet insertion points $z_i$ and boundary insertion points $x_i$ is easily determined using the global Ward identities. Nevertheless, the numerous subtleties that arise when discussing the KZ constraints for correlators of spectrally flowed operators make it instructive to discuss how Eq.~\eqref{KZ almost final} is satisfied in this context. We use this to compute the corresponding momentum-space residues of the string amplitude directly, and show that they match the results of \cite{Eberhardt:2021vsx} ---which also agrees with the holographic prediction--- obtained originally by a direct evaluation of the exact three-point function.

As discussed in Sec.~\ref{sec: covering maps}, for cases with $n=3$, the conditions
\begin{equation}
    \w_i \geq 1 \, , \qquad
    \sum_{j=1}^3\w_j \, \in \, 2\mathbb{Z}+1
    \, , \qquad
    \w_i - \sum_{j\neq i}\w_j < 1  \qquad \forall \, i = 1,2,3 \, ,
\end{equation}
are not only necessary but also sufficient conditions for the existence of an appropriate covering map satisfying $\Gamma(z\sim z_i)=x_i + a_i(z-z_i)^{\w_i}+ \cdots$ for $i=1,2,3$. The resulting function $\Gamma(z)$ can be derived analytically, see \cite{Lunin:2000yv}. In particular, one obtains
\begin{equation}
    a_i =
    \left(
    \begin{array}{c}
        \frac{\w_i+\w_{i+1}+\w_{i+2}-1}{2}  \\
        \frac{-\w_{i}+\w_{i+1}+\w_{i+2}-1}{2}
    \end{array}\right)
   \left(
    \begin{array}{c}
        \frac{-\w_{i}+\w_{i+1}-\w_{i+2}-1}{2} \\
        \frac{\w_{i}+\w_{i+1}-\w_{i+2}-1}{2}
    \end{array}\right)^{-1}
    \frac{x_{i,i+1}x_{i+2,i}}{
    x_{i+1,i+2}}
    \left(\frac{z_{i+1,i+2}}{
    z_{i,i+1}z_{i+2,i}}\right)^{\w_i} \,,
    \label{ai 3pt}
\end{equation}
where the subscripts should be understood mod 3.
Assuming that the $j$-constraint is satisfied, the ansatz \eqref{eq:localisation solution} reads
\begin{equation}
    \vev{\prod_{i=1}^3V_{j_ih_i}^{\w_i}(x_i,z_i)} = \prod_{i=1}^3 |a_i|^{-2h_i}|W_\Gamma(z_1,z_2,z_3)|^2 \, .
    \label{loc ansatz 3pt unfixed}
\end{equation}
Given that there are no additional insertions, we have $\Gamma_i - x_i = 0$ for all $i$, hence we can ignore all terms in the KZ equation \eqref{KZ almost final} containing the factor $\vev{J^+(z)\cdots}$. As anticipated above, this means that the third line in \eqref{KZ almost final} vanishes since Eq.~\eqref{ai 3pt} implies that
\begin{equation}
    \frac{\der_{z_i}a_i}{\w_i a_i}+\sum_{j \neq i} \frac{1}{z_{ij}} = 0 \, .
\end{equation}

In the second line of \eqref{KZ almost final}, however, we do have a factor $\sum_{j \neq i} z_{ij}^{-1}$ that must be taken into account. Inserting \eqref{loc ansatz 3pt unfixed} into the KZ constraint \eqref{KZ almost final} thus gives a differential equation for the function $W_\Gamma$ of the form
\begin{equation}
    \der_{z_i} \log \of{W_\G} = \sum_{j=1}^3 \off{\frac{k}{4}(\w_j-1)}\frac{\der_{z_i} a_j}{a_j} - \sum_{a=1}^N  \frac{k}{2}\frac{\der_{z_i} c_a}{c_a} + \frac{k}{2}\frac{\p_{z_i}C_\G}{C_\G} - \sum_{j \neq i} \frac{q_i q_j+1}{z_{ij}}  \, ,
\end{equation}
where $q_i$ was defined in \eqref{qi def}.
This is easily integrated, giving
\begin{equation}
\label{WGamma final 3pt}
    |W_\Gamma(z_1,\cdots,z_n)|^2 =
    \cN_3(j_i,\w_i) \Big|C_\Gamma^{\frac{k}{2}}\prod_{a=1}^N c_a^{-\frac{k}{2}} \prod_{i=1}^3 a_i^{\frac{k}{4}(\w_i-1)}  \prod_{i<j} z_{ij}^{-q_i q_j}
    \times (z_{12} z_{23} z_{13})^{-1}\Big|^2\, .
\end{equation}
The normalization constant $\cN_3$ must be chosen such that it matches the exact residue of the worldsheet correlator.  This was computed in \cite{Eberhardt:2021vsx}, and it leads to
\begin{equation}
    \cN_3(j_i,\w_i) = N_k\prod_{j=1}^3|\w_j^{-\frac{k}{4} (\w_j+1) +j_j}|^2\,,\label{3point normalization}
\end{equation}
with
\begin{equation}
    N_k = \frac{\nu^{\frac{k}{2}-2}}{2 \pi (k-2)^2 \gamma \left(\frac{k-1}{k-2}\right)} \,.\label{Nk factor}
\end{equation}
Notice that, in case we fix the worldsheet and spacetime insertion points at $(0,1,\infty)$, the normalization must be taken as
\begin{equation}
    \cN^{\infty}_3(j_i,\w_i)=N_k|\w_1^{-\frac{k}{4} (\w_1+1) +j_1}\w_2^{-\frac{k}{4} (\w_2+1) +j_2}\w_3^{\frac{k}{4} (\w_3+1) +j_3}|^2\,.\label{3point normalization infinity}
\end{equation}
In Eq.~\eqref{WGamma final 3pt} we have chosen to separate a factor that exactly cancels the usual $c$-ghost correlator
\begin{equation}
    \vev{c(z_1)\bar{c}{(\bz_1)}c(z_2)\bar{c}{(\bz_2)}c(z_3)\bar{c}{(\bz_3)}} = |z_{12}z_{23}z_{13}|^2 \, .
\end{equation}
This must be included when computing the full string correlator. Since there is no additional integration over worldsheet moduli, we can directly read off the full (residue of the) genus-zero string amplitude:
\begin{align}
     C_{S^2} N_k
    \Big|C_\Gamma^{\frac{k}{2}}\prod_{a=1}^N c_a^{-\frac{k}{2}} \prod_{i=1}^3 \w_i^{-\frac{k}{4}(\w_i+1)+j_i}a_i^{-h_i+\frac{k}{4}(\w_i-1)}  \prod_{i<j} z_{ij}^{-q_iq_j} \Big|^2\, ,
\end{align}
which matches the results from \cite{Eberhardt:2021vsx,Knighton:2024qxd}, see Eq.~\eqref{HCFT correlator m=0 unfixed}.

\subsection{Four-point functions}

We now discuss the case $n=4$, for which the localized ansatz \eqref{eq:localisation solution} gives
\begin{align}
\Bigl\langle \prod_{i=1}^4V_{j_ih_i}^{\w_i}(x_i,z_i)\Bigr\rangle
=\sum_{\Gamma} \prod_{i=1}^4 |a_i|^{-2h_i}  \delta^{(2)}(x_4-\Gamma_4) \, |W_\Gamma(z_1,\dots,z_4)|^2\ .
\end{align}
The function $W_\Gamma$ must be such that the KZ constraint is satisfied. For four-point functions there is only one non-zero factor  $(\Gamma_j-x_j)$: the one with $j=4$. Hence, Eq.~\eqref{KZ almost final} reads
\begin{equation}
	\begin{aligned}
		& \offf{\text{d}_i -
			\sum_{j=1}^4 \frac{k}{4}(\w_j-1)\frac{\der_{z_i} a_j}{a_j} + \sum_{a=1}^N  \frac{k}{2}\frac{\der_{z_i} c_a}{c_a} + \sum_{j \neq i} \frac{q_i q_j}{z_{ij}}-\of{\frac{k}{2}+1}\frac{\p_{z_i}C_\G}{C_\G}
		} \vev{\cdots}  \\
		& +
		\sum_{j \neq i} \frac{1}{z_{ij}}\vev{\cdots} +(\Gamma_4-x_4) \off{ \w_i \oint_{z_4} dz \frac{\vev{J^+(z)\cdots}}{z-z_i}
			- 2 \sum_{a=1}^N \der_{z_i}\lambda_a \oint_{z_4} dz \frac{\vev{J^+(z)\cdots}}{z-\lambda_a}}
		\\
		& = \frac{2j_i}{k-2} \off{\of{\frac{\der_{z_i}a_i}{\w_i a_i}+\sum_{j \neq i} \frac{1}{z_{ij}}}\vev{\cdots} +  (\Gamma_4-x_4) \oint_{z_4} dz \frac{\vev{J^+(z)\cdots}}{z-z_i}},
	\end{aligned}	\label{KZ almost final 4pt}
\end{equation}
for $i=1,\dots,4$. In order to turn this into a differential equation for $W_\Gamma$, we need to deal with the terms containing the function
\begin{equation}
\mathfrak{J}^+(z) \equiv    \vev{J^+(z) \prod_{p=1}^4 V_{j_p,h_p}^{\w_p}(x_p,z_p)} \, .
\end{equation}
\textit{A priori}, all we know about $\mathfrak{J}^+(z)$ is that it goes to zero as $1/z^2$ at infinity and  has poles of order $\w_p+1$ at each of the worldsheet insertion points. In principle, the local Ward identities reviewed in Sec.~\ref{sec:recursions} fix  $\mathfrak{J}^+(z)$ in terms of the descendant correlators $F^{+,j}_{\ell}$ defined in \eqref{defFin}. In practice, however, the explicit form and solutions of this complicated linear system are not known in full generality. Moreover, this becomes increasingly harder for $n$-point functions with $n>4$. We thus need a different strategy.

We first note that we only actually need to study the related  function
\begin{equation}
    K^{(4)}(z) \equiv \of{\Gamma_4 - x_4} \vev{J^+(z) \prod_{p=1}^4 V_{j_p,h_p}^{\w_p}(x_p,z_p)} \, .
\end{equation}
As it turns out, our results from the previous sections imply a considerable number of additional constraints that must be satisfied by this function, and that will allow us to fix it completely (as long as the primary four-point function satisfies the localization property \eqref{eq:localisation solution}). This  goes as follows.
\begin{itemize}
    \item As for $\mathfrak{J}^+(z)$, the fact that $J^+(z)$ is a holomorphic current of weight one guarantees that $K^{(4)}(z)$ must vanish at infinity (at least) as fast as $1/z^2$.

    \item The behavior of $\mathfrak{J}^+(z)$ near each of the poles of the covering map is fixed by the identities \eqref{J+ lambda} and \eqref{derJ+ at lambda final}. We have
    \begin{align}
     &   \vev{J^+(z \sim \lambda_a) \cdots}  \\
     & \qquad =  \frac{k}{c_a} \vev{\cdots} + \frac{z-\lambda_a}{c_a}  \off{ \sum_{j=1}^4\frac{k-2j_j}{z_j-\lambda_a}\vev{\cdots} - 2(\Gamma_4-x_4) \oint_4 \frac{\vev{J^+(z)\cdots}}{z-\lambda_a} } + \cdots \, , \nn
    \end{align}
   up to terms quadratic in $(z-\lambda_a)$. This implies that, when multiplied by $(\Gamma_4-x_4)$, the first two orders vanish since the corresponding coefficients are either given by primary correlators, which according to \eqref{eq:localisation solution} are proportional to the delta function $\delta(\Gamma_4-x_4)$, or already linear in $(\Gamma_4-x_4)$. It follows that $K^{(4)}(z)$ must have double zeros at $z=\lambda_a$  for all $a = 1,\dots, N$.

    \item The OPEs \eqref{JVxOPE} show that, for a given $j=1,\dots,4$,  near $z=z_j$ one can expand
    \begin{equation}
        \vev{J^+(z \sim z_j) \prod_{p=1}^n V_{j_p,h_p}^{\w_p}(x_p,z_p)} =
        \frac{h_j - \frac{k}{2}\w_j +j_j}{(z-z_j)^{\w_j+1}}\vev{ \prod_{p =1}^n V_{j_p, h_p+\delta_{jp}}^{\w_p}(x_p,z_p)} + \cdots\, ,
    \end{equation}
    up to higher-order terms in $(z-z_j)$. Now, since the coefficient of this higher order pole is proportional to a primary correlator, it will once again vanish when multiplied  by $(\Gamma_4-x_4)$. It follows that $K^{(4)}(z)$ can have at most poles of order $\w_j$ at the worldsheet insertion points $z_j$ (as opposed to poles of order $\w_j+1$) for all $j=1,\dots,4$.
\end{itemize}
Combined with the relation between the total number of poles $N$ and the spectral flow charges $\w_j$ given in Eq.~\eqref{Ndef}, this implies that $K^{(4)}(z)$ must take the form
\begin{equation}
\label{Ki(z) functions 1 4pt}
    K^{(4)}(z) = \frac{\prod_{a=1}^N(z-\la_a)^2}{\prod_{j=1}^{4}(z-z_j)^{\w_j}}P_{0} = \frac{C_\Gamma}{\der\Gamma(z)} \frac{P_0}{\prod_{j=1}^4 (z-z_j)}\, ,
\end{equation}
for some constant $P_0$. In the final equality we have used Eq.~\eqref{dGamma}.
Moreover, the coefficients of the poles of order $\w_j$ at $z=z_j$ are actually known. Near, say, $z=z_4$, we have
\begin{equation}
     K^{(4)} (z \sim z_4)  = \frac{\Gamma_4 - x_4}{(z-z_4)^{\w_4}}
        \vev{\of{J^+_{\w_4-1} V_{j_4h_4}^{\w_4}}(x_4,z_4)\prod_{p\neq 4} V_{j_p,h_p}^{\w_p}(x_p,z_p)}+\cdots \, .
\end{equation}
This descendant correlator can be expressed in terms of primary ones and their derivatives using \eqref{J+ term LHS}. Most of the resulting terms vanish upon multiplication  by $(\Gamma_4-x_4)$. The distributional identity in Eq.~\eqref{deltap identity} implies that the only term that survives is the one proportional to $\der_{x_4}$, giving
\begin{equation}
     K^{(4)} (z \sim z_4)  = \frac{(\w_4a_4)^{-1}\der_{z_4}\Gamma_4}{(z-z_4)^{\w_4}}
        \vev{\prod_{p=1}^4 V_{j_p,h_p}^{\w_p}(x_p,z_p)}+\cdots \, .
\end{equation}
Since $\der \Gamma(z \sim z_4) = \w_4 a_4(z-z_4)^{\w_4-1} + \cdots$, from \eqref{Ki(z) functions 1 4pt} we get
\begin{equation}
P_0 = z_{41}z_{42}z_{43} \of{C_\Gamma}^{-1}\der_{z_4}\Gamma_4 \vev{ \cdots}  \, ,
\end{equation}
where $\vev{ \cdots}$ stands for the relevant primary four-point function. In other words,
\begin{equation}
    K^{(4)}(z) = \hat{K}^{(4)} (z) \vev{\cdots}\,, \qquad \hat{K}^{(4)}(z) \equiv \frac{z_{41}z_{42}z_{43}}{(z-z_1)(z-z_2)(z-z_3)(z-z_4)}\frac{\der_{z_4}\G_4}{\der \G(z)}.
    \label{K4(z) final 4pt}
\end{equation}

We now use this result to prove that the third line in Eq.~\eqref{KZ almost final 4pt} vanishes.
From Eq.~\eqref{K4(z) final 4pt} we see that this is equivalent to proving that
\begin{equation}
\label{2j term equals 0 4pt}
    \frac{\der_{z_i}a_i}{\w_i a_i}+\sum_{j \neq i} \frac{1}{z_{ij}} +   \oint_{z_4} dz \frac{\hat{K}^{(4)}(z)}{z-z_i} = 0 \,.
\end{equation}
This is an identity which depends only on the structure of the covering maps themselves; all information related to the worldsheet theory \textit{per se} has now been stripped off. Although one can easily check numerically that \eqref{2j term equals 0 4pt} holds  on a case-by-case basis, say by using the ancillary code of \cite{Dei:2021yom}, we will provide an analytic proof.

It will be useful to consider the auxiliary function
\begin{equation}
    \hat{K}(z) \equiv \frac{z_{41}z_{42}z_{43}}{(z-z_1)(z-z_2)(z-z_3)(z-z_4)}\frac{\der_{z_4}\G(z)}{\der \G(z)}\,.
    \label{Ki(z) final 4pt}
\end{equation}
This has no residue at infinity since at large $z$ we have $\der \Gamma(z) \sim 1/z^2$ while $\der_{z_4}\G(z)\sim \der_{z_4}\Gamma_\infty$. Eqs.~\eqref{dGamma} and \eqref{dGammadzi near lambda} further imply that $\hat{K}(z)$ is regular at the poles of the covering map. This shows that
\begin{equation}
    \sum_{j=1}^4 \oint_{z_j} dz \frac{\hat{K}(z)}{z-z_i} = 0  \qquad \forall \, i=1,2,3,4.
    \label{no residue for K(z) 4pt}
\end{equation}
As for the worldsheet insertion points, Eqs.~\eqref{Gamma conditions at zi}, \eqref{dGammadzi near zj} and \eqref{dGammadzi near zi} imply that, up to linear order,  we have
\begin{equation}
    \off{\frac{\p_{z_4}\G(z)- \p_{z_4}\G_4}{\p\G(z)}}_{z \sim z_4} = - 1 + \frac{\p_{z_4}a_4}{\w_4a_4}(z-z_4)+\cdots \, ,
    \label{L(z) expansion 4pt near z4}
\end{equation}
and
\begin{equation}
    \off{\frac{\p_{z_4}\G(z)}{\p\G(z)}}_{z \sim z_j} =  \frac{\p_{z_4}a_j}{\w_ja_j}(z-z_j)+\cdots \, , \qquad \forall \, j = 1,2,3 \, .
    \label{L(z) expansion 4pt near z123}
\end{equation}
From now on, we need to distinguish the cases \(i=1,2,3\) from \(i=4\). We first consider the latter case. When $i=4$, the first three terms in the sum \eqref{no residue for K(z) 4pt} vanish. The only non-trivial contribution comes from the integral around $z=z_4$, which gives
\begin{align}
   \oint_{z_4} dz \frac{\hat{K}(z)}{z-z_4}  &= \frac{1}{z_{41}} + \frac{1}{z_{42}}+ \frac{1}{z_{43}} + \frac{\der_{z_4}a_4}{\w_4a_4} + \oint_{z_4} dz
     \frac{\hat{K}^{(4)}(z)}{z-z_4}\, .
\end{align}
Imposing that the RHS vanishes then implies \eqref{2j term equals 0 4pt}. The analysis for the cases with $i=1,2,3$ is slightly different. Let us take $i=1$ as an example. In this case the sum \eqref{no residue for K(z) 4pt} has two non-trivial contributions, namely those with $j=1$ and $j=4$. We obtain
\begin{align}
\begin{aligned}
    0 = \sum_{j=1}^4
    \oint_{z_j} dz \, \frac{\hat{K}(z)}{z-z_1}
    & =
     \frac{z_{42}z_{34}}{z_{21}z_{31}}\frac{\partial_{z_4} a_1}{\omega_1 a_1}+\frac{1}{z_{14}} + \oint_{z_4} dz
     \frac{\hat{K}^{(4)}(z)}{z-z_1} \\
    &=  \frac{\partial_{z_1} a_1}{\omega_1 a_1}
    + \sum_{j\neq 1} \frac{1}{z_{1j}} + \oint_{z_4} dz
     \frac{\hat{K}^{(4)}(z)}{z-z_1} \,,
\end{aligned}
\end{align}
where the final equality follows from \eqref{eq:chainrule_a1}. This shows that Eq.~\eqref{2j term equals 0 4pt} holds also for $i=1$. The cases with $i=2,3$ are treated analogously. This establishes Eq.~\eqref{2j term equals 0 4pt} for all covering maps relevant for any of the four-point functions under consideration.

We now move to the second line of Eq.~\eqref{KZ almost final 4pt}. Using the identity \eqref{dlogdGamma/dzi} in combination with the result we have just derived, this can be rewritten as
\begin{align}
\label{second term with J+ 4pt}
   & \sum_{j \neq i} \frac{1}{z_{ij}}\vev{\cdots} +  (\Gamma_4-x_4) \off{ \w_i \oint_{z_4} dz \frac{\vev{J^+(z)\cdots}}{z-z_i}
     - 2 \sum_{a=1}^N \der_{z_i}\lambda_a \oint_{z_4} dz \frac{\vev{J^+(z)\cdots}}{z-\lambda_a}} \\
     & \hspace{2cm}=
\off{\frac{\p_{z_i}C_\Gamma}{C_\Gamma} -\frac{\p_{z_i}a_i}{\w_ia_i}-\oint_{z_4} dz \, \hat{K}^{(4)}(z)\p_{z_i}\log \of{\p\G}} \vev{\cdots} \, . \nn
\end{align}
Using the Leibniz rule, we also have
\begin{align}
 \hat{K}^{(4)}(z) \der_{z_i} \log (\der \Gamma) =& - \der_{z_i} \of{\hat{K}^{(4)}(z)} + \der_{z_i}\off{\frac{z_{41}z_{42}z_{43}}{(z-z_1)(z-z_2)(z-z_3)(z-z_4)}} \frac{\der_{z_4}\G_4}{\der \Gamma}\nn\\&
+  \frac{z_{41}z_{42}z_{43}}{(z-z_1)(z-z_2)(z-z_3)(z-z_4)} \frac{\der_{z_i}\der_{z_4}\G_4}{\der \Gamma} \, . \label{eq: almost determinant}
\end{align}
Each of these terms can be integrated separately. First, Eq.~\eqref{L(z) expansion 4pt near z4} implies that
\begin{equation}
    0 =\oint_{z_4} dz\, \hat{K}(z) =  \oint_{z_4} dz\,\hat{K}^{(4)}(z)-1\,,
\end{equation}
and therefore
\begin{equation}
    \oint_{z_4} dz\,\der_{z_i}\off{\hat{K}^{(4)}(z)} = \der_{z_i} \off{\oint_{z_{4}} dz\,\hat{K}^{(4)}(z)} = 0 \, .
\end{equation}
The second term can be studied by replacing $\G_4 \to \G(z)$ in the numerator, and using that the resulting integral around infinity vanishes. On the one hand, if $i=4$, this leads to
 \begin{equation}
	\oint_{z_4}dz \p_{z_4}\of{\frac{z_{41}z_{42}z_{43}}{(z-z_1)(z-z_2)(z-z_3)(z-z_4)}}\frac{\p_{z_4}\G_4}{\p\G(z)} = - \frac{\p_{z_4}a_4}{\w_4 a_4}.
\end{equation}
On the other hand, if $z_i \in \{z_1,z_2,z_3\}$, one must again use the identities \eqref{eq:chainrule_a1}, \eqref{eq:chainrule_a2} and \eqref{eq:chainrule_a3}. For instance, for $i=1$ we get
 \begin{equation}
	\oint_{z_4}dz \p_{z_1}\of{\frac{z_{41}z_{42}z_{43}}{(z-z_1)(z-z_2)(z-z_3)(z-z_4)}}\frac{\p_{z_4}\G_4}{\p\G(z)} = - \frac{\p_{z_1}a_1}{\w_1 a_1} -
    \frac{1}{z_{12}} - \frac{1}{z_{13}}
    \,,
\end{equation}
and similarly for $i=2$ and $i=3$.
The net effect is twofold. In both cases, we find that the terms $\p_{z_i}a_i/(\w_ia_i)$ cancel with those already present in \eqref{second term with J+ 4pt}. Additionally, for $i=1,2,3$ this integral generates additional factors $z_{ij}$ that account for (the logarithmic derivative of) a multiplicative contribution of the form $(z_{12}z_{23}z_{13})^{-1}$ in $W_\Gamma$. As in the three-point case, this is precisely the factor needed to cancel the contributions from the $c$-ghosts. Finally,
the remaining term is given by
\begin{equation}
   \begin{aligned}
    \oint_{z_4}dz\frac{z_{41}z_{42}z_{43}}{(z-z_1)(z-z_2)(z-z_3)(z-z_4)} \frac{\der_{z_i}\der_{z_4}\G_4}{\der \Gamma} &= \p_{z_i}\ln\of{\p_{z_4}\G_4}\oint_{z_4}dz\hat{K}^{(4)}(z)\\
    &= \p_{z_i}\ln\of{\p_{z_4}\G_4}\,.
\end{aligned}
\end{equation}
This is precisely the term that accounts for the  Jacobian associated with the change of variables from $z_4$ to $\Gamma_4$. In the four-point case this is simply a scalar, namely $J = \p_{z_4}\G_4$. Once it is inserted into the final expression for $W_\Gamma$, this allows for a direct integration over the worldsheet modulus  $z_4$.

After the dust settles, we find that for correlators localized as in Eq.~\eqref{eq:localisation solution}, the KZ constraint for the vertex operator inserted at $z=z_4$ turns into a differential equation for the function $W_\Gamma$ of the form
\begin{equation}
	\begin{aligned}
		&\left\{\der_{z_4} -
			\sum_{j=1}^4 \frac{k}{4}(\w_j-1)\frac{\der_{z_4} a_j}{a_j} + \sum_{a=1}^N  \frac{k}{2}\frac{\der_{z_4} c_a}{c_a} -\frac{k}{2}\frac{\p_{z_4}C_\G}{C_\G}
            \right. \\
            & \hspace{4cm} \left.+ \sum_{j \neq 4} \frac{q_4 q_j}{z_{4j}}
		-\p_{z_4}\ln\of{\p_{z_4}\G_4}\right\} W_\Gamma(z_1,\cdots,z_4)=0 \,.
	\end{aligned}	\label{KZ  final}
\end{equation}
For  $z_i\in\{z_1,z_2,z_3\}$ we obtain a similar equation with an additional contribution, namely
\begin{equation}
	\begin{aligned}
		&
        \left\{\der_{z_i}-
			\sum_{j=1}^4 \frac{k}{4}(\w_j-1)\frac{\der_{z_i} a_j}{a_j} + \sum_{a=1}^N  \frac{k}{2}\frac{\der_{z_i} c_a}{c_a} -\frac{k}{2}\frac{\p_{z_i}C_\G}{C_\G} + \sum_{\substack{j\neq i \\ j\in\{1,2,3\}}}\frac{1}{z_{ij}}
            \right.\\
			&\hspace{4cm}
            \left.+ \sum_{j \neq i} \frac{q_i q_j}{z_{ij}} 
            -\p_{z_i}\ln\of{\p_{z_4}\G_4}\right\} W_\Gamma(z_1,\cdots,z_4)=0
	\end{aligned}	\label{KZ  final 123}
\end{equation}
Integrating thus leads to
\begin{equation}
\label{WGamma final 4pt}
    |W_\Gamma(z_1,\cdots,z_4)|^2 =
    \cN_4(j_i,\w_i) \Big|\frac{\p_{z_4}\G_4 }{z_{12} z_{23} z_{13}}\times C_\Gamma^{\frac{k}{2}}\prod_{a=1}^N c_a^{-\frac{k}{2}} \prod_{i=1}^4 a_i^{\frac{k}{4}(\w_i-1)}  \prod_{i<j} z_{ij}^{-q_i q_j}\Big|^2\, .
\end{equation}
with $\cN_4$ a normalization constant that we will fix in Sec. \ref{sec: normalization} below.

The spacetime correlator is again obtained by inserting this result into the correlator \eqref{eq:localisation solution}, including the ghost contribution, which cancels the factor $(z_{12}z_{23}z_{13})^{-1}$ in \eqref{WGamma final 4pt}, and further integrating over $z_4$. As anticipated above, this integral can easily be carried out thanks to the Jacobian factor $\p_{z_4}\G_4$ and the delta function $\delta(x_4-\Gamma_4)$. The residue of the spacetime correlator thus gives
\begin{align}
    C_{S^2}\cN_4(j_i,\w_i)\sum_{\Gamma}  \Big|\,C_\Gamma^{\frac{k}{2}} \prod_{a=1}^N c_a^{-\frac{k}{2}}\prod_{i=1}^4 a_i^{-h_i+\frac{k}{4}(\w_i-1)} \prod_{i<j} z_{ij}^{-q_iq_j} \Big|^2 \,,\label{eq: correlador localizado integrado 4pt}
\end{align}
which agrees with the holographic expectations \cite{Eberhardt:2021vsx} and the free-field computations of \cite{Hikida:2023jyc,Knighton:2024qxd} once the relative normalization is taken into account.

\subsection{KZ equations in flowed sectors: part III}

\label{sec: non-zero terms J+}

We finally tackle the general case of $n$-point functions. The important step is to obtain an  explicit expression for the terms proportional to $(\Gamma_i-x_i)$ when $n>4$.
As in the previous section, this can be done by considering the functions
\begin{equation}
    K^{(i)}(z) \equiv \of{\Gamma_i - x_i} \vev{J^+(z) \prod_{p=1}^n V_{j_p,h_p}^{\w_p}(x_p,z_p)} \, .
\end{equation}
The discussion is analogous to that of four-point functions: $K^{(i)}(z)$ must vanish at infinity (at least) as fast as $1/z^2$,  have double zeros at $z=\lambda_a$ for all $a$, and poles of order at most $\w_j$ at each of the worldsheet insertion points $z_j$ for $j=1,\dots,n$.
It follows that
\begin{equation}
\label{Ki(z) functions 1}
    K^{(i)}(z) = \frac{\prod_{a=1}^N(z-\la_a)^2}{\prod_{j=1}^{n}(z-z_j)^{\w_j}}P_{n-4}^{(i)}(z) = \frac{C_\Gamma}{\der\Gamma(z)} \frac{P_{n-4}^{(i)}(z)}{\prod_{j=1}^n (z-z_j)}\, ,
\end{equation}
for some polynomial $P_{n-4}^{(i)}(z)$ of degree at most $n-4$.
Moreover, the coefficients of the poles of order $\w_j$ at $z=z_j$  take the form
\begin{align}
\begin{aligned}
     K^{(i)} (z \sim z_j)  & = \frac{\Gamma_i - x_i}{(z-z_j)^{\w_j}}
        \vev{\of{J^+_{\w_j-1} V_{j_j,h_j}^{\w_j}}(x_j,z_j)\prod_{p\neq j} V_{j_p,h_p}^{\w_p}(x_p,z_p)}+\cdots \\
        & =\frac{(\w_ja_j)^{-1}\der_{z_j}\Gamma_i}{(z-z_j)^{\w_j}}
        \vev{\prod_{p= 1}^n V_{j_p,h_p}^{\w_p}(x_p,z_p)}+\cdots \, .
\end{aligned}
\end{align}
We thus get $n-3$ conditions of the form
\begin{equation}
P_{n-4}^{(i)}(z_j) = \of{C_\Gamma}^{-1}\der_{z_j}\Gamma_i\vev{ \cdots} \prod_{p \neq j}z_{jp} \qquad \forall \, j = 4,\dots,n \, ,
\end{equation}
which fix the polynomials $P_{n-4}^{(i)}(z)$ completely. Using the Lagrange interpolation method leads to
\begin{equation}
    K^{(i)}(z) = \hat{K}^{(i)}(z)\vev{\cdots} \, , \quad
    \hat{K}^{(i)}(z) \equiv \frac{1}{\der \G(z)}\off{\sum_{p=4}^n \frac{z_{p1}z_{p2}z_{p3}}{(z-z_1)(z-z_2)(z-z_3)(z-z_p)}\der_{z_p}}\G_i\,.
    \label{Ki(z) final}
\end{equation}

The function $K^{(i)}(z)$ appears in the integrand of various terms in Eq.~\eqref{KZ almost final}.
We first prove that the last line in that equation vanishes by showing that
\begin{equation}
\label{2j term equals 0}
    \frac{\der_{z_i}a_i}{\w_i a_i}+\sum_{j \neq i} \frac{1}{z_{ij}} + \sum_{j=4}^n  \oint_{z_j} dz
     \frac{\hat{K}^{(j)}(z)}{(z-z_i)} = 0 \, ,
\end{equation}
for all $i$. As before, this can be done by using that the function
\begin{equation}
    \hat{K}(z) = \frac{1}{\der \G(z)}\off{\sum_{p=4}^n \frac{z_{p1}z_{p2}z_{p3}}{(z-z_1)(z-z_2)(z-z_3)(z-z_p)}\der_{z_p}}\G(z)\, ,
    \label{K(z) final}
\end{equation}
has no residue at infinity nor at any of the poles of the covering map, hence it must satisfy
\begin{equation}
    \sum_{j=1}^n \oint_{z_j} dz \frac{\hat{K}(z)}{z-z_i} = 0 \qquad \forall \, i = 1,\dots,n \, .
\end{equation}
Near each of the insertion points, Eqs.~\eqref{Gamma conditions at zi}, \eqref{dGammadzi near zj} and \eqref{dGammadzi near zi} imply that
\begin{equation}
    \off{\frac{\p_{z_p}\G(z)- \p_{z_p}\G_j}{\p\G(z)}}_{z \sim z_j} = - \delta_{pj} + \frac{\p_{z_p}a_j}{\w_ja_j}(z-z_j)+\cdots \, ,
    \label{L(z) expansion}
\end{equation}
up to higher-order terms in $(z-z_j)$. In the following, the cases $i=1,2,3$ must be treated separately. Let us assume for now that $i\geq 4$. The expansion in Eq.~\eqref{L(z) expansion} implies that
\begin{align}
    \oint_{z_i} dz \frac{\hat{K}(z)}{z-z_i}  &= \frac{1}{z_{i1}} + \frac{1}{z_{i2}}+ \frac{1}{z_{i3}} + \frac{\der_{z_i}a_i}{\w_ia_i} + \oint_{z_i}  dz
     \frac{\hat{K}^{(i)}(z)}{z-z_i} \label{id1 around zi} \, ,
\end{align}
and
\begin{align}
    \oint_{z_j} dz \frac{\hat{K}(z)}{z-z_i} = \frac{1}{z_{ij}} + \oint_{z_j} dz
     \frac{\hat{K}^{(j)}(z)}{z-z_i}  \label{id1 around zj}\qquad \forall \, j\geq 4, j\neq i \,,
\end{align}
while the integrals around $z_j$ with $j=1,2,3$ vanish.
Combining Eqs.~\eqref{id1 around zj} and \eqref{id1 around zi}, we get the desired identity.  If, on the other hand,  $z_i\in \{z_1,z_2,z_3\}$, the same conclusion holds but the analysis is slightly different. For instance, if $i=1$, using identity \eqref{eq:chainrule_a1} we obtain
\begin{equation}
    \oint_{z_1} dz \, \frac{\hat{K}(z)}{z-z_1}
    = \oint_{z_1} dz \, \frac{\hat{K}^{(1)}(z)}{z-z_1}
    + \sum_{p=4}^n \frac{z_{p2}z_{3p}}{z_{21}z_{31}}\frac{\partial_{z_p} a_1}{\omega_1 a_1}
    = \frac{\partial_{z_1} a_1}{\omega_1 a_1}
    + \sum_{j=2,3} \frac{1}{z_{1j}}\,,
\end{equation}
where we have used that $\hat K^{(1)}(z)=0$ since $\Gamma_1=x_1$. The contributions from $j=2,3$ vanish again, while those coming from $j\geq 4$ work as in \eqref{id1 around zj} (the restriction $j \neq i$ trivializes). Summing over all poles then establishes the desired cancellation for $i=1$. The cases $i=2$ and $i=3$ are analogous. This proves the identity  Eq.~\eqref{2j term equals 0} in the general case.

We now move to the second line of Eq.~\eqref{KZ almost final}, which can be rewritten as
\begin{align}
\begin{aligned}
\label{second term with J+}
   & \sum_{j \neq i} \frac{1}{z_{ij}}\vev{\cdots} + \sum_{j=4}^n (\Gamma_j-x_j) \off{ \w_i \oint_{z_j} dz \frac{\vev{J^+(z)\cdots}}{z-z_i}
     - 2 \sum_{a=1}^N \der_{z_i}\lambda_a \oint_{z_j} dz \frac{\vev{J^+(z)\cdots}}{z-\lambda_a}} \\
     & \hspace{2cm}=
\off{(n-3)\frac{\p_{z_i}C_\Gamma}{C_\Gamma} -\frac{\p_{z_i}a_i}{\w_ia_i}-\sum_{j=4}^n\oint_{z_j} dz \, \hat{K}^{(j)}(z)\p_{z_i}\log \of{\p\G}} \vev{\cdots} \, .
\end{aligned}
\end{align}
Our goal is to find the relation between this expression  and the Jacobian appearing in Eq.~\eqref{Final WGamma m=0}. For this, and as in the previous section, we use the Leibniz rule to write
\begin{align}
\begin{aligned}
 \hat{K}^{(j)}(z) \der_{z_i} \log (\der \Gamma) =& - \der_{z_i} \of{\hat{K}^{(j)}(z)}
\label{int by parts dzi}\\
& +\sum_{p=4}^n \der_{z_i}\off{\frac{z_{p1}z_{p2}z_{p3}}{(z-z_1)(z-z_2)(z-z_3)(z-z_p)}} \frac{\der_{z_p}\G_j}{\der \Gamma}  \\
&+ \sum_{p=4}^n \frac{z_{p1}z_{p2}z_{p3}}{(z-z_1)(z-z_2)(z-z_3)(z-z_p)} \frac{\der_{z_i}\der_{z_p}\G_j}{\der \Gamma} \, .
\end{aligned}
\end{align}
The integral of the first term vanishes since \eqref{L(z) expansion} implies that
\begin{equation}
    0 = \sum_{j=4}^n \oint_{z_j} dz\, \hat{K}(z) = 3-n+ \sum_{j=4}^n \oint_{z_j} dz\,\hat{K}^{(j)}(z)\,.
\end{equation}

For the second term we proceed in analogy with the four-point case. Assume for now that $i\geq4$. Then the only non-trivial contribution is from the term with $p=i$.  Using \eqref{L(z) expansion} once again then gives
\begin{align}
    0 & = \sum_{j=4}^n \oint_{z_j} dz \offf{\der_{z_i}\off{\frac{z_{i1}z_{i2}z_{i3}}{(z-z_1)(z-z_2)(z-z_3)(z-z_i)}} \frac{\der_{z_i}\G}{\der \Gamma}}   \\
    &= \sum_{j=4}^n \oint_{z_j} dz \offf{\der_{z_i}\off{\frac{z_{i1}z_{i2}z_{i3}}{(z-z_1)(z-z_2)(z-z_3)(z-z_i)}} \frac{\der_{z_i}\G_j}{\der \Gamma}} + \frac{\der_{z_i}a_i}{\w_i a_i} \, . \nn
\end{align}
When $z_i\in\{z_1,z_2,z_3\}$, there are additional factors which account for the ghost contribution. Indeed, taking for instance $i=1$, we see that all the terms of the sum in (the second line of) \eqref{int by parts dzi} are non-vanishing, which leads to
\begin{align}
    0 & = \sum_{j,p=4}^n \oint_{z_j} dz \offf{ \der_{z_1}\off{\frac{z_{p1}z_{p2}z_{p3}}{(z-z_1)(z-z_2)(z-z_3)(z-z_p)}} \frac{\der_{z_p}\G}{\der \Gamma}}  \\
    &= \sum_{j,p=4}^n \oint_{z_j} dz \offf{\der_{z_1}\off{\frac{z_{p1}z_{p2}z_{p3}}{(z-z_1)(z-z_2)(z-z_3)(z-z_p)}} \frac{\der_{z_p}\G_j}{\der \Gamma}}+\sum_{p=4}^n \frac{z_{p2}z_{3p}}{z_{21}z_{31}}\frac{\p_{z_p}a_1}{\w_1a_1}\nn\\
    &= \sum_{j,p=4}^n \oint_{z_j} dz \offf{\der_{z_1}\off{\frac{z_{p1}z_{p2}z_{p3}}{(z-z_1)(z-z_2)(z-z_3)(z-z_p)}} \frac{\der_{z_p}\G_j}{\der \Gamma}} +\frac{\p_{z_1}a_1}{a_1\w_1} + \sum_{j=2,3}\frac{1}{z_{1j}}\, , \nn
\end{align}
and similarly for $i=2$ and $i=3$. Notice that this again cancels the factor $\der_{z_i}a_i/(\w_i a_i)$ and, additionally, adds exactly the necessary terms to account for the ghost contribution. In any case, for all values of $i$ the factor $\frac{\p_{z_i}a_i}{a_i\w_i}$ in \eqref{second term with J+} cancels.

We are left with a single contribution to analyze, namely the integral of the final term in Eq.~\eqref{second term with J+}. We now show that it gives the expected derivative of the  $(n-3)\times(n-3)$ Jacobian $J$ for the change of variables from $\{z_4,\dots,z_n\}$ to $\{\Gamma_4,\dots,\Gamma_n\}$. For this, we work in matrix notation by writing
\begin{equation}
     \sum_{j,p=4}^{n}  \oint_{z_j} dz \off{\frac{z_{p1}z_{p2}z_{p3}}{(z-z_1)(z-z_2)(z-z_3)(z-z_p)} \frac{\der_{z_i}\der_{z_p}\G_j}{\der \Gamma}} = \sum_{j,p=4}^{n}  M_{pj}\der_{z_i} J_{jp}=  \Tr \off{M \der_{z_i} J} \, ,
\end{equation}
where we have introduced two $(n-3)\times (n-3)$  matrices $M$ and $J$. The corresponding matrix elements are given by
\begin{equation}
    J_{jp} \equiv \der_{z_p} \Gamma_j \, , \qquad
    M_{pj} \equiv
    \oint_{z_j} dz \off{\frac{z_{p1}z_{p2}z_{p3}}{(z-z_1)(z-z_2)(z-z_3)(z-z_p)} \frac{1}{\der \Gamma}}.
\end{equation}
Here $J$ is the Jacobian matrix mentioned above. The matrix $M$ can actually be given a related interpretation: it is the inverse of this Jacobian, i.e.~$M = J^{-1}$. This can be seen by multiplying it by the Jacobian, which gives
\begin{align}
    \sum_{j=4}^n M_{pj}J_{jq} & =
    \sum_{j=4}^n
    \oint_{z_j} dz \off{\frac{z_{p1}z_{p2}z_{p3}}{(z-z_1)(z-z_2)(z-z_3)(z-z_p)} \frac{\der_{z_q}\Gamma_j}{\der \Gamma}} \nn \\
    &= \sum_{j=4}^n
    \oint_{z_j} dz \off{\frac{z_{p1}z_{p2}z_{p3}}{(z-z_1)(z-z_2)(z-z_3)(z-z_p)} \frac{\der_{z_q}\of{\Gamma_j-\Gamma}}{\der \Gamma}}  \\
    &= \sum_{j=4}^n
    \oint_{z_j} dz \off{\frac{z_{p1}z_{p2}z_{p3}}{(z-z_1)(z-z_2)(z-z_3)(z-z_p)} \delta_{qj}} = \sum_{j=4}^n \delta_{jp}\delta_{qj}=\delta_{pq} \, . \nn
\end{align}
Here we have used that, as in several of the computations above, the replacement of $\Gamma_j$ by $\Gamma(z)$ leads to a function with no residue at infinity. Combining this with the expansion \eqref{L(z) expansion}, we conclude that
\begin{equation}
     \sum_{j,p=4}^{n} \oint_{z_j} dz \frac{z_{p1}z_{p2}z_{p3}}{(z-z_1)(z-z_2)(z-z_3)(z-z_p)} \frac{\der_{z_i}\der_{z_p}\G_j}{\der \Gamma}= \Tr\off{J^{-1}\der_{z_i} J} = \der_{z_i} \ln\of{\det J} \, .
\end{equation}

The above results give explicit expressions for all KZ constraints of spectrally flowed $n$-point functions of the form \eqref{eq:localisation solution} in the form of differential equations for $W_\Gamma$. For vertex operators inserted at $z_i \notin \{z_1,z_2,z_3\}$ this reads
\begin{equation}
	\begin{aligned}
		& \offf{\der_{z_i} -
			\sum_{j=1}^n \frac{k}{4}(\w_j-1)\frac{\der_{z_i} a_j}{a_j} + \sum_{a=1}^N  \frac{k}{2}\frac{\der_{z_i} c_a}{c_a} -\frac{k}{2}\frac{\p_{z_i}C_\G}{C_\G}+ \sum_{j \neq i} \frac{q_i q_j}{z_{ij}}
		-\p_{z_i}\ln\of{\det J}} W_\Gamma=0 \,,
	\end{aligned}	\label{KZ  final n}
\end{equation}
while for $z_i\in\{z_1,z_2,z_3\}$ one gets an additional contribution needed to cancel the ghost contribution in the full string correlator, i.e.
\begin{equation}
	\begin{aligned}
		&\hspace{-2cm}
        \left\{\der_{z_i}-
			\sum_{j=1}^n \frac{k}{4}(\w_j-1)\frac{\der_{z_i} a_j}{a_j} + \sum_{a=1}^N  \frac{k}{2}\frac{\der_{z_i} c_a}{c_a} -\frac{k}{2}\frac{\p_{z_i}C_\G}{C_\G}+ \sum_{j \neq i} \frac{q_i q_j}{z_{ij}}
            \right.\\
			& \hspace{4cm}
            \left.+\sum_{\substack{j\neq i \\ j\in\{1,2,3\}}}\frac{1}{z_{ij}}-\p_{z_i}\ln\of{\det J}\right\} W_\G=0
	\end{aligned}	\label{KZ  final 123 n}
\end{equation}
This set of equations is easily integrated, giving
\begin{equation}
\label{Final WGamma m=0 2}
|W_\Gamma(z_1,\cdots,z_n)|^2 =
    \cN_n(j_i,\w_i) \Big|\frac{J_{\off{z_i \to \Gamma_i}}}{z_{12} z_{23} z_{13}}\times C_\Gamma^{\frac{k}{2}}\prod_{a=1}^N c_a^{-\frac{k}{2}} \prod_{i=1}^n a_i^{\frac{k}{4}(\w_i-1)}  \prod_{i<j} z_{ij}^{-q_i q_j}\Big|^2\, .
\end{equation}
 This is precisely the result anticipated in the introduction (see Eqs.~\eqref{Final WGamma m=0} and \eqref{qi def}). The normalization constant will be discussed in the following section.

As in the four-point case, the (residue of the) genus-zero string correlator is obtained by including the anti-holomorphic factors, inserting the ghost contribution, and integrating over the insertion points $z_i$ with $i\geq 4$. The presence of the Jacobian makes this a simple task. This yields a final result of the form
\begin{align}
    \vvev{\prod_{i=1}^n \cO_{q_i}^{\omega_i}(x_i)}
     =
     C_{S^2}\cN_n(j_i,\w_i)\sum_{\Gamma}  \Big|\,{C_\Gamma^{\frac{k}{2}}} \prod_{a=1}^N c_a^{-\frac{k}{2}}\prod_{i=1}^n a_i^{-h_i+\frac{k}{4}(\w_i-1)} \prod_{i<j} z_{ij}^{-q_iq_j} \Big|^2 \,.\label{eq: correlador localizado integrado}
\end{align}
Here, $q_i$ and $h_i$ should be thought of as functions of the spins $j_i$ and the spectral flow charges $\w_i$. This follows from the Virasoro condition and the fact that we are dealing with operators that are trivial in the internal sector. We are also using the double bracket notation to denote the residues of string correlators, as opposed to the worldsheet ones.  We conclude that the structure of \eqref{eq: correlador localizado integrado} matches the holographic prediction in Eq.~\eqref{HCFT correlator m=0 unfixed}.

\section{Normalization constants from the factorization limit}
\label{sec: normalization}

Having derived the functional form of the solutions to the KZ equations for $n$-point functions satisfying the $j$-constraint \eqref{j-constraint}, we now discuss their normalization constants. That is, we fix $\cN_n(j_i,\w_i)$.
In order to do this, we first identify the spacetime identity operator. We then provide a recursive derivation of $\cN_n(j_i,\w_i)$ by starting from three-point functions, which are known exactly \cite{Dei:2021xgh,Bufalini:2022toj,Eberhardt:2021vsx}, and imposing the correct factorization structure. Our discussion is closely related to that of \cite{Dei:2019iym} for symmetric orbifold CFTs.

\subsection{An aside on the spacetime identity operator}

Here we argue that, in the context of the computation of the residues of spacetime correlators,
\begin{equation}
    \cI(x) = \int d^2z \, \cO_1(x,z)\,, \qquad \cO_1(x,z) \equiv \frac{\eta_k}{N_k} V^{\w=1}_{\frac{k}{2}-1,h=0}(x,z)\,.\label{Identity operator}
\end{equation}
acts as the spacetime identity operator up to a factor counting the number of sheets of the covering map (see also \cite{Eberhardt:2021vsx}).
Here $\eta_k$ is some $k$-dependent normalization factor, relative to the constant $N_k$ defined in Eq.~\eqref{Nk factor} when discussing three-point functions.

The worldsheet vertex operator $\cO_1(x,z)$ has conformal weight $\Delta=1$, and spacetime weight $h=0$. Its integrated version commutes with all the currents $J^a$. This can easily be seen by noting that
\begin{equation}
    \Oo_1(x,z) = \frac{1}{2-k}\frac{\eta_k}{N_k} \of{J^{-}_{-1}V^{1}_{\frac{k}{2}-1,1}}(x,z)
\end{equation}
with  $V^{1}_{\frac{k}{2}-1,1}$ the primary field with $\w=1$, $j=\frac{k}{2}-1$ and $h=1$, hence it has $m=-j$.
The KZ equation for this vertex operator reads
\begin{equation}
    \of{J^+_0 \cO_1}(x,z)= -\frac{\eta_k}{N_k} \p_zV^{1}_{\frac{k}{2}-1,1}(x,z) \,.
\end{equation}
The $\mathfrak{sl}(2,\RR)_k$ algebra also implies that
\begin{equation}
    \of{J^+_1 \cO_1}(x,z) = -\frac{\eta_k}{N_k} V^{1}_{\frac{k}{2}-1,1}(x,z)\,.
\end{equation}
Looking back at \eqref{JVxOPE}, we see that the OPE of $\Oo_1$ with the current $J^+(z)$ takes the form
\begin{equation}
    J^+(z)\cO_1(x,w) \sim -\p_w\of{\frac{\eta_k}{N_k} \frac{V^{1}_{\frac{k}{2}-1,1}(x,w)}{z-w}} \,.
\end{equation}
A similar formula shows that the OPEs with $J^{-}(z)$ and $J^3(z)$ are also regular up to total derivatives.  This means that $\Ii(x)$ can be inserted in any correlation function without changing its conformal properties. As we will see shortly, its only effect is to count the number of sheets of the covering space \cite{Giveon:1998ns,Kim:2015gak}. Indeed, a suitable choice for the constant $\eta_k$ leads to
\begin{equation}
    \int \prod_{p=1}^{\tilde n} d^2z_p \vev{\prod_{i=1}^n V_{j_i,h_i}^{\w_i}(x_i,z_i)\prod_{p=1}^{\tilde n}\cO_1(x_p,z_p)} = N^{\tilde n}\vev{\prod_{i=1}^n V_{j_i,h_i}^{\w_i}(x_i,z_i)}\,,
    \label{Identity correlator}
\end{equation}
for correlators satisfying \eqref{j condition intro}. Here $N = 1+\frac{1}{2}\sum_{i=1}^n(\w_i-1)$ is the degree of the covering maps contributing to the correlator on the RHS, which remains unchanged upon inserting additional singly-flowed operators. 
Indeed, if the original spins $j_i$ satisfy \eqref{j condition intro}, the LHS correlator also localizes since
\begin{equation}
    \sum_{i=1}^n j_i +\sum_{p=1}^{\tilde n}\of{\frac{k}{2}-1} = 1 +\frac{k-2}{2}(n+{\tilde n}-2) \Leftrightarrow \sum_{i=1}^n j_i = 1 +\frac{k-2}{2}(n-2)\,.
\end{equation}
The localized solution \eqref{eq:localisation solution} for the correlator appearing on the LHS of \eqref{Identity correlator}  takes the form
\begin{align}
    &\vev{\prod_{i=1}^n V_{j_i,h_i}^{\w_i}(x_i,z_i)\prod_{p=1}^{\n}\cO_1(x_p,z_p)} = \nn \\
    &\hspace{3cm}\sum_{\G} \prod_{i=1}^n |a_i|^{-2h_i}\prod_{i=4}^{n}\delta^{(2)}(x_i-\G_i)\prod_{p=1}^\n\delta^{(2)}(x_p-\G_p) |W_\G(z_i,z_p)|^2\,.
\end{align}
Now, inserting additional singly-flowed operators does not impose any extra constraints on the map $\G(z)$, hence the sum goes over the same covering maps as before. Furthermore, since $\cO_1$ has $q\of{\frac{k}{2}-1} =0$, the only dependence of the functions $W_\G$ on $z_p$ enters through the Jacobian, yielding
\begin{align}
   |W_\G(z_i,z_p)|^2 = \cN_{n+\n}(j_I,\w_I)\left|\frac{J_{\off{z_I \to \Gamma_I}}}{z_{12} z_{23} z_{13}}\times C_\Gamma^{\frac{k}{2}}\prod_{a=1}^N c_a^{-\frac{k}{2}} \prod_{i=1}^{n} a_i^{\frac{k}{4}(\w_i-1)}  \prod_{i<j} z_{ij}^{-q_i q_j}\right|^2\,.
\end{align}
Here the collective index $I$ is such that $(j_I,\w_I) = (j_i,\w_i)$ if $I\leq n$ and $(j_I,\w_I) = (\frac{k}{2}-1,1)$ if $I > n$. The Jacobian $J_{\off{z_I \to \Gamma_I}}$ accounts for both changes of variables, namely $z_i \to \G_i$ and $z_p\to \G_p$. This factorizes since the covering map $\G(z)$ does not depend explicitly on $z_p$, hence its derivatives satisfy
\begin{equation}
    \frac{\p \G_i}{\p z_p} = 0 \,\quad \text{and}  \,\quad \frac{\p \G_p}{\p z_{p'}} = \delta_{p,p'} \p_p \G_p\,.
\end{equation}
This shows that
\begin{equation}
   |W_\G(z_i,z_p)|^2 = \prod_{p=1}^\n |\p_p \G_p|^2 |W_\G(z_i)|^2\,,
\end{equation}
where the function $W_\G(z_i)$ is the one associated with the residue of the correlator on the RHS of Eq.~\eqref{Identity correlator}.
The factors $|\p_p \G_p|^2$ are then absorbed by performing the integrals over the variables $z_p$ using the corresponding delta functions. Each of these integrals receives one contribution from each pre-image of $x_p$, giving 
\begin{equation}
    \int d^2z_p\, \delta^{(2)}(x_p-\G(z_p))\,|\p_p\G_p|^2 = N\,.
\end{equation}
 It follows that \eqref{Identity correlator} holds as long as the normalization constants satisfy 
\begin{equation}
    \cN_{n+\tilde{n}}(j_I,\w_I) \of{\frac{\eta_k}{N_k}}^{\tilde{n}} = \cN_{n}(j_i,\w_i)\,.
\end{equation}

In the following section we will provide a recursive argument for computing $\cN_{n}(j_i,\w_i)$. The initial steps will be provided by constants appearing in spacetime two- and three-point functions. As discussed above, the latter are known exactly. The former, defined as
\begin{equation}
    \label{spacestime 2pt 0}\vev{\cO^{\w}_{j,{h}}(0)\cO^{\w}_{1-j,{h}}(\infty)} = \cN^{\infty}_2(j,1-j,\w,\w) \, ,
\end{equation}
are notoriously subtle in string theory.
We will bypass this issue by using Eq.~\eqref{Identity correlator} with $n=2$ and $\bar{n}=1$ as a definition, thus rewriting the two-point function in terms of a three-point one. The only relevant covering map is $\Gamma(z)=z^\w$, hence the identity insertion simply produces a factor $\w$. Using \eqref{3point normalization infinity} we get
\begin{equation}
    \label{spacestime 2pt} \cN^{\infty}_2(j,1-j,\w,\w) = \frac{1}{\w}\,\frac{\eta_k}{N_k}\,\cN^\infty_{3}\Big(j,\frac{k}{2}-1,1-j,\w,1,\w\Big) = \eta_k\, \w \, .
\end{equation}
This is consistent with the fact that for this two-point function the corresponding covering map data trivializes, and agrees with the expression obtained in \cite{Eberhardt:2021vsx}.

\subsection{Spacetime factorization from the worldsheet}\label{sec: factorization}

Consider the contribution to a given $n$-point function satisfying the $j$-constraint \eqref{j-constraint} in the factorization limit $x_1\to x_2$. On the localization locus, two kinds of contributions emerge in this limit. The first possibility is that the worldsheet insertion points also collide, $z_1\to z_2$, and the monodromy of the resulting covering map is of the single-cycle type. The second possibility is that $z_1$ and $z_2$ remain apart, which happens when the intermediate state corresponds to a multi-cycle operator. We discuss explicit examples of both situations in Appendix \ref{app:factorization examples}. Here we focus on the single-cycle channels, which are the relevant ones for fixing the normalization constants ${\cal{N}}_n(j_i,\w_i)$.

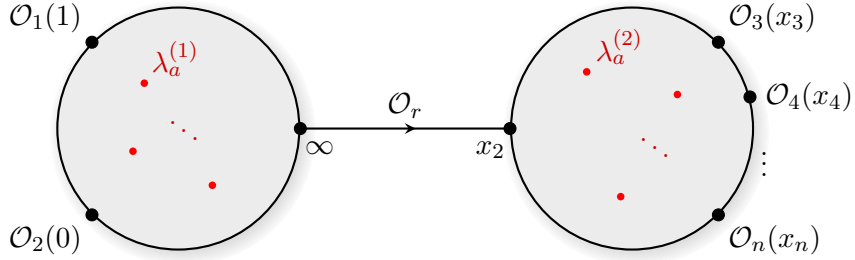
\begin{figure}[h!]
    \centering
\begin{tikzpicture}[
    scale=1.5,
    flow/.style={
        thick,
        decoration={markings, mark=at position 0.55 with {\arrow{stealth}}},
        postaction={decorate}
    },
    cftblob/.style={
        circle,
        draw=black,
        thick,
        fill=gray!15,
        minimum size=3.2cm, 
        circular drop shadow={shadow xshift=0.3ex, shadow yshift=-0.3ex, opacity=0.3}
    },
    pole/.style={
        circle,
        fill=red,
        inner sep=1pt
    }
]

    \node[cftblob] (bubbleL) at (-2.0, 0) {};

    \node[cftblob] (bubbleR) at (2.0, 0) {};

    \filldraw (bubbleL.135) circle (1.5pt);
    \filldraw (bubbleL.225) circle (1.5pt);
    \filldraw (bubbleL.0) circle (1.5pt);

    \node[above left] at (bubbleL.135) {$\cO_{1}(1)$};
    \node[below left] at (bubbleL.225) {$\cO_{2}(0)$};

    \node[below right=1pt and -2pt] at (bubbleL.0) {$\infty$};

    \node[pole] (pL1) at (-2.3, 0.4) {};
    \node[pole] (pL2) at (-1.7, -0.5) {};
    \node[pole] (pL3) at (-2.4, -0.2) {};
    \node[above right=-1pt and -1pt, text=red!80!black] at (pL1) {$\lambda_a^{\rm in}$};
    \node[text=red!80!black] at (-1.95, 0.05) {$\ddots$}; 

    \draw[flow] (bubbleL.0) -- (bubbleR.180) node[midway, above] {$\cO_r$};

    \filldraw (bubbleR.180) circle (1.5pt);
    \node[below left=1pt and -2pt] at (bubbleR.180) {$x_2$};

    \filldraw (bubbleR.45) circle (1.5pt);
    \filldraw (bubbleR.15) circle (1.5pt);
    \filldraw (bubbleR.-45) circle (1.5pt);

    \node[above right] at (bubbleR.45) {$\cO_{3}(x_3)$};
    \node[right=2pt] at (bubbleR.15) {$\cO_4(x_4)$};
    \node[below right] at (bubbleR.-45) {$\cO_{n}(x_{n})$};

    \path (bubbleR.15) -- (bubbleR.-45) node[midway, right=6pt] {$\vdots$};

    \node[pole] (pR1) at (1.6, 0.5) {};
    \node[pole] (pR2) at (2.4, 0.3) {};
    \node[pole] (pR3) at (1.9, -0.6) {};
    \node[above left=-1pt and -25pt, text=red!80!black] at (pR1) {$\lambda_a^{\rm out}$};
    \node[text=red!80!black] at (2.2, -0.1) {$\ddots$}; 

\end{tikzpicture}
    \caption{Factorization structure of the $n$-point correlation function in the limit $x_1 \to x_2$.}
    \label{fig:factorization_npoint}
\end{figure}

Let us focus on the contribution of a given covering map $\Gamma$, and assume that\footnote{In this limit it is convenient to use the global Ward identities to fix $z_2$, $z_3$ and $z_4$ instead of $z_1$, $z_2$, $z_3$, so that $z_1$ is determined by $\Gamma(z_1)=x_1$. The number of covering maps contributing for fixed $x_i$ is, of course, gauge independent.} $z_1$ approaches $z_2$. The locations of the corresponding poles, namely the values of $\la_a$ for $a=1,\dots,N$, get divided into two sets according to their scaling behavior. We have  
\begin{equation}
    \la_a -z_2 = \cO(1)\qquad \text{or} \qquad \la_a -z_2 = \cO(z_{21})\,.\label{eq:lambda_scaling}
\end{equation}
This follows from the scattering equations \eqref{scattering equations}. In other words, we may define new coefficients $\la_a^{\rm in}$ and $\la_a^{\rm out}$ such that 
\begin{subequations}
\label{lambda scalings}
    \begin{align}
    \la_a &\sim z_2 + z_{12} \la_a^{\rm in}\,,\qquad a=1,\dots,r\,,\label{eq:lambda_group1}\\
    \la_a &= \la_a^{\rm out}\,,\qquad\qquad\quad\,\, a=r+1,\dots, N\,.\label{eq:lambda_group2}
\end{align}
\end{subequations}
for some fixed non-negative integer $r$. The value of the latter depends on the details of the covering map. 
In the limit $z_{12}\to 0$, $\la_a^{\rm in}$ and $\la_a^{\rm out}$  satisfy  two independent sets scattering equations, namely 
\begin{equation}
    \frac{\w_2-1}{\la_a^{\rm in}}+\frac{\w_1-1}{\la_a^{\rm in}-1}-\sum_{\substack{b \leq r\\ b \neq a }}\frac{2}{\la_a^{\rm in}-\la_b^{\rm in}} = 0\, 
\qquad \forall \, a=1,\dots,r \,,
\label{eq:scattering_1}
\end{equation}
and 
\begin{equation}
    \\
    \frac{\w_r-1}{\la_a^{\rm out}-z_2}+\sum_{i=3}^n\frac{\w_i-1}{\la_a^{\rm out}-z_i}-\sum_{\substack{b > r\\b \neq a  }}\frac{2}{\la_a^{\rm out}-\la_b^{\rm out}}=0
    \, 
\qquad \forall \, a=r+1,\dots,N \,,\label{eq:scattering_2}
\end{equation}
where we have defined 
\begin{equation}
    \w_r=\w_1+\w_2-2r-1 \, . 
    \label{def wr}
\end{equation}
The solutions of \eqref{eq:scattering_1} define a covering map $\G^{\rm in}$ with three branch points located at $(0,1,\infty)$, the corresponding twists being $(\w_2,\w_1,\w_r)$. As discussed in Sec.~\ref{sec: covering maps}, this map is unique up to global transformations. Moreover, it exists for all allowed values of $r$, since $\w_1+\w_2+\w_r$ is odd and $|\w_1-\w_2|+1\leq\w_r\leq\w_1+\w_2-1$.  Similarly, the solutions to Eq.~\eqref{eq:scattering_2} determine maps $\G^{\rm out}$ with $n-1$ insertions at $(z_2,\dots,z_n)$,  and winding numbers $(\w_r,\w_3,\dots,\w_n)$. In contrast to $\G^{\rm in}$, for $n-1\geq4$ there can be several such maps $\G^{\rm out}$. The total number is given by the corresponding Hurwitz number. This structure, which underlies the factorization properties of correlation functions in symmetric orbifold CFTs, is also the key to describe the spacetime factorization of the integrated worldsheet correlators considered in this paper. Fig. \ref{fig:factorization_npoint} provides a schematic representation of this procedure: on the left we get a three-point function involving the vertex operators $\cO_{1}$, $\cO_{2}$ and $\cO_{r}$, computed using  the covering map $\G^{\rm in}$, while on the right we obtain an $(n-1)$-point function which can be studied using the covering map $\G^{\rm out}$, and where the  operator $\cO_{2}$, which was originally inserted on the AdS$_3$ boundary at $x_2$, has been replaced by $\cO_{r}$.

We would like to find a set of normalization constants  $\cN_n(j_i,\w_i)$ that are consistent with this factorization procedure. For this, it is necessary to determine the leading-order expressions for each of the factors appearing on the RHS of Eq.~\eqref{eq: correlador localizado integrado}.
We start by considering the derivatives of the covering maps $\G^{\rm in}$ and $\G^{\rm out}$, which are given by
\begin{align}
    &\p \G^{\rm in}(z) = C_\G^{\rm in}\frac{z^{\w_2-1}(z-1)^{\w_1-1}}{\prod_{a=1}^r\of{z-\la_a^{\rm in}}^2}\,, \label{derGamma1}\\
    &\p \G^{\rm out}(z) = C_\G^{\rm out}\frac{(z-z_2)^{\w_r-1}\prod_{i=3}^n(z-z_i)^{\w_i-1}}{\prod_{a=r+1}^N\of{z-\la_a^{\rm out}}^2}\,. \label{derGamma2}
\end{align}
In the case of $\G^{\rm out}$, the expressions for the relevant coefficients $a_i^{\rm out}$ and residues $c_a^{\rm out}$ are thus analogous to those of Eqs.~\eqref{ai expression} and \eqref{ca expression}, respectively. On the other hand, for $\G^{\rm in}$ this works slightly differently because one of the branch points is located at infinity. As it was briefly mentioned  in Sec.~\ref{sec: covering maps}, the coefficient $a_r^{\rm in}$ is identified from the asymptotic expansion
\begin{equation}
    \G^{\rm in}(z\to\infty) \sim \frac{(-1)^{\w_r+1}}{a_r^{\rm in}}z^{\w_r}\,.
\end{equation}
This shows that $a_r^{\rm in}$ is directly related to the constant $C_\G^{\rm in}$ appearing in Eq.~\eqref{derGamma1}:
\begin{equation}
   a_r^{\rm in} = (-1)^{\w_r+1}\frac{\w_r}{C_\G^{\rm in}}\,.
\end{equation}
On the other hand, for the original  covering map the constant $C_\G$ was formally defined by a condition of the form
\begin{equation}
    x_{ij} = \int_{z_j}^{z_i}dz \p\G(z) \,,\label{eq:CG_condition}
\end{equation}
for any choice of $i$ and $j$. Since a similar equation holds for $C_\Gamma^{\rm out}$, it  follows from \eqref{lambda scalings} that, for $(i,j)\neq (1,2)$ and at leading order in $z_{12}$, one has
\begin{equation}
   \frac{1}{C_\G} = \int_{z_j}^{z_i}dz \frac{\prod_{p=1}^n (z-z_p)^{\w_p-1}}{x_{ij}\prod_{a=1}^N(z-\la_a)^2}\approx \int_{z_j}^{z_i}dz \frac{(z-z_2)^{\w_r-1}\prod_{p=3}^n (z-z_p)^{\w_p-1}}{x_{ij}\prod_{a=r+1}^N\of{z-\la^{\rm out}_a}^2}= \frac{1}{C_{\G}^{\rm out}}\,.\label{eq:CG_leading}
\end{equation}
Here we have used that
\begin{equation}
    \prod_{a=1}^N(z-\la_a)^2 = \prod_{a=1}^r(z -\la_a)^2\prod_{a=r+1}^N(z-\la_a)^2\approx (z-z_2)^{2r}\prod_{a=r+1}^N\of{z-\la^{\rm out}_a}^2\,.
\end{equation}
In other words, in the factorization limit we have $C_\G = C_{\G}^{\rm out}$. The leading behavior of each of the coefficients $c_a$ and $a_i$ can be determined analogously from Eqs.~\eqref{ca expression} and \eqref{ai expression}. For $a_1$ for instance, we start from
\begin{align}
    \w_1 a_1 = C_\G \frac{\prod_{i\neq 1}(z_1-z_i)^{\w_i-1}}{\prod_{a=1}^N(z_1-\la_a)^2}\, ,
\end{align}
and use that
\begin{equation}
    \prod_{a=1}^N(z_1-\la_a)^2 = \prod_{a=1}^r(z_1-\la_a)^2\prod_{a=r+1}^N(z_1-\la_a)^2\approx z_{12}^{2r}\prod_{a=1}^r\of{1-\la^{\rm in}_a}^2\prod_{a=r+1}^N\of{z_2-\la^{\rm put}_a}^2\,,
\end{equation}
hence
\begin{align}
    \w_1 a_1 \approx  C^{\rm put}_\G \frac{z_{12}^{\w_2-2r-1}\prod_{i>2}(z_2-z_i)^{\w_i-1}}{\prod_{a=1}^r\of{1-\la^{\rm in}_a}^2\prod_{a=r+1}^N\of{z_2-\la^{\rm out}_a}^2}=\frac{z_{12}^{\w_r-\w_1}}{C_{\G}^{\rm in}}\w_1a_1^{\rm in}\w_ra_r^{\rm out}\,.
\end{align}
The other coefficients can be studied similarly, giving
\begin{equation}
    a_1 = \frac{z_{12}^{\w_r-\w_1}}{C_{\G}^{\rm in}}a_1^{\rm in}\w_ra_r^{\rm out}\,,
    \qquad
    a_2 = \frac{z_{12}^{\w_r-\w_2}}{C_{\G}^{\rm in}}a_2^{\rm in}\w_r a_r^{\rm out}\,,
    \qquad
    a_j = a_j^{\rm out}
    \quad \forall \, j\geq3\,,
\end{equation}
and
\begin{equation}
    c_{a} = \frac{z_{12}^{\w_r+1}}{C_{\G}^{\rm in}} c_a^{\rm in}\w_ra_r^{\rm out} \quad \forall \, a\leq r \, ,
    \qquad
    c_{a} = c_{a}^{\rm out}\quad \forall \, a>r\,.
    \label{ca factorization}
\end{equation}
By considering Eq.~\eqref{eq:CG_condition} for $(i,j)=(1,2)$ one can also derive a relation between the worldsheet and boundary differences $z_{12}$ and $x_{12}$. At leading order, this takes the form
\begin{equation}
    x_{12} =C_{\G}\int_{z_2}^{z_1}dz \frac{\prod_{i=1}^n (z-z_i)^{\w_i-1}}{\prod_{a=1}^N(z-\la_a)^2} \approx z_{12}^{\w_r}\frac{C_{\G}^{\rm out}}{C_{\G}^{\rm in}} \frac{\prod_{j=3}^n z_{2j}^{\w_j-1}}{\prod_{a=r+1}^{N}(z_2-\la_a^{\rm out})^2} =(-1)^{\w_r+1}z_{12}^{\w_r}a_r^{\rm in}a_r^{\rm out}\, .
    \label{eq:x21_relation}
\end{equation}
The power of $z_{12}$ in Eq.~\eqref{eq:x21_relation} suggests that, in this regime, covering maps come in groups of $\w_r$. 
Indeed, $a_r^{\rm in}$ and $a_r^{\rm out}$ are independent of $z_1$, hence, for fixed boundary data and small $x_{12}$, there are exactly $\w_r$ solutions $z_1^{(s)}$ near $z_{12}=0$.
Each of these solutions defines a \textit{different} covering map $\Gamma$ with $\Gamma(z_i)=x_i$, i.e.~a different term in the original sum over covering maps. All such terms contribute in \textit{exactly the same} way in the factorization limit, since they lead to the same pair $(\Gamma^{\rm in},\Gamma^{\rm out})$. This degeneracy must be taken into account when fixing the normalization constant we are interested in. 
More details can be found in App.~\ref{app:factorization examples}.
Combining the expressions in
Eqs.~\eqref{lambda scalings}-\eqref{eq:x21_relation} then shows that, at first non-trivial order in powers of $z_{12}$, and for a particular value of $r$, one can rewrite the contribution from the covering map $\Gamma$ in \eqref{eq: correlador localizado integrado} as
\begin{align}
    \begin{aligned}
    &C_\G^{\frac{k}{2}} \prod_{a=1}^N c_a^{-\frac{k}{2}} \prod_{i=1}^n a_i^{-h_i+\frac{k}{4}(\w_i-1)}  \prod_{i<j} z_{ij}^{-q_iq_j} \approx x_{12}^{h_r-h_1-h_2}  \prod_{3\leq i} z_{2i}^{-q_i(q_1+q_2)} \prod_{3\leq i<j} z_{ij}^{-q_iq_j} \\
    & \qquad \times \prod_{a=1}^r \of{c_a^{\rm in}}^{-\frac{k}{2}} \prod_{i=1,2} \of{a_i^{\rm in}}^{-h_i+\frac{k}{4}(\w_i-1)} \of{a_r^{\rm in}}^{-h_r+\frac{k}{4}(\w_r-1)} \\
    & \qquad \times  \of{C_\G^{\rm out}}^{\frac{k}{2}} \prod_{a=r+1}^N \of{c_a^{\rm out}}^{-\frac{k}{2}}  \of{a_r^{\rm out}}^{-h_r+\frac{k}{4}(\w_r-1)} \prod_{i\geq 3}^n \of{a_i^{\rm out}}^{-h_i+\frac{k}{4}(\w_i-1)} \, ,
    \label{factorized structure 1 map}
    \end{aligned}
\end{align}
where the constant $C_\G^{\rm in}$ does not appear explicitly since one of the corresponding insertions is located at infinity. This is precisely the factorized structure we expect from the point of view of the holographic CFT \cite{Dei:2019iym,Eberhardt:2021vsx}.
As discussed above, each channel comes with a multiplicity factor $\w_r$.

Moreover, even though we already knew that, in this  channel, this intermediate state had spectral flow charge $\w_r$, the power of $z_{2i}$ in the first line of \eqref{factorized structure 1 map} indicates that it must also have $q=q_1+q_2$, i.e.
\begin{equation}
    j = j_1+j_2+1-\frac{k}{2} \, .
\end{equation}
Consistently, this ensures that the three-point function and the ($n-1$)-point function involved in this computation both satisfy the $j$-constraint separately, and hence localize as well.
The value of the parameter $h_r$ that we get is then consistent with the corresponding spacetime weight, as defined by the Virasoro condition:
\begin{equation}
    h_r = \frac{1}{\w_r}\of{-\frac{j(j-1)}{k-2}+\frac{k}{4}\w_r^2-1}\,.\label{eq:hr_definition}
\end{equation}
Hence, by re-inserting the antiholomorphic dependence and the normalization factors, and  summing over the allowed values of $r$ (see for instance \cite{Eberhardt:2018ouy,Lunin:2000yv,Dei:2019iym}), we find that, in the $x_1 \to x_2$ limit, the string correlator behaves as
\begin{align}
&\hspace{-2cm}\vvev{\prod_{i=1}^n \cO^{\w_i}_{j_i,h_i}(x_i)} \sim \sum_{\w_r} |x_{12}|^{2(h_r-h_1-h_2)}\hat{\cN}_r \times  \, \label{eq:correlator_factorization}\\
&\frac{\vvev{ \cO^{\w_2}_{j_2,h_2}(0)\cO^{\w_1}_{j_1,h_1}(1)\cO^{\w_r}_{1-j,{h_r}}(\infty)}\vvev{\cO^{\w_r}_{j,{h_r}}(x_2)\prod_{i=3}^n \cO^{\w_i}_{j_i,h_i}(x_i)}}{\vvev{\cO^{\w_r}_{j,{h_r}}(0)\cO^{\w_r}_{1-j,{h_r}}(\infty)}}\,, \nn 
\end{align}
where
\begin{equation}
    \hat{\cN}_r=\frac{\cN_n(j_i,\w_i)\eta_k \w_r^2}{\cN_3^{\infty}(j_1,j_2,1-j,\w_1,\w_2,\w_r)\cN_{n-1}(j,j_3,\dots,j_n,\w_r,\w_3,\dots,\w_n)}\,.\label{eq:Nr_definition}
\end{equation}
The sum over $\w_r$ in \eqref{eq:correlator_factorization} runs from $\w_r=|\w_1-\w_2|+1$ to $\w_r=\w_1+\w_2-1$, and is restricted to those channels for which the $(n-1)$-point covering map $\G^{\rm out}$, with twists $(\w_r,\w_3,\dots,\w_n)$, exists. As explained above, here we include only the single-cycle contributions.
For later convenience, here we have multiplied and divided by the two-point function \eqref{spacestime 2pt}, discussed in the previous section. Note that the factor $\w_r^2$ in \eqref{eq:Nr_definition} has a two-fold origin: one power of $\w_r$ comes from the two-point function, while the other one is the multiplicity factor discussed above.

To ensure a consistent spacetime OPE structure, we must then impose $\hat{\cN}_r = 1$. This implies a set of recursion relations of the form
\begin{equation}
    \cN_n(j_i,\w_i) = \eta_k^{-1}\w_r^{-2}\cN^{\infty}_3(j_1,j_2,1-j,\w_1,\w_2,\w_r)\cN_{n-1}(j,j_3,\dots,j_n,\w_r,\w_3,\dots,\w_n)\,.\label{eq:Nn_recursive}
\end{equation}
For instance, for $n=4$ this reads
\begin{equation}
    \cN_4(j_i,\w_i) = \eta_k^{-1}\w_r^{-2}\cN^{\infty}_3(j_1,j_2,1-j,\w_1,\w_2,\w_r)\cN_{3}(j,j_3,j_4,\w_r,\w_3,\w_4)\,.\label{eq:N4_recursive}
\end{equation}
A useful consistency check of our final result will be that the dependence on $j$ and $\w_r$ on the RHS of this equation cancels. As discussed in Sec.~\ref{sec: three-point functions KZ}, the exact three-point function of \cite{Dei:2021xgh,Bufalini:2022toj} leads to
\begin{equation}
    \cN_3(j_i,\w_i)=N_k\prod_{j=1}^3\w_j^{-\frac{k}{2} (\w_j+1) +2j_j}\,,
\end{equation}
and
\begin{equation}
    \cN^{\infty}_3(j_i,\w_i)=N_k\prod_{j=1}^2\w_j^{-\frac{k}{2} (\w_j+1) +2j_j}\w_3^{\frac{k}{2} (\w_3+1) +2j_3}\,.
\end{equation}
where the precise value of the $k$-dependent coefficient $N_k$ was given in Eq.~\eqref{Nk factor}. Hence,
\begin{equation}
    \cN_4(j_i,\w_i) = \eta_k^{-1}
    N_k^2\prod_{j=1}^4\w_j^{-\frac{k}{2} (\w_j+1) +2j_j} \,.
\end{equation}
Iterating this procedure gives
\begin{equation}
    \cN_n(j_i,\w_i) =\eta^{3-n}_k N^{n-2}_k \prod_{i=1}^n \w_i^{-\frac{k}{2}(\w_i+1)+2j_i}\qquad \forall \, n \geq 2 \, .\label{eq:Nn_solution}
\end{equation}
This finalizes our computation of string correlators satisfying the $j$-constraint \eqref{j-constraint}
\footnote{The constant $\eta_k = 2(k-2)$ was computed in \cite{Eberhardt:2021vsx}, see also Sec.~\ref{sec: extension to m and matching}.}. Once this normalization is included, and the differences in conventions are accounted for, the final result coincides exactly with those of \cite{Hikida:2023jyc} and \cite{Knighton:2023mhq,Knighton:2024qxd}. The holographic matching  will be presented in the following section, after extending the computation of string residues to correlators satisfying the less restrictive condition \eqref{j condition intro with m}.

\section{Extension to correlators with $\m \neq 0$ and holographic matching}
\label{sec: extension to m and matching}

Let us now consider SL(2,$\R$) $n$-point functions
$\vev{\prod_{i=1}^n V_{j_i,h_i}^{\w_i}(x_i,z_i)}$
satisfying
\begin{equation}
\label{j constraint with m}
    \sum_{i=1}^n j_i = 1+\frac{k-2}{2}(n-\m-2) \, ,
\end{equation}
for any non-negative integer $\m$. From the holographic perspective of \cite{Eberhardt:2021vsx}, AdS$_3$ string correlators are expected to develop isolated poles at these points in momentum space\footnote{These are not the only poles one expects. We will come back to this later on.}, for which the corresponding residues can be computed perturbatively. The result, written in integral form, was given in Eq.~\eqref{HCFT residues def}.
So far, we have dealt only with the $\m=0$ case. We have described how the corresponding correlators localize according to Eq.~\eqref{eq:localisation solution}, and used the KZ constraints to derive  the function $W_\Gamma$ which was left unfixed in \cite{Eberhardt:2019ywk}. We now argue that the localization argument can be extended to correlators satisfying \eqref{j constraint with m} with any $\m >0$.

Consider the operator
\begin{equation}
    \Oo_2(x,z) \equiv V_{j=k-2,h=1}^{\w=2}(x,z) \, .
\end{equation}
As discussed in Sec.~\ref{sec: string correlators def}, this is nothing but the worldsheet avatar of the twist-two marginal operator which, in the proposal of \cite{Eberhardt:2021vsx}, deforms the holographic CFT away from the symmetric orbifold point. The operator $\Oo_2(x,z)$, used originally in \cite{Eberhardt:2021vsx} and more recently in \cite{Hikida:2023jyc} from a free field perspective, has unit worldsheet and spacetime weights. Furthermore, it can be written as
\begin{equation}
    \Oo_2(x,z) = \frac{1}{4-2k}\of{J^{-}_{-2}V^{2}_{k-2,2}}(x,z)
\end{equation}
Here $V^{2}_{k-2,2}$ is a primary with $\w=2$, $j=k-2$ and $h=2$, hence it has $m=-j$.
The KZ equation for the vertex $V^{2}_{k-2,2}(x,z)$ ensures that
\begin{equation}
    2\of{J^+_1 \cO_2}(x,z)= -\p_zV^{2}_{k-2,2}(x,z) \,.
\end{equation}
On the other hand, we can compute the action of $J^+_2$ over $\cO_2$ using the $\mathfrak{sl}(2,\RR)_k$ algebra as
\begin{equation}
    \of{J^+_2 \cO_2}(x,z) = -V^{2}_{k-2,2}(x,z)\,.
\end{equation}
Looking back at \eqref{JVxOPE}, this implies that the OPE of $\Oo_2$ with the current $J^+(z)$ takes a form
\begin{equation}
    J^+(z)\cO_2(x,w) \sim -\p_w\of{\frac{V^{2}_{k-2,2}(x,w)}{2(z-w)^2}} + \p_x\of{\frac{\cO_2(x,w)}{z-w}}\,.
\end{equation}
In other words, this is regular up to total derivatives with respect to the worldsheet and spacetime insertion points.
The same conclusion holds for the OPEs with $J^3(z)$ and $J^-(z)$, which respectively give
\begin{equation}
    J^3(z)\cO_2(x,w) \sim -\p_w\of{\frac{xV^{2}_{k-2,2}(x,w)}{2(z-w)^2}} + \p_x\of{\frac{x\cO_2(x,w)}{z-w}}\,,
\end{equation}
and
\begin{equation}
    J^-(z)\cO_2(x,w) \sim -\p_w\of{\frac{x^2V^{2}_{k-2,2}(x,w)}{2(z-w)^2}} + \p_x\of{\frac{x^2\cO_2(x,w)}{z-w}}\,.
\end{equation}
This means that $\cO_2$ can be used as a screening operator, i.e.~we can always insert
\begin{equation}
    \int d^2x d^2z \Oo_2(x,z)
\end{equation}
in any correlation function without changing its conformal properties. 
In analogy with the Coulomb gas treatment of Liouville theory, this implies that, for the residues at the poles \eqref{j constraint with m}, the net effect is only to multiply the correlator by an overall factor independent of the quantum numbers of the corresponding vertex operators. In other words, we have 
\begin{align}
\label{F with m extra O2 insertions}
  &\mathop{\mathrm{Res}}_{\sum_{i=1}^n j_i = 1+\frac{k-2}{2}(n-\m-2) }  \vev{\prod_{i=1}^n V_{j_i,h_i}^{\w_i}(x_i,z_i)} = \\
  & \qquad \frac{g^\m}{\m!} \int \prod_{p=1}^\m d^2\chi_p d^2\zeta_p  \mathop{\mathrm{Res}}_{\sum_{i=1}^n j_i + \m(k-2) = 1+\frac{k-2}{2}(n+\m-2) } \vev{\prod_{i=1}^n V_{j_i,h_i}^{\w_i}(x_i,z_i)
   \prod_{p=1}^\m \Oo_2(\chi_p,\zeta_p)} \, , \nn
\end{align}
for some constant $g$, which depends only on $k$, and where we have included a conventional symmetry factor.

This identity has a crucial consequence. If the correlator on the LHS of Eq.~\eqref{F with m extra O2 insertions} satisfies the modified condition \eqref{j constraint with m}, one can easily check that the correlator appearing in the integrand on the RHS of this equation localizes! Indeed, defining a new index $I$ that runs over the $n$ original insertions as well as the $\m$ new insertions, we have
\begin{equation}
    \sum_{I=1}^{n+\m} j_I = \sum_{i=1}^n j_i + \m (k-2) = 1+ \frac{k-2}{2}(n - \m - 2) + \m (k-2) = 1 + \frac{k-2}{2}(n+\m-2)  \, .
\end{equation}
Hence, using our results from the previous sections gives
\begin{equation}
\label{localized solution with O2}
    \vev{\prod_{i=1}^n V_{j_i,h_i}^{\w_i}(x_i,z_i)
   \prod_{p=1}^{\m} \Oo_2(\chi_p,\zeta_p)}   = \sum_{\Gamma} \prod_{I=1}^{n+\m} |a_I|^{-2h_I} \prod_{I=4}^{n+\m} \delta^{(2)}(x_I-\Gamma_I) \, |W_\Gamma|^2 \, ,
\end{equation}
with
\begin{equation}
\label{Final WGamma m 2}
|W_\Gamma|^2 =
    \cN_{n+\m}(j_I,\w_I) \Big|\frac{J_{\off{z_I \to \Gamma_I}}}{z_{12} z_{23} z_{13}}\times C_\Gamma^{\frac{k}{2}}\prod_{a=1}^N c_a^{-\frac{k}{2}} \prod_{I=1}^{n+\m} a_I^{\frac{k}{4}(\w_I-1)}  \prod_{I<J} z_{IJ}^{-q_I q_J}\Big|^2\, .
\end{equation}
Here the sum runs over the discrete set of connected genus-zero  covering maps satisfying not only the conditions \eqref{Gamma conditions at zi} ---with $x_i$ replaced by $\Gamma_i$ for $i \geq 4$--- but also
\begin{equation}
    \Gamma(z \sim \zeta_p) = \tilde{\Gamma}_p + \tilde{a}_p (z-\zeta_p)^2 + \cdots \qquad
    \forall \quad p = 1,\dots,\m \,
\end{equation}
for some coefficients $\tilde{a}_p$ and $\tilde{\Gamma}_p$.
We have also introduced the shorthands $z_I = (z_i,\zeta_p)$, $x_I = (x_i,\chi_p)$, $a_I = (a_i,\tilde{a}_p)$, etc. In particular, when $I>n$ we have $\w_I = 2$, $h_I=1$ and $q_I = \alpha$, see Eq.~\eqref{def alpha}. Finally, the number of poles is now given by
\begin{equation}
   N=1+ \frac{1}{2}\sum_{i=1}^n(\w_i-1) + \frac{\m}{2}\, .
\end{equation}
Furthermore, the normalization constant appearing in Eq.~\eqref{F with m extra O2 insertions} can be fixed by comparing the correlator given by Eq. \eqref{localized solution with O2} with $n=3$ and $\m=1$ with the residue of the associated exact three-point function. This was computed in \cite{Eberhardt:2021vsx}, and we obtain
\begin{equation}
    \label{g factor}
    g = \frac{4(k-2)}{\pi}\frac{\eta_k}{N_k}\nu^{1-\frac{k}{2}}\,.
\end{equation}

The computation of the residues of string correlators for $n$-point functions satisfying \eqref{j constraint with m} is now straightforward. Including the overall normalization, the final expression can be written as
\begin{align}
	\mathop{\mathrm{Res}}_{\sum_{i=1}^n j_i = 1+\frac{k-2}{2}(n-\m-2) }\vvev{\prod_{i=1}^n \cO_{j_i,h_i}^{\w_i}(x_i)} &= C_{S^2} \sum_{\G} \frac{g^\m}{\m!} \, \cN_{n+\m}(j_I,\w_I)\times \\
	&\hspace{-6cm}\int \prod_{p=1}^\m d^2\chi_p \Big|C^{\frac{k}{2}}_\G \prod_{a=1}^N c_a^{-\frac{k}{2}}\prod_{p=1}^\m \tilde{a}_p^{-1+\frac{k}{4}}\prod_{i=1}^n a_i^{-h_i+\frac{k}{4}(\w_i-1)}\prod_{i<j}z_{ij}^{-q_iq_j}\prod_{i,p}(z_i-\zeta_p)^{-q_i\alpha}\prod_{p<r}\zeta_{pr}^{-\alpha^2}\Big|^2\,.\nn
\end{align}
This  holds for any values of $n$, $\w_i$ and $\m$. It also matches the free-field results of \cite{Hikida:2023jyc,Knighton:2023mhq,Knighton:2024qxd}, as well as the holographic predictions of \cite{Eberhardt:2021vsx}, namely Eq.~\eqref{HCFT residues def}, as long as we impose that the relative normalizations between worldsheet and boundary path integrals and vertex operators, as defined in \eqref{HCFT correlator m=0 unfixed}, are given by
\begin{align}
\label{final CS2 and CC}
    C_{S^2} &= {\rm N}\frac{N_k^2}{\eta_k^3}
    \, , \quad
    \cC(j_i,\w_i)= {\rm N}^{1/2}\frac{N_k}{\eta_k}\, \w_i^{2j_i-\frac{1}{2}} \, ,
\end{align}
and relate $g$ with (minus) the boundary coupling constant $\mu$ by means of
\begin{equation}
    \mu = - g \, \cC(k-2,2) = -\frac{2^{2k-\frac{5}{2}}(k-2)}{\pi} \text{N}^{\frac{1}{2}}\nu^{1-\frac{k}{2}} \, .
\end{equation}
Moreover, in \cite{Eberhardt:2021vsx}, an explicit computation of the exact two-point function led to the corresponding normalization constant being $\eta_k = 2(k-2)$. Inserting this value, our result \eqref{final CS2 and CC}  matches their Eq.~(3.2) exactly.

Finally, let us stress that the  analysis carried out in this paper can easily be extended to all singularities related to \eqref{j constraint with m} by replacements of the form $j_i \to 1-j_i$ by means of the worldsheet reflection symmetry of Eq.~\eqref{Reflection flowed}.

\section{Discussion and outlook}
\label{sec: conclusions}

\subsection{Summary of results}

Over the last few years, our understanding of string correlators in AdS$_3$ has sharpened considerably. The results of this paper establish a direct connection between the two main routes that have emerged for studying string correlators in AdS\(_3\): the algebraic analysis of the SL(2,\(\mathbb R\)) WZW model and the free-field description of strings near the boundary. The first of these approaches exploits the symmetry constraints to determine the correlation functions of the SL(2,$\R$) WZW model, the main building block of the worldsheet theory. So far, this has been successful for two- and three-point functions \cite{Dei:2021xgh,Bufalini:2022toj}, which have been derived in exact form. A similar formula has been conjectured for four-point functions \cite{Dei:2021yom}. In this context, holomorphic covering maps play an important role, but, in a sense, not a crucial one: SL(2,$\R$) correlators can be non-zero even for configurations where such maps do not exist.
Extending this analysis to $n$-point functions with arbitrary $n$ is, however, technically challenging.

For the purpose of establishing the AdS$_3$/CFT$_2$ holographic duality, however, it has also proven useful to adopt a different perspective. SL(2,$\R$) correlators depend on many quantum numbers: the worldsheet insertion points $z_i$ and the ones on the boundary, namely $x_i$, the unflowed spins $j_i$, the spacetime weights $h_i$, and the spectral flow charges $\w_i$. From the string perspective, however, some of these are redundant. More precisely, $j_i$ gets related to  $h_i$ and $\w_i$ once the Virasoro condition is imposed, and the worldsheet moduli are to be integrated over. It is conceivable that studying the string path integral directly might lead to exact expressions for the spacetime correlators, at least at genus zero.

The authors of \cite{Knighton:2023mhq,Knighton:2024qxd} build on this idea by focusing on the near-boundary region of the geometry, where the worldsheet theory admits a free-field description in terms of the Wakimoto fields $\phi$, $\gamma$ and $\beta$, and the interaction term can be ignored. String $n$-point functions can then be computed by using the Coulomb gas method. As in the Liouville case, one can only access configurations that satisfy certain charge conservation conditions in this way. This translates into the $j$-constraint in Eq.~\eqref{j constraint with m}. The resulting formulas compute the residues of the exact string amplitudes associated with the corresponding poles in momentum space. Crucially, the path integral over the holomorphic field $\gamma(z)$ only gets contributions from the locus where suitable covering maps exist.

The holographic CFT has long been known to be related to a symmetric orbifold CFT, but its precise definition has proved elusive \cite{Seiberg:1999xz,Eberhardt:2019qcl}. Recent input from the string side has provided important information on complementary aspects of the holographic CFT. The concrete, albeit perturbative proposal of \cite{Eberhardt:2021vsx}, given in Eq.~\eqref{HCFT def}, matches the expected momentum-space residues of string amplitudes, and has been argued to follow from the perspective of long strings \cite{Knighton:2024pqh}. The exact results of \cite{Dei:2021xgh,Dei:2021yom,Bufalini:2022toj} hint at possible non-perturbative completions of the boundary theory.

In this paper, we have constructed a bridge between these two ways of studying strings in AdS$_3$ with NSNS fluxes. The derivation builds on the localization property derived originally in \cite{Eberhardt:2019ywk}: correlators satisfying the $j$-constraint \eqref{j condition intro} take the form in Eq.~\eqref{eq:localisation solution}, where the function $W_\Gamma(z_1,\dots,z_n)$ was left unfixed. In particular, such correlators vanish whenever  the worldsheet and boundary insertion points are  \textit{not} related by a suitable covering map as in $\Gamma(z_i)=x_i$. We presented a streamlined derivation of the fact that the Ansatz \eqref{eq:localisation solution} solves the local Ward identities for any $W_\Gamma$ in Sec.~\ref{sec: localization m=0}. The main result of this paper is that the Knizhnik--Zamolodchikov equations of the SL(2,$\R$) model completely fix the functional form of $W_\Gamma(z_i)$. The solution is given in \eqref{Final WGamma m=0}.
The normalization is subsequently fixed by studying the factorization properties of the resulting worldsheet correlators.
The presence of the Jacobian factor appearing in \eqref{Final WGamma m=0} further simplifies the integration over worldsheet moduli needed to compute the string correlator.

In a second step, we have shown how to extend the analysis to correlation functions satisfying the less restrictive condition \eqref{j condition intro with m}. The procedure mirrors the perturbative expansion of the boundary theory, and the exact results for the residues of string correlators are fully consistent with holographic expectations and previous results.
No free-field description, nor any restriction to the near-boundary region, is needed to derive these results.

\subsection{Future directions}

The results of this paper also suggest a number of  directions for future research, several of which are under current investigation.

\medskip

\noindent \textbf{Derivation in the $m$-basis  and Liouville/$H_3^+$ correspondence.} In this paper we have worked in the so-called $x$-basis, which is the appropriate one to describe operators that are local on the AdS$_3$ boundary. However, we believe that it should be possible to obtain an equivalent result by taking an appropriate limit to the $m$-basis, and making use of the parafermionic decomposition \cite{Maldacena:2001km}, along the lines of \cite{Hikida:2023jyc} (but without resorting to free-field descriptions). This is suggested by the explicit expressions of the main building blocks of the solution for $W_\Gamma$ given in Eq.~\eqref{Final WGamma m=0}: the factors describing the covering map data and the powers of $z_{ij}$ associated with a correlation function computed in the untwisted sector of the boundary theory. Both of these factors can be suitably re-interpreted in $m$-basis language. Roughly speaking, the first one corresponds to a correlator involving exponentials of the free field that bosonizes the Cartan current $J^3(z)$, while the second one takes the form of a Liouville correlator \cite{Giribet:2011xf}. Formalizing this intuition should help clarify the relation between the original
Liouville/$H_3^+$ duality \cite{Teschner:1997ft,Ribault:2005wp,Hikida:2007tq} and the appearance of a Liouville-like sector in the holographic CFT. Note that, in the proposal of \cite{Eberhardt:2021vsx}, this is non-trivially intertwined with the symmetric orbifold structure since the marginal operator that generates the  exponential wall belongs to the twist-two sector.

\medskip

\noindent \textbf{Screening operators in the SL(2,$\R$) model.}
In the $x$-basis free-field computations of \cite{Dei:2023ivl,Knighton:2023mhq,Knighton:2024qxd}, the authors made use of  screening operators which are written in the $m$-basis and belong to the $\w=-1$ sector of the theory. Essentially, these operators encode the singularities of the relevant covering maps, hence the negative values of the spectral flow charge.
Including these operators in the analysis will be crucial in constructing the $m$-basis derivation of the KZ solution mentioned in the previous point. Conversely, the $m$-basis approach could help clarify the relation between these screening operators and those employed much earlier in \cite{Giribet:2000fy,Iguri:2007af,Iguri:2009cf,Giribet:2011xf}.

\medskip

\noindent \textbf{Remaining poles of string correlators.}
The exact expressions available for three- and four-point functions show that the set of momentum-space poles of worldsheet correlators goes beyond Eq.~\eqref{j condition intro with m}. As discussed at the end of Sec.~\ref{sec: string correlators def} in the language of the holographic CFT, this must be extended to the two-parameter set defined by \eqref{charge conservatio two integers}, where both $\m$ and $\n$ are arbitrary non-negative integers.
Interestingly, the two free-field approaches mentioned in the previous point seem complementary: the more recent one allows for the computation of correlators satisfying the charge conservation condition \eqref{charge conservatio two integers} with $\n = 0$, while the more traditional one is best suited for studying the effect of the Wakimoto interaction term, corresponding to spin configurations with $\n >0$ (and $\m=0$). However, extending their results to the full set of poles seems complicated. The methods of \cite{Giribet:2000fy,Iguri:2007af,Iguri:2009cf,Giribet:2011xf} are formulated in the $m$-basis, while the $x$-basis path-integral computation of
\cite{Dei:2023ivl,Knighton:2023mhq,Knighton:2024qxd} breaks down away from the near-boundary regime.
One of the main motivations for the present work was to develop a method for computing string residues  based directly on the symmetry properties of the exact SL(2,$\R$) model, which allows one to treat all the above poles on the same footing, at least in principle. So far, we have shown that the KZ equations allow one to recover all known results for correlators with arbitrary spectral flow charges and all values of $\m$, while keeping $\n=0$. This was done by assuming initially that the constraint \eqref{j condition intro} is valid, and then relaxing it by including insertions of the appropriate screening operator, which turns out to be the worldsheet avatar of the boundary operator $\Oo_\alpha^2$ appearing in the perturbative definition of the holographic CFT.  The extension to cases with $\n >0$ by inserting an appropriate second screening operator is currently under investigation.

\medskip

\noindent \textbf{Exact SL(2,$\R$) $n$-point function and non-perturbative definition of the holographic CFT.}
Gaining control over the full set of poles of string amplitudes in AdS$_3$ might provide valuable insights for extending the analysis of SL(2,$\R$) correlators for generic values of the unflowed spins $j_i$ beyond four-point functions ---and actually proving the conjecture of \cite{Dei:2021yom}---bringing us closer to solving the model. This might also help provide a more precise definition of the boundary theory that goes beyond conformal perturbation theory. Indeed, as can be read off from \eqref{charge conservatio two integers}, the two families of poles are exchanged under $\alpha\to\alpha^{-1}$, reminiscent of the Liouville inversion $b\to 1/b$. In the context of Liouville theory, for instance, invariance under an analogous inversion, combined with the free-field results, ultimately provided enough information to fix the three-point functions exactly. Although this inversion is not an exact symmetry of the  boundary theory considered here, the relation between the two families of poles (together with the existence of the exact SL(2,$\R$) correlators) suggests that a remnant of this mechanism may nevertheless constrain the non-perturbative completion.

\medskip

\noindent \textbf{Extension to higher genera.}  Here we have focused on tree-level correlators for simplicity. These leading contributions are written in terms of genus-zero covering maps. We expect that our derivation can be generalized to higher genus, although the resulting analysis might be technically challenging. The starting point is given by an analogous localization property, which was derived in \cite{Eberhardt:2020akk}, see also \cite{Knighton:2020kuh} for a related discussion in the tensionless limit. Additionally, one could also consider situations where the boundary itself is a higher-genus surface \cite{Eberhardt:2020bgq}.

\medskip

\noindent \textbf{Structural aspects of the HCFT from the worldsheet.} It is a long-standing problem to describe the OPE structure of the holographic CFT directly from the string worldsheet point of view \cite{Kutasov:1999xu,Aharony:2004xn, Aharony:2006th,Aharony:2007fs,Sriprachyakul:2025ubx}. The non-conventional aspects of spectrally flowed representations render the analysis rather intricate in the context of strings in AdS$_3$. The need to integrate over the worldsheet moduli further complicates the story \cite{Maldacena:2001km,Iguri:2024yhb}.  We view the discussion presented in Sec.~\ref{sec: normalization} regarding the spacetime factorization of string amplitudes in AdS$_3$ (or more precisely, of their residues) and the manifest parallels with the OPE structure of symmetric orbifold CFTs as a step in this direction. This might provide important clues regarding how to derive the spacetime OPE when the $j$-constraint is not satisfied. Furthermore, given that, on the boundary, multi-cycle operators can be obtained from suitable collision  limits of single-cycle fields, we hope that controlling such limits from the dual string perspective might help describe multi-particle states in the worldsheet language.

\medskip

\noindent \textbf{Supersymmetric models.}
A natural next step is to adapt our techniques to supersymmetric models such as  AdS$_3\times S^3\times T^4$ or AdS$_3\times S^3\times S^3\times S^1$. The former case is more under control, and was studied previously for instance in \cite{Eberhardt:2021vsx,Iguri:2022pbp,Iguri:2023khc,Barone:2025vww,Sriprachyakul:2024gyl,Yu:2024kxr,Yu:2025qnw}, and more recently in \cite{Knighton:2026wva}. On the holographic side, though, the precise form of the marginal perturbation has not been fleshed out. From the worldsheet perspective, in the RNS formalism the AdS$_3$ sector factorizes into a bosonic SL(2,$\R$) model at level $k=n_5+2$ and a free fermion one with level $-2$ \cite{Giveon:1998ns}, where $n_5$ is the number of NS5-brane sources. Traditionally, correlation functions in this context have been computed by combining the contributions from these two sectors. If one follows the prescription of \cite{Iguri:2023khc}, the results of the present paper can be applied to both SL(2,$\R$) models. However, it was argued recently in \cite{Knighton:2026wva} that working in this factorized way might obscure the analysis, at least for the discussion of the relevant residues of string correlators. A superspace treatment may therefore provide a more natural framework for extending the present analysis to the supersymmetric backgrounds.

\medskip

\noindent \textbf{BTZ black holes and black hole microstates}. The SL(2,$\R$) WZW model plays a prominent role in theories describing string propagation in black hole backgrounds and related geometries such as coherent black hole microstates in the context of the Fuzzball program. Important examples include the two-dimensional Euclidean black hole, i.e.~the cigar theory \cite{Witten:1991yr}, and the BTZ geometry \cite{Banados:1992wn,Banados:1992gq}, as well as the two- and three-charge black holes discussed recently in \cite{Massai:2025nci}.  In all cases, the worldsheet theory can be constructed by taking orbifolds of AdS$_3$ or considering closely related coset models. It would be interesting to exploit our understanding of SL(2,$\R$) correlators to describe how the presence of horizons and/or singularities is encoded in the analytic structure of string amplitudes in these non-trivial geometries. Similar questions can be asked in the context of the  microstate models studied in \cite{Martinec:2017ztd,Martinec:2018nco,Bufalini:2021ndn,Bufalini:2022wyp,Bufalini:2022wzu,Martinec:2025xoy}.

\medskip

\noindent \textbf{Holography beyond AdS.} It is suspected that the notion of holography extends beyond the realm of the AdS/CFT correspondence. The asymptotically flat linear-dilaton regime of NS5-F1 solutions provides a concrete incarnation of these ideas, originally considered in the language of Little String Theory \cite{Giveon:1999tq,Kutasov:2001uf,Israel:2003ry,Georgescu:2024iam}. The boundary theory has recently been reinterpreted in terms of  $T\bar{T}$-deformations of two-dimensional CFTs \cite{Zamolodchikov:2004ce,Cavaglia:2016oda,Smirnov:2016lqw}. More precisely, in the AdS$_3$ context one considers the so-called single-trace $T\bar{T}$ deformations \cite{Giveon:2017nie,Chakraborty:2019mdf,Apolo:2019zai,Chakraborty:2020yka,Apolo:2021wcn,Georgescu:2022iyx,Chakraborty:2023wel}. In the worldsheet description, this translates into a marginal deformation, which is believed to be solvable \cite{Giveon:2017myj,Asrat:2017tzd}. Strings in linear dilaton backgrounds thus provide an ideal framework to sharpen the discussion, for instance by matching observables defined unambiguously on both sides of a non-AdS/non-CFT holographic duality. In particular, recent progress in this direction was  made in the tensionless limit \cite{Dei:2024uyx,Dei:2024sct,Dei:2026bnp}. It would be very interesting to adapt our techniques to the relevant gauged models in order to understand how the residues of string amplitudes behave in the linear dilaton region, and describe what role the covering maps play in this context.

\acknowledgments

The work of the authors is supported by CONICET-Argentina and UBACyT project 20020250400031BA.

\appendix 
\section{Explicit examples of the factorization limit}
\label{app:factorization examples}

In this appendix we illustrate two features of the factorization structure  discussed in Sec.~\ref{sec: normalization}. We discuss a couple of four-point functions for which all relevant covering maps are known, and show that, in the spacetime factorization limit  $x_1\to x_2$, the following situations can occur.  
\begin{enumerate}
\item[(i)] Single-cycle channels (in the language of the holographic CFT) are  characterized by the fact that $x_1 \to x_2$ implies $z_1\to z_2$. Moreover, if the exchanged state has spectral flow charge $\w_r$, the corresponding factorization has degeneracy $\w_r$, in the sense that $\w_r$ covering maps $\Gamma$ generate the same pair\footnote{To be precise, this is the case for four-point functions, where the factorization is in terms of three-point functions, for which the covering maps are unique up to global transformations. For $n$-point functions with $n\geq 5$, one of these maps is not unique, hence it may also happen that a given channel with spectral flow charge $\w_r$ contains more than one pair of factorized maps, each of which comes with a degeneracy factor $\w_r$.} $(\G^{\rm in},\G^{\rm out})$. 

\item[(ii)] There also exist contributions associated with covering maps for which  $z_1$ and $z_2$ remain apart. These were not included in Eq.~\eqref{eq:correlator_factorization} since they correspond to multi-cycle contributions \cite{Maldacena:2001km}, which are, in general, subleading. 
\end{enumerate}
Throughout this appendix we take   
\begin{equation}
(z_2,z_3,z_4)=(0,1,\infty)\,,\qquad (x_2,x_3,x_4)=(0,1,\infty)\,,
\label{eq:app gauge}
\end{equation}
so that $z_1$ is determined by $\G(z_1)=x_1$. The contribution of a given covering map $\Gamma(z)$ to the residue of the four-point function reads $\cN_4^\infty\,|F_\G|^2$, where, according to Eq.~\eqref{eq: correlador localizado integrado}, 
\begin{equation}
F_\G=\prod_{a=1}^{N-\w_4} c_a^{-\frac k2}\prod_{i=1}^4 a_i^{-h_i+\frac k4(\w_i-1)}\prod_{1\le i<j\le 3}z_{ij}^{-q_iq_j}\,,
\label{eq:app F}
\end{equation}
Since one of the operators is inserted at infinity, the actual number of poles is $N-\w_4$, with $N$ defined in \eqref{Ndef}. 
The constant $\cN_4^\infty$ is obtained from \eqref{eq:Nn_solution} by flipping the sign of the term $\frac k2(\w_4+1)$,  as was done for $\cN_3^\infty$ in \eqref{3point normalization infinity}, i.e.~ 
\begin{equation}
    \cN_4^{\infty}=\eta^{-1}_k N^{2}_k \prod_{i=1}^3 \w_i^{-\frac{k}{2}(\w_i+1)+2j_i} \w_4^{\frac{k}{2}(\w_4+1)+2j_4} \, .
    \label{eq:N4inf}
\end{equation}
As discussed in Sec.~\ref{sec: factorization}, in the factorization limit the spin of the exchanged states is given by $j=j_1+j_2+1-\frac k2$, so that $q_r\equiv q(j)=q_1+q_2$, see Eq.~\eqref{qi def}.

\subsection{Four-point function with spectral flow charges (2,2,2,2)}
\label{app:2222}
Let us focus on the correlator with spectral flow charges $(\w_1,\w_2,\w_3,\w_4)=(2,2,2,2)$. There is only a single pole, which we place at $z=\lambda$. Imposing $\G(0)=\p\G(0)=\p\G(z_1)=\p\G(1)=0$, $\G(1)=1$, and $\G(z\to\infty)\sim z^2$ implies that the relevant covering maps must take the form
\begin{equation}
\G(z)=\frac{z^2\big[(2\lambda-1)z+2-3\lambda\big]}{z-\lambda}\,,\qquad
z_1=\frac{\lambda(3\lambda-2)}{2\lambda-1}\,, 
\label{eq:app Gamma 2222}
\end{equation}
such that 
\begin{equation}
\p\G(z)=2(2\lambda-1)\frac{z\,(z-1)(z-z_1)}{(z-\lambda)^2} \, .
\end{equation}
The remaining condition $\G(z_1)=x_1$  reads
\begin{equation}
x_1(2\lambda-1)=\lambda\,(3\lambda-2)^3\,,
\label{eq:app x of lambda}
\end{equation}
such that, for generic values of $x_1$,  the different solutions to this quartic equation define exactly four distinct covering maps. In all cases, the resulting $a_i$ coefficients can be written as 
\begin{equation}
a_1=\frac{(2\lambda-1)(3\lambda-2)(3\lambda-1)}{\lambda(\lambda-1)}\,,\quad
a_2=\frac{3\lambda-2}{\lambda}\,,\quad
a_3=\frac{3\lambda-1}{1-\lambda}\,,\quad
a_4=\frac{1}{1-2\lambda}\,,
\label{eq:app a 2222}
\end{equation}
where $a_4$ follows from $\G(z\to\infty)\sim -a_4^{-1}z^2$. 
The residue of $\G$ at $z=\lambda$ is  
$c=2\lambda^2(\lambda-1)^2$.  Hence, the residue of the string four-point function satisfying the $j$-constraint is given by 
\begin{equation}
    \cN_4^\infty\,\sum_{\lambda}\Big|c^{-\frac k2}\,a_1^{\frac k4-h_1}a_2^{\frac k4-h_2}a_3^{\frac k4-h_3}a_4^{\frac k4-h_4}\,z_1^{-q_1q_2}(z_1-1)^{-q_1q_3}\Big|^2 \, ,
 \label{eq:app 4pt closed 2222}
\end{equation}
where the sum runs over the four solutions of \eqref{eq:app x of lambda}.

In the factorization limit, namely $x_1\to0$, three of these solutions merge at $\lambda=\frac23$, while the fourth one approaches $\lambda=0$. Eq.~\eqref{eq:app Gamma 2222} then shows that, in this example, the limit $x_1\to 0$ implies $z_1\to 0$ in all cases. 
We are interested in the first order expansion around $x_1=0$, where the different values of $z_1$ take the form 
\begin{equation}
z_1^{(s)}=e^{\frac{2\pi is}{3}}\of{4x_1}^{1/3}\,,\qquad s=0,1,2\,,\qquad \mathrm{and}\qquad z_1^{(3)}=\frac{x_1}{4}\,.
\label{eq:app roots 2222}
\end{equation}
We now argue that the first three solutions correspond to a factorization of the $(2,2,2,2)$ covering map into configurations we can schematically denote as $(2,2,3)(3,2,2)$, describing a channel where the exchanged state has spectral flow charge $\w_r=3$. The remaining solution  corresponds to a channel with $\w_r=1$, and will thus give rise to covering maps compatible with the factorized structure $(2,2,1)(1,2,2)$. 
 
\paragraph{Single-cycle channel with  \texorpdfstring{$\w_r=3$}{wr=3}.}
For the three roots $z_1^{(s)}$ with $s=0,1,2$, the single pole  remains at a finite location in the $x_1 \to 0$ limit. In the language of Sec.~\ref{sec: factorization}, this corresponds to a case with $r=0$.
We will obtain two covering maps with three ramification points, located in both cases at $(0,1,\infty)$. The first one, denoted as $\G^{\rm out}$, is associated with $(\w_r,\w_3,\w_4)=(3,2,2)$. It must then have a single pole since $N_{\rm out}-2=1$. The second one, denoted as $\G^{\rm in}$, is associated with $(\w_2,\w_1,\w_r)=(2,2,3)$, and will have no poles since  $N_{\rm in}-3=0$. Both maps are unique up to global conformal transformations.

The outer region is defined by $|z-z_2|\gg|z_{12}|$, with $z_2=0$ in the present example. Keeping $z$ fixed and sending $\lambda\to\frac23$ in \eqref{eq:app Gamma 2222} gives
\begin{equation}
\G^{\rm out}(z)=\frac{z^3}{3z-2}\,,\qquad \p\G^{\rm out}(z)=\frac{6z^2(z-1)}{(3z-2)^2}\,.
\label{eq:app Gamma2 2222}
\end{equation}
The relevant data thus reads 
\begin{equation}
    a_r^{\rm out}=-\frac12\, \qquad 
    a_3^{\rm out}=3\, \qquad 
    a_4^{\rm out}=-3\, \qquad 
    c^{\rm out}=\frac{8}{81}\, .
\end{equation}
The inner region corresponds to $|z-z_2|\sim|z_{12}|\ll 1$. We can zoom into this regime by defining the rescaled variable $y\equiv z/z_1$. We also need to rescale the map itself, dividing it by $x_1$. The inner map is then obtained from \eqref{eq:app Gamma 2222} by taking $x_1,z_1,z \to 0$ while keeping $y$ and $x_1/z_1^3$ fixed, giving
\begin{equation}
\G^{\rm in}(y) = \lim_{z_1 \to 0}
\frac{\G(z_1y)}{x_1} = \lim_{\lambda \to \frac{2}{3}}\frac{y^2(1-\lambda y)} {(2\lambda-1)-(3\lambda-2)\,y} = 3y^2-2y^3\,, \qquad 
\der\G^{\rm in}(y) = -6 y(y-1)\, .
\label{eq:app Gamma1 2222}
\end{equation}
This gives  
\begin{equation}
    a_1^{\rm in}=-3\, \qquad 
    a_2^{\rm in}=3\, \qquad 
    a_r^{\rm in}=-\frac{1}{2}\, .
\end{equation}
Both descriptions must agree in the overlap region, but we have to be careful with the rescalings introduced above. Matching the behavior of $\G^{\rm out}(z)$ at the origin with that of $x_1 \G^{\rm in}(z/z_1)$ at infinity implies 
\begin{equation}
    a_r^{\rm out} = \frac{x_1}{a_r^{\rm in}z_1^3}
\end{equation}
Given that $z_1=z_{12}$ and $x_1=x_{12}$ in this example, we can write this as 
\begin{equation}
x_{12}=z_{12}^{3}a_r^{\rm in}a_r^{\rm out}=\tfrac14\,z_{12}^3\,,
\label{eq:x21_relation 2222 example}
\end{equation}
which is nothing but a specific instance of  Eq.~\eqref{eq:x21_relation}. 
Summarizing, we have started from three distinct  four-point maps, and shown that they produce the same pair ($\G^{\rm in},\G^{\rm out}$). At leading order they are related by the $\mathbb{Z}_3$ action $z_{12}\to e^{2\pi i/3}z_{12}$, which does not affect \eqref{eq:x21_relation 2222 example}.
 
Let us discuss the leading contributions to the four-point function. From the point of view of the original map \eqref{eq:app a 2222},
the $\lambda \to \frac{2}{3}$ limit gives 
\begin{equation}
a_1\sim-\tfrac34z_1\,,\qquad a_2\sim \tfrac34z_1\,,\qquad a_3\to3\,,\qquad a_4\to-3\,,\qquad c\to\tfrac{8}{81}\,.
\end{equation}
Further taking $|z_{13}|=|z_1-1|\to 1$ shows that, at first order, each map gives 
\begin{equation}
|F_{\G^{(s)}}|^2\simeq\of{\tfrac{8}{81}}^{-k}\of{\tfrac34}^{2(\frac k2-h_1-h_2)}3^{2(\frac k2-h_3-h_4)}\Big|z_1^{(s)}\Big|^{2(\frac k2-h_1-h_2-q_1q_2)}\;\,,
\end{equation}
with $s=0,1,2$. The different phases of $z_1^{(s)}$ are clearly irrelevant, hence all three maps give the same contribution in the factorization limit. As expected, this can be recovered from the inner and outer maps derived above. Applying a formula analogous to Eq.~\eqref{eq:app F} to these maps gives
\begin{equation}
|F_{\G^{\rm in}}|^2=3^{2(\frac k2-h_1-h_2)}\,2^{2(h_r-\frac k2)}\,,\qquad
|F_{\G^{\rm out}}|^2=\of{\tfrac{8}{81}}^{-k}\,2^{2(h_r-\frac k2)}\,3^{2(\frac k2-h_3-h_4)}\,,
\end{equation}
Finally, using that $|z_1^{(s)}|^3\simeq 4|x_1|$ and inserting the quantum numbers of the exchanged state indeed shows that, at first order,  
\begin{equation}
|F_{\G^{(s)}}|^2=|x_{12}|^{2(h_r-h_1-h_2)}\,|F_{\G^{\rm in}}|^2\,|F_{\G^{\rm out}}|^2\,,
\label{eq:app per map 2222}
\end{equation}
provided 
\begin{equation}
\tfrac k2-h_1-h_2-q_1q_2=3\,(h_r-h_1-h_2)\,,
\end{equation}
which follows from the respective Virasoro conditions \eqref{eq:hr_definition} and the chosen value of the intermediate spin $j$. 

Taking the factorization limits thus requires us to include  a degeneracy factor $\w_r=3$.
We now show that this is consistent with the normalization factors computed in Sec.~\ref{sec: factorization}. From Eqs.~\eqref{3point normalization infinity} and \eqref{eq:N4inf} we have 
\begin{equation}
\cN_4^\infty=3^{-2}\eta_k^{-1}\;\cN_3^\infty\!\big(j_2,j_1,1-j;2,2,3\big)\;\cN_3^\infty\!\big(j,j_3,j_4;3,2,2\big)\,.
\end{equation}
Hence summing \eqref{eq:app per map 2222} over the three original maps gives
\begin{equation}
\sum_{s=0}^{2}\cN_4^\infty|F_{\G^{(s)}}|^2\simeq
3 \times  |x_{12}|^{2(h_r-h_1-h_2)} \frac{
\vvev{\cO^{2}_{j_2}(0)\cO^{2}_{j_1}(1)\cO^{3}_{1-j}(\infty)}\vvev{\cO^{3}_{j}(0)\cO^{2}_{j_3}(1)\cO^{2}_{j_4}(\infty)}}{9\,\eta_k}\,,
\end{equation}
which is nothing but \eqref{eq:correlator_factorization} with $\hat\cN_r=1$ provided the two-point function takes the form given in \eqref{spacestime 2pt}, namely $\cN_2^\infty=\w_r\eta_k=3\eta_k$. 
 
\paragraph{Single-cycle channel with  \texorpdfstring{$\w_r=1$}{wr=1}.}
For the fourth root, $\lambda \simeq \frac{x_1}{8}$ and $z_1\simeq \frac{x_1}{4}$ at first order, hence the original pole approaches $z_2=0$ together with $z_1$. In the language of Sec.~\ref{sec: factorization}, this corresponds to a case with $r=1$. Keeping $z$ fixed while sending $\lambda\to0$ in \eqref{eq:app Gamma 2222}, we find that the pole cancels against one of the zeros at $z=0$, giving 
\begin{equation}
\G^{\rm out}(z)=z(2-z)\,,\qquad \der \G^{\rm out}(z) = 2(1-z)\,,
\label{eq:app Gamma2 wr1}
\end{equation}
which has twists $(\w_r,\w_3,\w_4)=(1,2,2)$ at $(0,1,\infty)$, and 
\begin{equation}
    a_r^{\rm out}=2\, \qquad 
    a_3^{\rm out}=-1\, \qquad 
    a_4^{\rm out}=1\, .
\end{equation}
In the inner region, we take 
\begin{equation}
\G^{\rm in}(y) = \lim_{z_1 \to 0}
\frac{\G(z_1y)}{x_1}= \frac{y^2}{2y-1}\,, \qquad 
\der\G^{\rm in}(y) = \frac{2y(y-1)}{(2y-1)^2}\, ,
\label{eq:app Gamma1 2222b}
\end{equation}
which has twists $(\w_2,\w_1,\w_r)=(2,2,1)$ at $(0,1,\infty)$, and a single pole at $y=\frac{1}{2}$. This leads to 
\begin{equation}
    a_1^{\rm in}=1\, \qquad 
    a_2^{\rm in}=-1\, \qquad 
    a_r^{\rm in}=2\, \qquad 
    c^{\rm in}=\frac{1}{8} \, .
\end{equation}
A consistent overlap gives  $x_{12}=a_r^{\rm out}a_r^{\rm in}z_{12}$, i.e.~$x_1= 4 z_{1}$, which agrees with  \eqref{eq:app roots 2222}. 
In the factorization limit, from \eqref{eq:app a 2222} we have
\begin{equation}
    a_1\simeq \frac{4}{z_1} \, \qquad
    a_2\simeq -\frac{4}{z_1} \, \qquad
    a_3\to -1 \, \qquad
    a_4\to 1 \, \qquad
    c\simeq \frac{z_1^2}{2} \, .
\end{equation}
so that
\begin{equation}
|F_{\G^{(3)}}|^2\simeq 2^{3k-4h_1-4h_2} |z_1|^{2h_1+2h_2-3k-2q_1q_2}\,.
\end{equation}
This is easily seen to coincide with 
\begin{equation}
    |x_{12}|^{2(h_r-h_1-h_2)}\,|F_{\G^{\rm in}}|^2\,|F_{\G^{\rm out}}|^2
\end{equation}
using the explicit expressions of $a_i^{\rm out}$, $a_i^{\rm in}$ and $c^{\rm in}$, together with  $|x_{12}| = |x_1|=4 |z_1|$ and the relation 
\begin{equation}
 h_1+h_2-\frac{3k}{2}-q_1q_2=h_r-h_1-h_2\,,
\end{equation}
which follows from the Virasoro conditions. 
In this case there is no degeneracy, hence 
\begin{equation}
\cN_4^\infty=\eta_k^{-1}\;\cN_3^\infty\!\big(j_2,j_1,1-j;2,2,1\big)\;\cN_3^\infty\!\big(j,j_3,j_4;1,2,2\big)\,,
\end{equation}
leads to 
\begin{equation}
\cN_4^\infty|F_{\G^{(3)}}|^2\simeq
 |x_{12}|^{2(h_r-h_1-h_2)} \frac{
\vvev{\cO^{2}_{j_2}(0)\cO^{2}_{j_1}(1)\cO^{1}_{1-j}(\infty)}\vvev{\cO^{1}_{j}(0)\cO^{2}_{j_3}(1)\cO^{2}_{j_4}(\infty)}}{\eta_k}\,,
\end{equation}
which reproduces \eqref{eq:correlator_factorization} with $\hat\cN_r=1$ provided $\cN_2^\infty=\w_r\eta_k=\eta_k$, as in \eqref{spacestime 2pt}. 

\subsection{ Four-point function with spectral flow charges (2,2,2,4)}
\label{app:(2,2,2,4)}

Let us now briefly discuss the correlator with spectral flow charges $(\w_1,\w_2,\w_3,\w_4)=(2,2,2,4)$. This has no poles since   $N-4=0$. 
Imposing $\G(0)=0$, $\p\G(0)=0$, $\G(1)=1$, $\p\G(1)=0$, $\p\G(z_1)=0$ and $\G(z\to \infty) \sim z^4$ implies that the relevant maps must take the form 
\begin{equation}
\G(z)=\frac{3z^4-4(1+z_1)z^3+6z_1z^2}{2z_1-1}\,,\qquad
\p\G(z)=\frac{12\,z\,(z-1)(z-z_1)}{2z_1-1}\,.
\label{eq:app Gamma 2224}
\end{equation}
The remaining condition $\G(z_1)=x_1$ then reads
\begin{equation}
x_1(2z_1-1)=z_1^3\,(2-z_1)\,.
\label{eq:app x of z1}
\end{equation}
This is a quartic equation for $z_1$, so for generic values of $x_1$ there are, once again,  exactly four covering maps. In all cases,  the relevant coefficients read 
\begin{equation}
a_1=\frac{6z_1(z_1-1)}{2z_1-1}\,,\qquad
a_2=\frac{6z_1}{2z_1-1}\,,\qquad
a_3=\frac{6(1-z_1)}{2z_1-1}\,,\qquad
a_4=-\frac{2z_1-1}{3}\,.
\label{eq:app a 2224}
\end{equation}
The string residue can thus be written as 
\begin{equation}
\vvev{\cO_1(x_1)\cO_2(0)\cO_3(1)\cO_4(\infty)}=
\cN_4^\infty\!\!\sum_{z_1}\Big|a_1^{\frac k4-h_1}a_2^{\frac k4-h_2}a_3^{\frac k4-h_3}a_4^{\frac{3k}4-h_4}\,z_1^{-q_1q_2}(z_1-1)^{-q_1q_3}\Big|^2 ,
\label{eq:app 4pt closed}
\end{equation}
where the sum runs over the different  solutions of \eqref{eq:app x of z1}.

In the factorization limit $x_1 \to 0$, one of the solutions of Eq.~\eqref{eq:app x of z1} approaches $z_1=2$, while the other three all approach   $z_1=0$. 
At first order, these four solutions read 
\begin{equation}
   z_1^{(s)} = e^{\frac{2\pi i s}{3}}\of{-\frac{x_1}{2} }^{1/3}\, , \qquad s=0,1,2\, , \qquad \mathrm{and} \qquad z_1^{(3)}= 2-\frac{3}{8}x_1 \, .
\end{equation}
While the first three cases give $z_{12}=z_1\to 0$, the fourth one does \textit{not}. 
On the one hand, the first three solutions correspond to a factorization of the map with twists $(2,2,2,4)$ into a product of the form $(2,2,3)(3,2,4)$, i.e.~with $\w_r=3$. This channel works analogously to that of the previous section, so we will not discuss it further. On the other hand, naively the remaining solution would  correspond to $\w_r=1$, but no three-point map with twists $(1,2,4)$ exists. We will argue it corresponds to a double-cycle channel. 

\paragraph{The non-colliding branch.}
\label{app:noncolliding}
The isolated solution has $x_1\to 0=x_2$ while $z_1\to2\neq z_2$. Here the worldsheet does not degenerate: the covering map converges to a regular degree-four cover,
\begin{equation}
\G^*(z)=\lim_{z_1\to2}\G(z)=z^2(z-2)^2\,,
\label{eq:app Gamma star}
\end{equation}
which has two second-order ramification points, one at $z=0$ and the other one at $z=2$. We also get $\Gamma^*(0) = \Gamma^*(2) = x_2$. The monodromy around $x_2$ is of the type $(2)(2)$, i.e.~it corresponds to the product of two disjoint twist-2 operators. Moreover, at leading order we have 
\begin{equation}
a_1\to4\,,\qquad a_2\to4\,,\qquad a_3\to-2\,,\qquad a_4\to-1\,,\qquad z_{12}\to2\,,\qquad z_{13}\to1\,.
\end{equation}
Hence, up to subleading terms proportional to  positive powers of $|x_1|$,  $|F_\G|^2$ becomes
\begin{equation}
|F_{\G^*}|^2 = 4^{2(\frac k2-h_1-h_2)}\,2^{2(\frac k4-h_3)}\,2^{-2q_1q_2}\,.
\end{equation}
This contribution is regular as $x_1\to x_2$. In other words, it scales as $|x_{12}|^{2(h_{12}-h_1-h_2)}$ with $h_{12}=h_1+h_2$. This is the behavior expected for the contribution associated with a double-cycle operator $:\!\cO^{\w_1}_{j_1}\cO^{\w_2}_{j_2}\!:$ appearing in the spacetime OPE. We expect that, once the correct normalization is taken into account, $\cN^\infty_4|F_{\G^*}|^2$ gives the covering-space expression for $\vvev{:\!\cO_1\cO_2\!:(0)\,\cO_3(1)\cO_4(\infty)}$ computed with the map \eqref{eq:app Gamma star}. 

Except in some pathological cases (see for instance \cite{Pakman:2009ab}), such double-cycle contributions are, in general, subleading. This is the reason why we have chosen not to include them in the discussion around Eq.~\eqref{eq:correlator_factorization} in Sec.~\ref{sec: factorization}.

\bibliographystyle{JHEP}
\bibliography{refs}

\end{document}